\documentclass[aps,floatfix,nofootinbib,twocolumn]{revtex4}

\usepackage{amsmath,amssymb}

\usepackage{mathrsfs}

\usepackage{epsfig,psfrag}
\usepackage{array,dcolumn}

\begin{document}
\newcommand{\mrs}{\mathscr}

\newcommand{\vecb}[1]{\mbox{\boldmath$#1$}}

\title{ Nucleon-nucleon interaction in the $J$-matrix inverse
scattering approach and few-nucleon systems}
\author{A. M. Shirokov}\email{shirokov@nucl-th.sinp.msu.ru}
\address{ Skobeltsyn Institute of Nuclear Physics, Moscow State University,
Moscow, 119992, Russia}
\author{A. I. Mazur} \email{mazur@hpicnit.khstu.ru}    
\author{ S. A. Zaytsev}\email{zaytsev@fizika.khstu.ru}
\address{Physics Department, Khabarovsk State Technical University,
Tikhookeanskaya 136, Khabarovsk 680035, Russia}
\author{J. P. Vary}\email{jvary@iastate.edu} 
\author{T. A. Weber}\email{taweber@iastate.edu}
\address{
Department of Physics and Astronomy,
Iowa State University, Ames, Ia 50011-3160, USA}
\begin{abstract}\centerline{\parbox{.92\textwidth}{\hspace{10pt}
The nucleon-nucleon interaction is constructed by means of the
$J$-matrix version of inverse scattering theory.
Ambiguities
of the interaction
are
eliminated by postulating
tridiagonal and quasi-tridiagonal forms of the potential matrix in the
oscillator basis in uncoupled and
coupled waves, respectively. The obtained interaction is very accurate in
reproducing the $NN$ scattering  data and deuteron properties. The
interaction is used in the no-core shell model calculations of $^3$H and
$^4$He nuclei. 
The resulting binding energies of $^3$H and $^4$He are very
close to experimental values.
}}
\end{abstract}
\maketitle

\section{Introduction}

The  nucleon-nucleon ($NN$) interaction is one of the most important
ingredients in  microscopic nuclear structure studies. 
Ideally,
the $NN$ interaction should be derived from the quark-gluon
theory. The modern status of QCD, however, makes it possible to predict
reaction cross sections only at high enough energies; 
the QCD-based
derivation of the $NN$ potential describing the nucleon-nucleon
interaction at the low energies that are of primary importance for
nuclear physics  
applications is impossible at the moment.

Nucleon-nucleon potentials
conventionally referred to as
`{\em realistic}', are derived from the  meson exchange theory. Modern
realistic $NN$ potentials
like Bonn~\cite{Bonn}, 
Argonne~\cite{Argonne},
Nijmegen~\cite{Stocks}, etc., are carefully
fitted to the existing experimental data on  $NN$ scattering and
deuteron properties.
Unfortunately, none of the known $NN$
interactions provides a
completely
satisfactory description of the trinucleon and
other light nuclei. To overcome this deficiency, meson exchange
\cite{TM} or 
phenomenological~\cite{Argonne-3} three-nucleon forces 
are usually introduced.
Impressive
progress was achieved recently in the description of
the trinucleon and $^4$He binding energies with realistic $NN$ and
three-nucleon  forces \cite{alpha-3}. However, the three-nucleon
force parameters in such studies are sometimes fitted to the trinucleon
binding and some of them
may not be consistent
with the
parameters of the two-body interaction.
In one very detailed study, when
the three-body interaction
parameters
were
chosen consistently with the two-body
parameters,
the three-nucleon force contribution to the
triton binding energy was shown to be negligible
\cite{trinucl}.

Impressive progress using effective field theory has recently been reported
(see review in Ref.~\cite{vanKolck}).  
The versions that provide the most accurate fit to the
nucleon-nucleon properties~\cite{Refs:Machleidt} use a momentum-space
cutoff and are still quite strong at short distances. The match with the
nuclear many body model space cutoff is unclear; 
additional renormalization is required for typical model spaces that
are feasible. We aim in 
this paper to have
high quality descriptions of the phase shifts with softer potentials whose
cutoff is well-matched to the anticipated application in many-body systems.

Various microscopic models have been designed for the studies of
few-body systems. It was demonstrated in Ref.~\cite{Bench}  that all
modern
realistic
microscopic models provide approximately the same results for
the $^4$He ground state.  The
no-core shell model~\cite{Vary,Vary3}
is one of
these models.
This model can be used not only
for the few-body nuclear applications
but also, with  modern computer
facilities,  for
microscopic studies  of heavier nuclei with the number of nucleons  $A$ up to
$A\sim 12$ \cite{Vary3}. The no-core shell
model is based
on a
wave function expansion in
a
many-body oscillator function series with the aim
to describe bound states and narrow resonances treated as bound states.
%

The oscillator basis matrices of the modern realistic $NN$ potentials
are very large and cannot be directly used without a severe truncation
in the many-body no-core shell model calculations. As a result, the
convergence of the calculations appears to be slow. This
deficiency is conventionally addressed by
constructing
the so-called {\em effective $NN$ interaction} (see, e.~g.,
\cite{Vary}). Ideally the effective $NN$ interaction should reproduce
in the finite model space the results of the infinite model space
calculation. In a realistic application, the construction of the
effective $NN$ interaction is a complicated problem involving various
approximations.

In this contribution, we construct the $NN$ interaction by means of
the $J$-matrix version of
inverse scattering theory
\cite{Ztmf1,Zret,Zret2}. The matrix of the $NN$ potential in the
oscillator basis is obtained
for each partial wave independently.
Therefore, in our approach we
derive the $NN$ interaction as a set of potential matrices for
different partial waves.
We
reproduce the experimental $NN$ scattering data and
deuteron properties with small
potential matrices. Our $NN$ interaction can be treated as
an
effective interaction since its
matrix can be directly used in the no-core shell model calculations
without additional truncation. However, our effective $NN$ interaction
reproduces
the energy spectrum and other
observables in a many-body system as well as
deuteron
properties and $NN$ scattering data.
 From this point of view, our $NN$
interaction can be treated as a realistic one as well. Our interaction is not
related to the meson exchange theory, however we shall see that we
obtain the deuteron and scattering wave functions that are very close
to the ones obtained with
realistic meson exchange potentials.

The potential derived by the $J$-matrix inverse scattering approach is
ambiguous. The ambiguity originates from the phase-equivalent
transformation suggested in Ref.~\cite{PHT} (see also \cite{Halo-b,Halo-c}
and references therein). The ambiguity is eliminated in the present
approach by a phenomenological ansatz that the potential matrix in the
uncoupled partial waves is tridiagonal. Therefore our potentials are
{\em Inverse Scattering Tridiagonal
Potentials} (ISTP). The non-central nature of the
$NN$ interaction 
is manifested
%
in the coupling of some partial waves, and the
tridiagonal potential ansatz should be extended to allow for the
coupling of these partial  waves.
We postulate phenomenologically the
simplest generalization of the tridiagonal form of the
potential matrix in this case;
however, we refer to our potentials as ISTP in the cases of both uncoupled
and coupled partial waves (though, strictly speaking, it is not
correct in the later case).
It is just the tridiagonal ansatz that brings us
to the scattering wave functions which are very close to the ones
provided by the  meson exchange realistic $NN$ potentials. However, in
the case of the coupled $sd$ waves we perform a phase equivalent
potential transformation to improve the description of the deuteron
properties.

The suggested ISTP are used in the no-core shell model calculations of
$^3$H and $^4$He.
We shall see that the predicted $^3$H and
$^4$He binding energies are very close to the experimental values. We
do not use three-nucleon interactions, yet
our predictions of the $^3$H and $^4$He bindings
are approximately of the same accuracy as the predictions based
on the best realistic meson exchange two-nucleon
plus
three-nucleon forces.

Here we would like to mention some recent papers where other
approaches to the problem of constructing high-quality effective
interaction were utilized. The authors of Refs.~\cite{Dole,Plessas}
added phenomenological non-local terms to  a cut-off
Yukawa tail of the realistic $NN$
potentials. The obtained interaction reproduces the $^3$H binding
energy. The additional non-local terms do not reduce the rank of the
potential energy matrix in the oscillator basis of the underlying
realistic $NN$ interaction.  Therefore the use of this
interaction in the shell model studies requires the construction of
the shell model effective interaction. 

A very interesting approach is the construction of
the  low momentum $NN$
potential $V_{low-k}$ 
from the realistic $NN$ interactions (see the review in 
Ref.~\cite{Kuo-rev}). 
The use of
$V_{low-k}$ in the shell model applications still requires  the construction of
the shell model effective interaction but this problem is 
simplified. The effective   interaction obtained from $V_{low-k}$ was
used  successfully in various shell model applications (see, e.~g.,
\cite{Kuo2}). Unfortunately it is still unclear whether this
interaction provides the correct binding of three-body and four body
nuclear systems. Contrary to $V_{low-k}$, our ISTP is designed for the
direct use 
in the shell model applications. 

The paper is organized as follows. In the next Section we present the
single channel $J$-matrix inverse scattering approach, derive ISTP in
the uncoupled partial waves and discuss their properties. The
derivation and discussion of the ISTP properties in the coupled
partial waves can be
found in the Section~III. The results of the  $^3$H and $^4$He
calculations are presented in the Section~IV. A short summary of the
results can be found in Section~V.

\section{Single channel $J$-matrix inverse scattering approach and
ISTP in uncoupled $NN$ partial waves}

The $J$-matrix formalism in the quantum scattering theory was
initially proposed in atomic physics \cite{Yamani}. Within the
$J$-matrix formalism, the continuum spectrum wave function 
 is expanded in an infinite series of $L^2$ functions. This approach was
shown  to be one of
the most efficient and precise methods in calculations of
photoionization \cite{Broad,St-Uzh,St-izv} and electron scattering by atoms
\cite{Konov}. In nuclear physics the same approach has been developed
independently \cite{Fil,NeSm} as the method of the harmonic oscillator
representation of scattering theory. This method  has been
successfully used in various  nuclear applications allowing for the
 two-body continuum, e.~g. 
nucleus-nucleus scattering  has been studied in the algebraic
version of RGM based on  the $J$-matrix formalism  (see the review papers
\cite{FVCh,Rev}); the effect of $\Lambda$ and neutron decay channels in
hypernuclei production reactions has been investigated in Refs.
\cite{M1,M2}, etc. The approach was extended to the case of true
few-body scattering in \cite{SmSh} and utilized in the studies of 
the monopole excitations of the $^{12}$C nucleus in the $3\alpha$
cluster model in Ref.~\cite{Mikh}.  It was also used in the studies of
double-$\Lambda$ hypernuclei in Ref.\cite{Zeit} and of weakly bound nuclei in
the three-body cluster model in Refs.~\cite{PHT,Halo-b,Halo-c}.

The $J$-matrix version of the inverse scattering theory was suggested
in Refs.~\cite{Ztmf1,Zret,Zret2}. The discussion of the  general
formalism
below follows the ideas of  Refs.~\cite{Ztmf1,Zret,Zret2}, however,
some formulas
are presented here in a manner that should
be more convenient for the current application. The tridiagonalization
of the interaction obtained by the inverse scattering methods
have not previously
been discussed in the literature,
hence the
corresponding
theory and results are new.

The oscillator-basis $J$-matrix formalism is discussed in detail elsewhere
(see, e.~g., \cite{Yamani,Bang}). We present here only some relations
needed for understanding  the inverse scattering $J$-matrix
approach.

The 
Schr\"odinger equation  in the partial wave with 
orbital angular momentum $l$  reads
 \begin{equation}
  H^l \,\Psi_{lm}(E,\vecb{r})=E\,\Psi_{lm}(E,\vecb{r}).
  \label{eq:Sh}
  \end{equation}
The wave function is given by
\begin{equation}
\Psi_{lm}(E,\vecb{r})=\frac1r\,u_{l}(E,r)\,Y_{lm}(\hat{\vecb{r}}) ,
\end{equation}
where $Y_{lm}(\hat{\vecb{r}})$ is the spherical function.
Within the  $J$-matrix formalism, the radial wave function
$u_{l}(E,r)$ is expanded in an oscillator function  series
  \begin{equation}
  u_{l}(E,r) = \sum_{n=0}^{\infty} a_{nl}(E)\, R_{nl}(r),
  \label{eq:row}
  \end{equation}
where
\begin{multline}
 R_{nl}(r)=(-1)^n\
\sqrt{ \frac{2 n!}{r_0 
\,\Gamma(n +l+3/2) } }
               \left(\frac{r}{r_0}\right)^{l+1}\\
\times\exp\!{\left(-\frac {r^2}{2r_0^2}\right)}\:
        L_n^{l+\frac12}\!\left(\frac {r^2}{r_0^2}\right),
\label{e24}
\end{multline}
$L^{\alpha}_n(x)$ is the associated Laguerre polynomial, the
oscillator radius  $r_{0}=\sqrt{\hbar / m \omega}$, and $m$ is the reduced
mass. All energies are given in the units of the oscillator basis
parameter  $\hbar\omega$.

The wave function in the oscillator representation $a_{nl}(E)$
is a solution of the infinite set of
algebraic equations
  \begin{equation}
  \sum_{n'=0}^{\infty}\,(H_{nn'}^{l}-\delta_{nn'}E)\,
  a_{n'l}(E)=0,
  \label{eq:Infsys}
  \end{equation}
where the Hamiltonian matrix elements
$H_{nn'}^l =T_{nn'}^l +V_{nn'}^{l}$, the kinetic energy  matrix elements
\begin{subequations}
\label{Ekin}
\begin{align}
T^{l}_{n, \, n-1} &= - \frac12 \sqrt{n(n+l+1/2)},
\label{Ekin-a} \\[1.5ex]
T^{l}_{n,\, n}&= \frac12\,(2n+l + 3/2),
\label{Ekin-b}\\[1.5ex] 
T^{l}_{n, \, n+1}&=- \frac12 \sqrt{(n+1)(n+l+3/2)},
\label{Ekin-c}
\end{align}
\end{subequations}
and the potential energy $V^l$ within the $J$-matrix formalism is
approximated by the truncated matrix with elements
  \begin{equation}
    \widetilde{V}_{nn'}^l =
\left\{
    \begin{array}{lllcl}
           V_{nn'}^l \ \ \ \ \ &\text{if} \ \ & n\ &\text{and}\ & n'\leq N;\\
          \ 0 &\text{if} & n &\text{or} & n'>N .
     \end{array} \right.
  \label{trunc}
  \end{equation}
In the inverse scattering $J$-matrix approach, the potential energy is
constructed in the form of the finite matrix of the type
(\ref{trunc}); therefore the $J$-matrix solutions with such
an interaction are exact.
%

In the {\em external part of the model space} spanned by the functions
(\ref{e24}) with $n\geq N$, Eq.~(\ref{eq:Infsys}) takes the form of a
three-term recurrence relation
  \begin{multline}
  T_{n,\,n-1}^l \,a_{n-1,\,l}(E) +
  (T_{nn}^l - E )\,a_{nl}(E) \\
+ T_{n,\,n+1}^l \,a_{n+1,\,l}(E) = 0 .
  \label{eq:TRS}
  \end{multline}
Any solution of Eq.~(\ref{eq:TRS}) is a superposition of the
fundamental regular $S_{nl}(E)$ and irregular $C_{nl}(E)$
solutions~\cite{Yamani,Bang}, 
\begin{gather}
\label{aN1-delta}
a_{nl}(E)=\cos\delta(E)\;S_{nl}(E)+\sin\delta(E)\;C_{nl}(E),
\end{gather}
where
\begin{equation}
  S_{nl}(E) = 
  \sqrt{ \frac{\pi\, r_{0}\,n!} {\Gamma (n+l+3/2)}}\;
q^{l+1}\, \exp \left(-\frac{q^2}{2}\right) \,
   L_n^{l+\frac12}(q^2),
  \label{eq:Snl}
\end{equation}
\begin{multline}
  C_{nl}(E) = (-1)^l \,
  \sqrt{ \frac{\pi\, r_{0}\,n!}{\Gamma (n+l+3/2)} }\;
  \frac{q^{-l}}{\Gamma (-l+1/2)} \\
 \times
\exp \left(-\frac{q^2}{2} \right)
\Phi (-n-l-1/2,\, -l+1/2;\, q^2),
  \label{eq:Cnl}
  \end{multline}
$\Phi (a,\, b;\, z)$ is a confluent hypergeometric function~\cite{Abram},
$q=\sqrt{2E}$, and
$\delta(E)$ is the scattering phase shift.

The wave function in the oscillator
representation $a_{nl}(E)$
in the {\em internal part of the model space} spanned by the functions
(\ref{e24}) with $n\leq N$, can be expressed through the external
solution  $a_{N+1,l}(E)$:
  \begin{equation}
  a_{nl}(E) = {\mrs G}_{nN}\;T^{l}_{N,\,N+1}\:a_{N+1,\, l}(E) .
  \label{cnin}
  \end{equation}
The matrix elements,
  \begin{equation}
  {\mrs G}_{nn'} = -\sum_{\lambda ^{\prime}=0}^{N}
  \frac{ \langle n|\lambda ^{\prime}\rangle
    \langle\lambda ^{\prime} |n'\rangle }
  { E_{\lambda ^{\prime}}-E } ,
  \label{oscrm}
  \end{equation}
are expressed through the eigenvalues $E_\lambda$ and eigenvectors
$\langle n|\lambda\rangle$ of the truncated Hamiltonian matrix, i.~e.
$E_\lambda$ and $\langle n|\lambda\rangle$ are obtained by solving the
algebraic problem
\begin{equation}
\sum_{n'=0}^N H_{nn'}^l \langle n'|\lambda\rangle
= E_\lambda\langle n|\lambda\rangle,
\qquad n\leq N.
\label{Alge}
\end{equation}
The matrix element ${\mrs G}_{NN}$
is of primary importance in the
calculation of the phase shift  $\delta(E)$:
  \begin{gather}
  \label{osctg}
\tan \delta (E) =  - \frac{ S_{Nl}(E)-{\mrs G}_{NN}\, T^{l}_{N,\,N+1}\,
S_{N+1,\,l}(E) }
  { C_{Nl}(E)-{\mrs G}_{NN}\, T^{l}_{N,\,N+1}\, C_{N+1,\,l}(E) } .
  \end{gather}

In the 
direct  $J$-matrix approach, we first solve
Eq.~(\ref{Alge}) and next calculate the phase shift  $\delta(E)$ by
means of Eq.~(\ref{osctg}). In the inverse scattering $J$-matrix
approach, the phase shift  $\delta(E)$ is
taken
to be known at any
energy $E$ and,
instead of solving (\ref{Alge}),
we
extract the eigenvalues
$E_\lambda$ and the eigenvectors $\langle n|\lambda\rangle$  from this
information.

First we
assign some value to $N$, the rank of the desired potential
matrix [see Eq.~(\ref{trunc})]. Generally, with a finite rank potential
matrix  it is possible to reproduce the phase shift $\delta(E)$
only in
a
finite energy interval;
larger $N$ supports a larger energy interval.
However, from the point
of view of many-body applications, it is desirable to have $N$ as
small as possible.

The components $a_{nl}(E)$ of the wave function in the oscillator
representation, should be finite at arbitrary energy $E$. This is seen
from Eqs.~(\ref{cnin})--(\ref{oscrm}) to be
possible at the energies $E=E_\lambda$, $\lambda=0,$ 1, ...~, $N$
only if
\begin{gather}
a_{N+1,\,l}(E_\lambda)=0.
\label{Elam}
\end{gather}
Knowing the phase shift, we can calculate
$a_{N+1,\,l}(E)$ at any energy $E$ using Eq.~(\ref{aN1-delta}).
Therefore we can solve numerically the transcendental equation~(\ref{Elam})
and find the eigenvalues $E_\lambda$, $\lambda=0,$ 1, ...~, $N$.

Due to Eq.~(\ref{Elam}),
\begin{gather}
\label{a-Talor}
a_{N+1,\,l}(E)\mathop{\longrightarrow}_{E\to E_\lambda}
   \alpha^{\lambda}_l (E-E_\lambda),
\intertext{where}
\label{alpha}
 \alpha^{\lambda}_l=\left.\frac{d\,a_{N+1,\,l}(E)}{d\,E}\right|_{E=E_\lambda}.
\end{gather}
Now it is easy to derive from Eqs.~(\ref{cnin})--(\ref{oscrm}) the
following equation:
\begin{gather}
\label{Eq:eigen}
a_{Nl}(E_\lambda)=\left|\langle N|\lambda\rangle\right|^2
           \alpha_l^{\lambda}\;T_{N,\,N+1}^l,
\intertext{or, equivalently,}
\label{Nlabda}
|\langle N|\lambda\rangle|^2=
\frac{a_{Nl}(E_\lambda)}{\alpha_l^{\lambda}\;T_{N,\,N+1}^l}.
\end{gather}
Within the $J$-matrix formalism, both $a_{Nl}(E)$ and $a_{N+1,\,l}(E)$
fit  Eq.~(\ref{aN1-delta}) and can be calculated using this equation
at any energy $E$.
Hence, one can also calculate~$\alpha_l^{\lambda}$ by means of
Eq.~(\ref{alpha}). Therefore the components $\langle N|\lambda\rangle$
can be obtained from Eq.~(\ref{Nlabda}) (the sign of the components
$\langle N|\lambda\rangle$ is of no importance).

Equations   (\ref {Elam}) and    (\ref{Nlabda}) provide the general
solution of the $J$-matrix inverse scattering problem: solving
these equations we obtain the sets of $E_\lambda$ and
$\langle N|\lambda\rangle$, and these quantities completely determine
the phase shifts $\delta(E)$. However $\langle N|\lambda\rangle$ are
supposed to be 
the components of the eigenvectors $\langle n|\lambda\rangle$ of the
truncated Hamiltonian matrix [see Eq.~(\ref{Alge})] that should fit the
completeness relation 
\begin{gather}
\label{complete}
\sum_{\lambda = 0}^N \langle n|\lambda\rangle\langle\lambda|n'\rangle
 = \delta _{nn'} ,
 \end{gather}
hence we should have 
\begin{gather}
\label{completeN}
\sum_{\lambda = 0}^N \langle N|\lambda\rangle\langle\lambda|N\rangle
 =1 .
 \end{gather}
Generally the set of $\langle N|\lambda\rangle$ obtained by means of
Eq.~(\ref{Nlabda})  violates the completeness
 relation~(\ref{completeN}). Therefore this set of
 $\langle N|\lambda\rangle$ ideally describing the phase shifts, cannot
 be treated as the set of last components of the normalized eigenvectors
 $\langle n|\lambda\rangle$ of any truncated Hermitian Hamiltonian
 matrix; in  other words, the set of $\langle N|\lambda\rangle$
 violating Eq.~(\ref{completeN}) cannot be used to construct  a
 Hermitian Hamiltonian matrix. 

To overcome this difficulty, we fit Eq.~(\ref{completeN}) by changing
the value of the component  $\langle N|\lambda=N\rangle$ corresponding
to the highest eigenvalue $E_{\lambda=N}$. This modification spoils
the description of the phase shifts $\delta(E)$ at energies $E$
different from $E_\lambda$, $\lambda=0$, 1,~... , $N$. We  restore the
phase shift  description 
in the energy interval $[0,E_{\lambda=N-1}]$ by 
variation of
$E_{\lambda=N}$. From the above consideration it is clear that larger 
$N$ values make it possible to reproduce phase shifts in larger
energy intervals   $[0,E_{\lambda=N-1}]$. 


There is an
ambiguity in determining 
the potential matrix describing the given phase shifts $\delta(E)$: 
any of the phase equivalent transformations
discussed in Refs.~\cite{PHT,Halo-b,Halo-c} [see also
Eqs.~(\ref{Unitary})--(\ref{PHTV}) below]
that do not change the
truncated Hamiltonian eigenvalues $E_\lambda$ and respective eigenvector
components~$\langle N|\lambda\rangle$, results in a potential matrix
that brings us to 
the same phase shifts $\delta(E)$ at any energy~$E$. Additional model
assumptions are needed to resolve this ambiguity. As was already
mentioned,
we assume
the tridiagonal form of
the potential matrix.
We now discuss
the construction
of the tridiagonal potential matrix supposing  $N$ and the sets of
$E_\lambda$ and $\langle N|\lambda\rangle$ to be known.

If the potential matrix is tridiagonal, the equations~(\ref{Alge})
can be rewritten as
\begin{subequations}
 \label{Syst1}
\begin{align}
& H^l_{00}\langle 0|\lambda\rangle +H^l_{01}\langle 1|\lambda\rangle
         =E_\lambda\langle 0|\lambda\rangle,
 \label{Syst1a}
\\[1.5ex] 
& H^l_{n,\,n-1} \langle n-1|\lambda\rangle
+H^l_{nn} \langle n|\lambda\rangle
+H^l_{n,\,n+1} \langle n+1|\lambda\rangle \notag \\
&\hspace{1.2cm}=E_\lambda\langle n|\lambda\rangle
\qquad\quad(n=1,\,2,\,\ldots\, ,\, N-1),
 \label{Syst1b}
\\[1.5ex] 
& H^l_{N,\,N-1}\langle N-1|\lambda\rangle +
H^l_{NN}\langle N|\lambda\rangle =E_\lambda\langle
N|\lambda\rangle.
 \label{Syst1c}
\end{align}
\end{subequations}
The unknown quantities in Eq.~(\ref{Syst1c})  are
the component $\langle N-1|\lambda\rangle$ and the Hamiltonian matrix
elements $H^l_{N,\,N-1}$ and $H^l_{NN}$. We multiply
Eq.~(\ref{Syst1c}) by $\langle\lambda|N\rangle$, sum the result over
$\lambda$, and use the 
completeness relation~(\ref{complete})
to obtain the formula for the calculation of $H^l_{NN}$:
\begin{equation}
H^l_{NN} = \sum_{\lambda =0}^N E_{\lambda}
\langle N|\lambda\rangle ^2 .
\label{hNN}
\end{equation}

The Hermitian conjugate of Eq.~(\ref{Syst1c}) reads:
\begin{gather}
\langle\lambda| N-1\rangle
H^l_{N,\,N-1} +
\langle\lambda| N\rangle H^l_{NN}
=\langle\lambda|N\rangle E_\lambda .
 \label{Syst1c-c}
\tag{\ref{Syst1c}-c}
\end{gather}
We multiply Eq.~(\ref{Syst1c}) by Eq.~(\ref{Syst1c-c}), sum the result over
$\lambda$, and use the completeness relation (\ref{complete}) to obtain
the following expression for the calculation of~$H^l_{N,\,N-1}$:
\begin{equation}
H^l_{N,\,N-1} = -\,\sqrt{\displaystyle\sum_{\lambda =0}^N E_{\lambda}^2
\langle N|\lambda\rangle ^2 -
\left({H^l_{NN}}\right)^2 }.
\label{hN1N}
\end{equation}
Generally, the sign in the right-hand-side of Eq.~(\ref{hN1N}) is
arbitrary. Here we use an additional assumption
that the
off-diagonal Hamiltonian matrix elements $H^l_{n,\,n\pm 1}$ are
dominated by the kinetic energy
so that
the sign of these matrix
elements is the same as
the kinetic energy matrix elements
$T^l_{n\,n\pm 1}$ [see Eqs.~(\ref{Ekin})]. This assumption brings us to
the minus sign in the right-hand-side of Eq.~(\ref{hN1N}).

Now equation (\ref{Syst1c}) can be used to calculate the last unknown
quantity,
\begin{gather}
\langle N-1|\lambda\rangle = \frac{1}{\,  
H^l_{N,\,N-1}\, }
\left(  E_{\lambda} \langle N|\lambda\rangle -
 H^l_{NN}\langle N|\lambda\rangle \right).
\label{gN1N}
\end{gather}

We now
turn to equation (\ref{Syst1b})
with
$n=N-1$. This equation
contains one more  term~than  Eq.~(\ref{Syst1c}), however this term
does not include unknown quantities. We perform with
Eq.~(\ref{Syst1b}) exactly the same manipulations to obtain
expressions for 
$H^l_{N-1,\,N-1}$,
$H^l_{N-2,\,N-1}$ and  $\langle N-2|\lambda\rangle$.
Setting $n=N-2$ in Eq.~(\ref{Syst1b}),
we obtain the expressions for 
$H^l_{N-2,\,N-2}$,
$H^l_{N-3,\,N-2}$ and  $\langle N-3|\lambda\rangle$,
etc.
Equation~(\ref{Syst1a}) is needed only to calculate the last matrix
element $H^l_{00}$.
As a result,
we obtain the following
generalization of Eq.~(\ref{hNN}) valid at $n=N$, $N-1$, ...~,~0:
\begin{gather}
H^l_{nn}=
\sum_{\lambda =0}^N E_{\lambda} \langle n|\lambda\rangle ^2 .
\label{hii} 
\end{gather}
The equations
\begin{equation}
H^l_{n,\,n-1}=
-\sqrt{\displaystyle\sum_{\lambda =0}^N
E_{\lambda}^2 \langle n|\lambda\rangle ^2 -
\left(H^l_{nn}\right)^2 -\left(H^l_{n,\,n+1}\right)^2} 
\label{hi1i} 
 \end{equation}
and
 \begin{multline}
\langle n-1|\lambda\rangle = \frac{1}{
\, H^l_{n,\,n-1} \, }
\left( \sum_{\lambda =0}^N E_{\lambda} \langle n|\lambda\rangle 
\right.\\
\left.\vphantom{\sum_{\lambda =0}^N} -
 H^l_{nn} \langle n|\lambda\rangle  -
H^l_{n,\,n+1} \langle n+1|\lambda\rangle \right)
\label{gi1i}
\end{multline}
are valid at $n=N-1$, $N-2$,...~,~1. Equations
(\ref{hN1N})--(\ref{gi1i}) make it possible to calculate all unknown
quantities. After calculating the Hamiltonian matrix elements
$H^l_{nn'}$, we derive the ISTP 
matrix elements by the obvious equations
\begin{subequations}
\label{pot1ch}
\begin{align}
&V^l_{nn}=H^l_{nn}-T^l_{nn},
\label{pot1ch-a}\\[1.5ex]
&V^l_{n,\,n\pm 1}=H^l_{n,\,n\pm 1}-T^l_{n,\,n\pm 1}.
\label{pot1ch-b}
\end{align}
\end{subequations}

The above theory is used to construct the $NN$ ISTP matrix elements in
uncoupled partial waves. We use as input the $np$ scattering phase
shifts reconstructed from the experimental data by the Nijmegen group
\cite{Stocks}. The oscillator basis parameter
$\hbar\omega=40$~MeV. Usually in the shell model calculations, the
complete $\varkappa\hbar\omega$ model space is used, i.~e. all
many-body oscillator
basis states (configurations) with $\sum_i\varkappa_i\leq\varkappa$ where
the 
single-particle state
oscillator quanta $\varkappa_i=2n_i+l_i$,
are included 
in the calculation.
Thus, to be applicable to all $p$-shell nuclei in accessible model
spaces, we
suggest
the $8\hbar\omega$ and $7\hbar\omega$  ISTP, i.~e. the rank of the
ISTP matrix $N$ is
chosen so that $2N+l=8$ in the partial waves with even orbital angular
momentum $l$  and $2N+l=7$ in the partial waves with odd orbital angular
momentum~$l$.

The non-zero matrix elements of the obtained ISTP in uncoupled partial
waves are presented in Tables~\ref{pot1s0}--\ref{pot3f3} (in
$\hbar\omega=40$~MeV units).

\extrarowheight=3pt
\begin{table}
\caption{Non-zero matrix elements in $\hbar\omega$ units  of the
$8\hbar\omega$ ISTP matrix in the $^1s_0$ partial wave.}
\begin{ruledtabular}
\begin{tabular}{>{$}c<{$}>{$}c<{$}>{$}c<{$}} 
n & V_{nn}^l  & \mbox{$V_{n,\,n+1}^l=V_{n+1,\,n}^l$}  \\ \hline
 0& -0.3706925910512869 & 0.1340546812405571\\
 1& -0.1599160886224698 & 0.01647436916961609\\
 2&  0.1395932055925835 & -0.1334461921366397 \\
 3&   0.2668242073073204 &  -0.07869019612934602\\
 4&  0.0414909332158313 & \\ 
\end{tabular}
\label{pot1s0}

\caption{Non-zero matrix elements in $\hbar\omega$ units  of the
$7\hbar\omega$ ISTP matrix in the $^1p_1$ partial wave.}
\begin{tabular}{>{$}c<{$}>{$}c<{$}>{$}c<{$}} 
 n & V_{nn}^l  & V_{n,\,n+1}^l=V_{n+1,\,n}^l   \\ \hline
 0&  0.10619936477245400  &  -0.09441150969281098 \\
 1&  0.32183202739863150  &  -0.19861423056402480 \\
 2&  0.38227890301930240  &  -0.12529300192170380 \\
 3&  0.08818666274780007  &  \\ 
\end{tabular}\label{pot1p1}

 \caption{Non-zero matrix elements in $\hbar\omega$ units  of the
$8\hbar\omega$ ISTP matrix in the $^1d_2$ partial wave.}
\begin{tabular}{>{$}c<{$}>{$}c<{$}>{$}c<{$}} 
 n & V_{nn}^l  & V_{n,\,n+1}^l=V_{n+1,\,n}^l   \\ \hline
 0  & -0.04182464628865646 &  0.03831247883572108 \\
 1  & -0.1129604626451339  &  0.06873518464832284 \\
 2  & -0.1276115098155470  &  0.04042212068310880 \\
 3  & -0.02554669840505408 &    \\ 
\end{tabular}\label{pot1d2}
\end{ruledtabular}
\end{table}

\begin{table}
\caption{Non-zero matrix elements in $\hbar\omega$ units  of the
$7\hbar\omega$ ISTP matrix in the $^1f_3$ partial wave.}
\begin{ruledtabular}
\begin{tabular}{>{$}c<{$}>{$}c<{$}>{$}c<{$}} 
 n & V_{nn}^l  & V_{n,\,n+1}^l=V_{n+1,\,n}^l   \\ \hline
 0&    0.04238710037363047 & -0.02790556099208952 \\
 1&   0.07474001110555983 & -0.02815383549650696  \\
 2&   0.02511618088977574 &  \\ 
\end{tabular}\label{pot1f3}

\caption{Non-zero matrix elements in $\hbar\omega$ units  of the
$7\hbar\omega$ ISTP matrix in the $^3p_0$ partial wave.}
\begin{tabular}{>{$}c<{$}>{$}c<{$}>{$}c<{$}} 
 n & V_{nn}^l  & V_{n,\,n+1}^l=V_{n+1,\,n}^l   \\ \hline
 0&  -0.13674752057396140 &  0.01511502604716686 \\
 1&  0.08786870226069166  & -0.10590497118041760 \\
 2&  0.23624887864971810  & -0.08040102075340183 \\
 3&   0.04909915603358606  & \\ 
\end{tabular}\label{pot3p0}

 \caption{Non-zero matrix elements in $\hbar\omega$ units  of the
$7\hbar\omega$ ISTP matrix in the $^3p_1$ partial wave.}
\begin{tabular}{>{$}c<{$}>{$}c<{$}>{$}c<{$}}
 n & V_{nn}^l  & V_{n,\,n+1}^l=V_{n+1,\,n}^l   \\ \hline
 0&  0.08893328127606703  & -0.09288011075116410 \\
 1&  0.33899943058663640  & -0.21111518227376910 \\
 2&  0.36158649481733020  & -0.09828565221995666 \\
 3&  0.05167268571103811  &  
\end{tabular}\label{pot3p1}

\caption{Non-zero matrix elements in $\hbar\omega$ units  of the
$8\hbar\omega$ ISTP matrix in the $^3d_2$ partial wave. }
\begin{tabular}{>{$}c<{$}>{$}c<{$}>{$}c<{$}} 
 n & V_{nn}^l  & V_{n,\,n+1}^l=V_{n+1,\,n}^l   \\ \hline
 0  & -0.20024057805517500 &  0.1193321938724192 \\
 1  & -0.28898789873267020 &  0.1463047726434774 \\
 2  & -0.25522202901437920 &  0.0792277802117809 \\
 3  & -0.05421394437761595 &    \\
\end{tabular}
\label{pot3d2}
\end{ruledtabular}
\end{table}

\begin{table}
 \caption{Non-zero matrix elements in $\hbar\omega$ units  of the
$7\hbar\omega$ ISTP matrix in the $^3f_3$ partial wave.}
\begin{ruledtabular}
\begin{tabular}{>{$}c<{$}>{$}c<{$}>{$}c<{$}} 
 n & V_{nn}^l  & V_{n,\,n+1}^l=V_{n+1,\,n}^l   \\ \hline
 0&   0.02629214811765302 & -0.01394097030190910 \\
 1&   0.03463672270715756 & -0.01259217885072639 \\
 2&  0.01119624135204766 & \\ 
\end{tabular}\label{pot3f3}
\end{ruledtabular}
\end{table}

\begin{figure}
\centerline{\psfig{figure=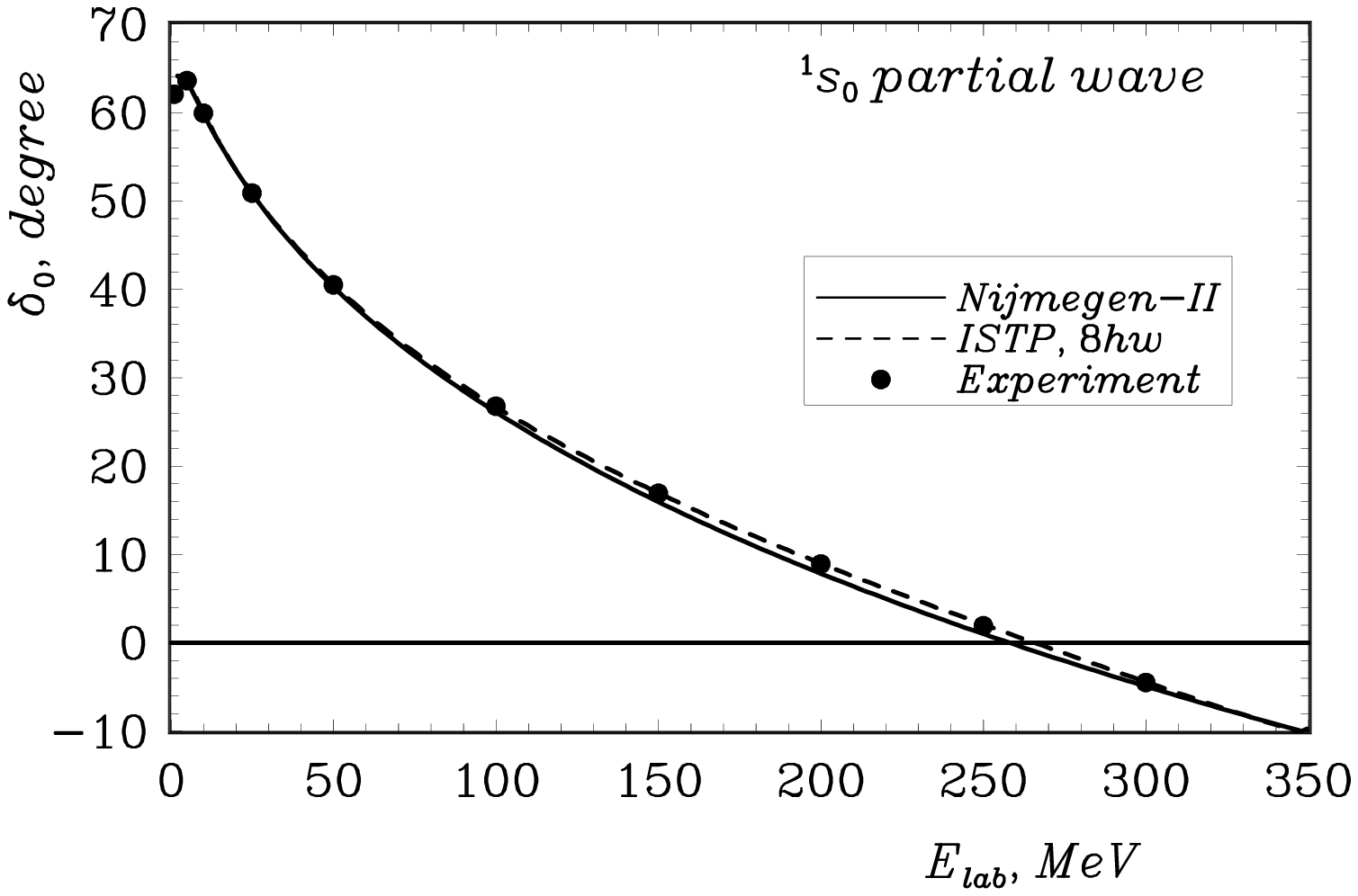,width=0.5\textwidth}}
\caption{ $^1s_0$ $np$ scattering phase shifts. Filled circles ---
experimental data of Ref.~\cite{Stocks}; solid line --- realistic
meson exchange Nijmegen-II potential~\cite{Stocks} phase shifts;
dashed line~--- ISTP phase shifts.} \label{ph1s0}
\vspace{3ex}
\centerline{\psfig{figure=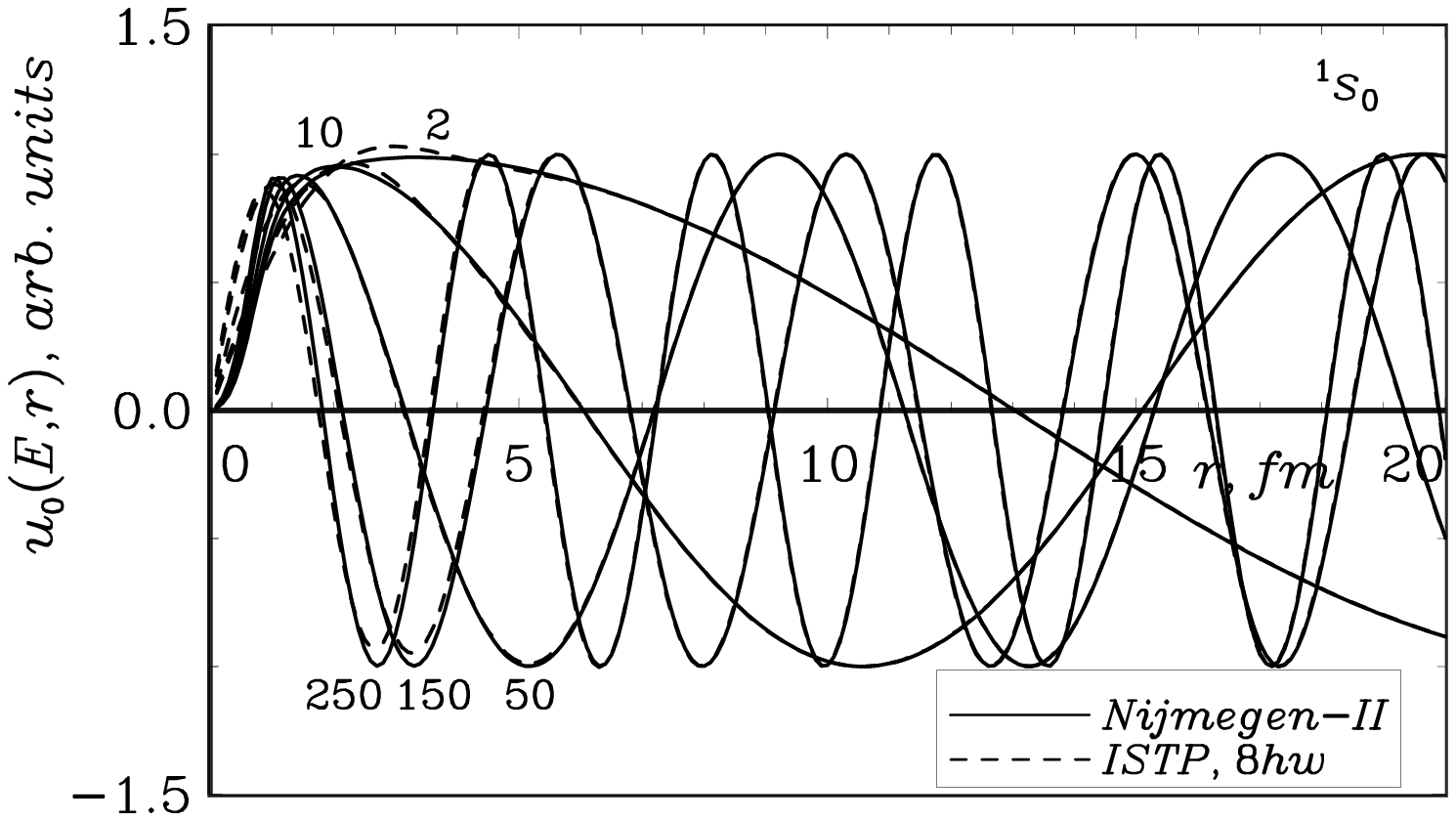,width=0.5\textwidth}}
\caption{ $^1s_0$ $np$ scattering wave functions
at the laboratory energies
$E_{\rm lab}= 2$, 10, 50, 150, 250~MeV. Solid line --- realistic
meson exchange Nijmegen-II potential~\cite{Stocks} wave functions;
dashed line~--- ISTP wave functions.} \label{wf1s0}
\end{figure}

\begin{figure}
\centerline{\psfig{figure=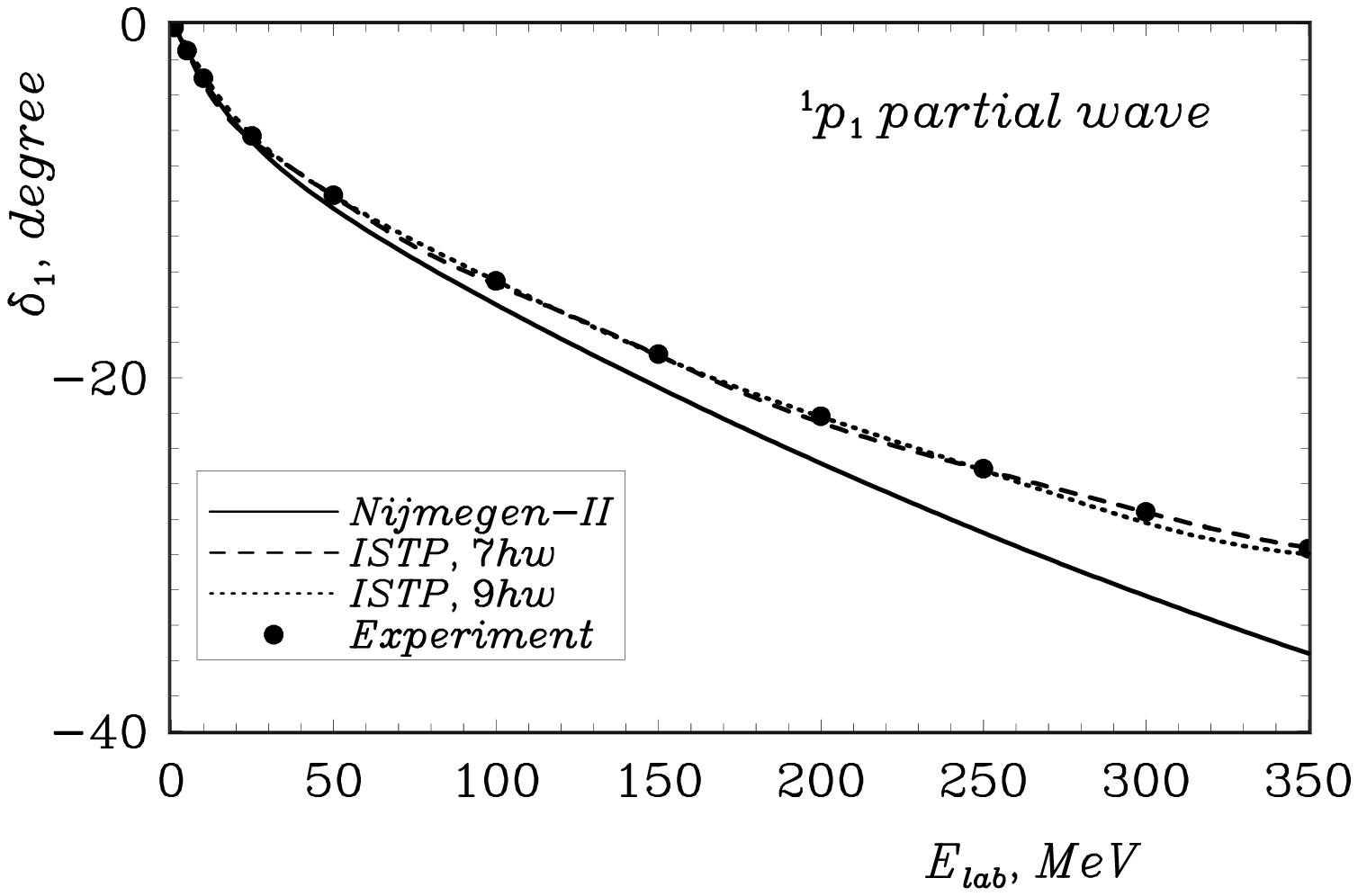,width=0.5\textwidth}}
\caption{$^1p_1$ $np$ scattering  phase shifts. Filled circles ---
experimental data of Ref.~\cite{Stocks}; solid line --- realistic
meson exchange Nijmegen-II potential~\cite{Stocks} phase shifts;
dashed line~--- $7\hbar\omega$ ISTP phase shifts; dotted line~---
$9\hbar\omega$ ISTP phase shifts.}
\label{ph1p1}
\vspace{3ex}
\centerline{\psfig{figure=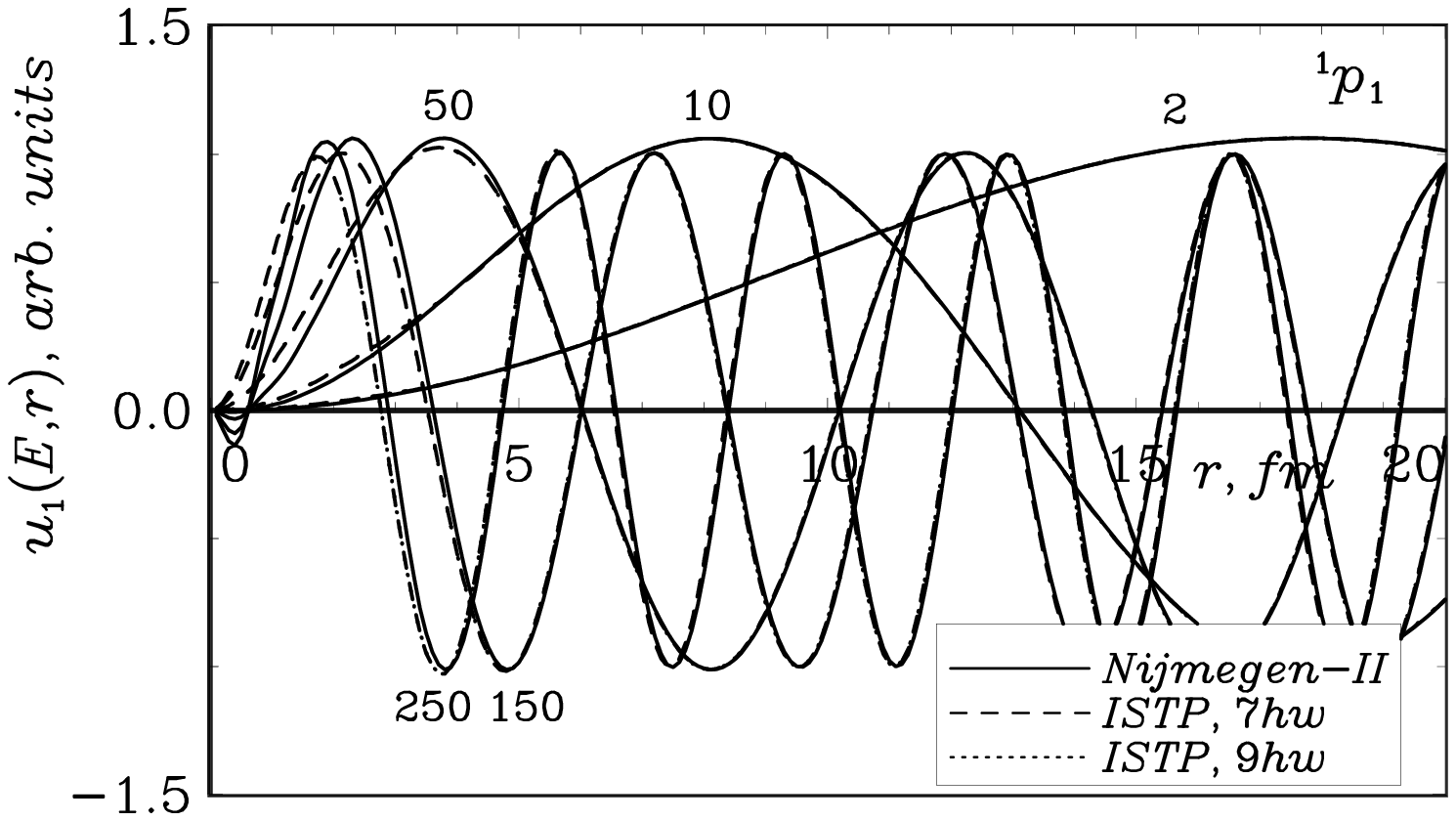,width=0.5\textwidth}}
\caption{$^1p_1$ $np$ scattering wave functions
at the laboratory energies
$E_{\rm lab}= 2$, 10, 50, 150, 250~MeV. Solid line --- realistic
meson exchange Nijmegen-II potential~\cite{Stocks} wave functions;
dashed line~---  $7\hbar\omega$ ISTP wave functions;  dotted line~---
$9\hbar\omega$ ISTP wave functions.}
\label{wf1p1}
\end{figure}

\begin{figure}
\centerline{\psfig{figure=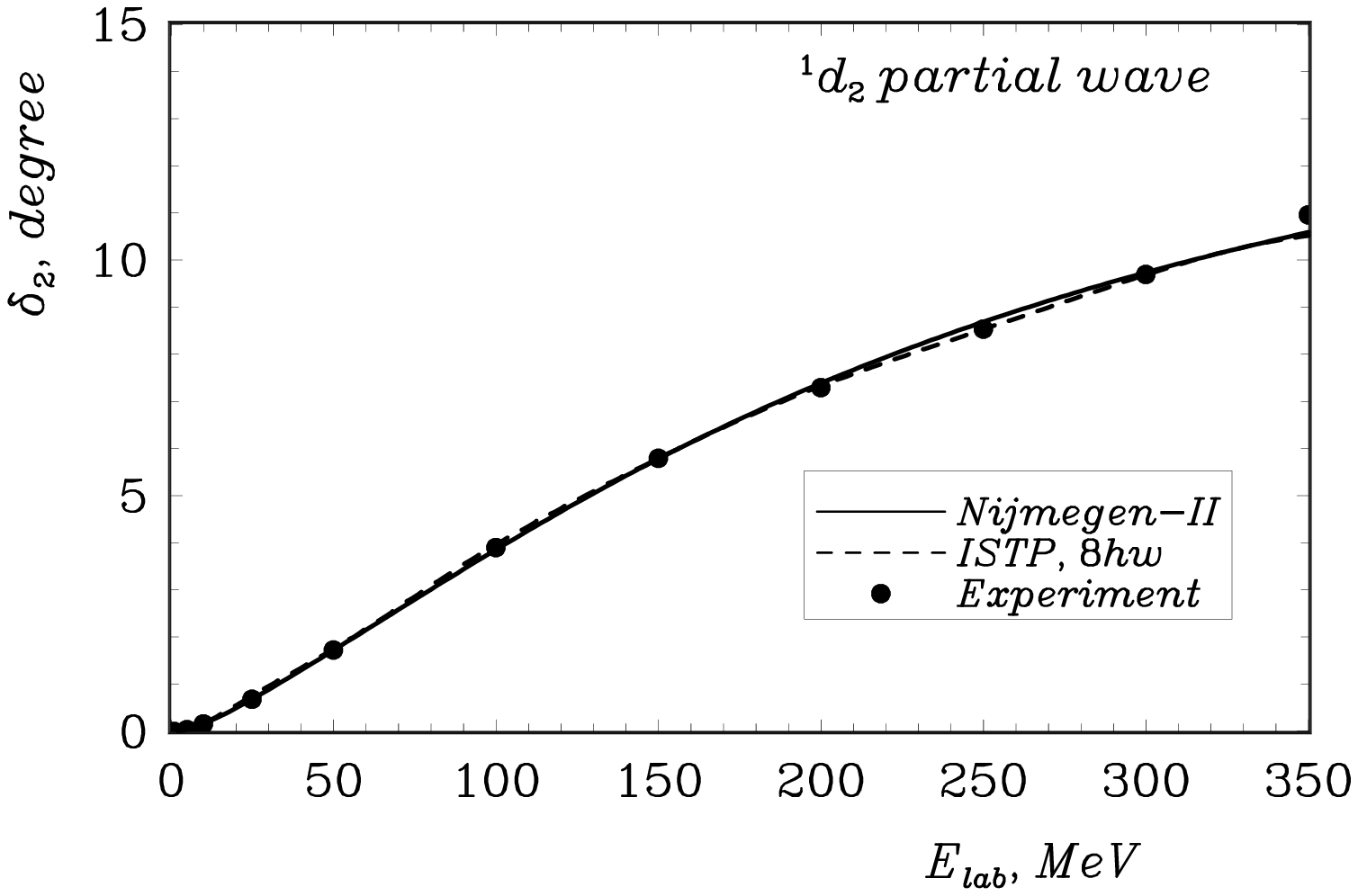,width=0.5\textwidth}}
\caption{$^1d_2$ $np$ scattering phase shifts.
See  Fig.~\ref{ph1s0} for details.} \label{ph1d2}
\vspace{3ex}
\centerline{\psfig{figure=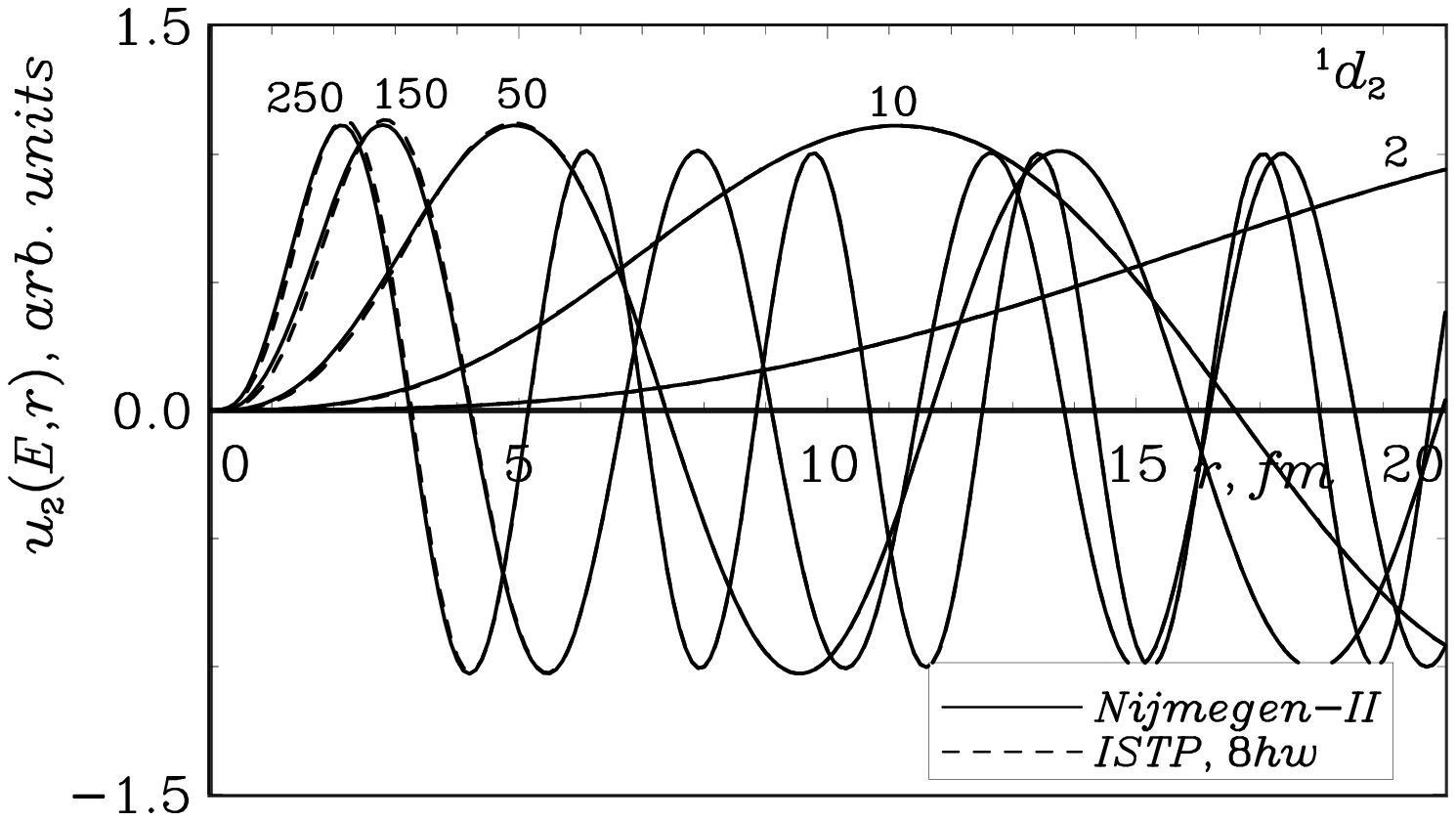,width=0.5\textwidth}}
\caption{$^1d_2$ $np$ scattering wave functions
at the laboratory energies
$E_{\rm lab}= 2$, 10, 50, 150, 250~MeV. See Fig.~\ref{wf1s0} for details.}
\label{wf1d2}
\end{figure}

\begin{figure}
\centerline{\psfig{figure=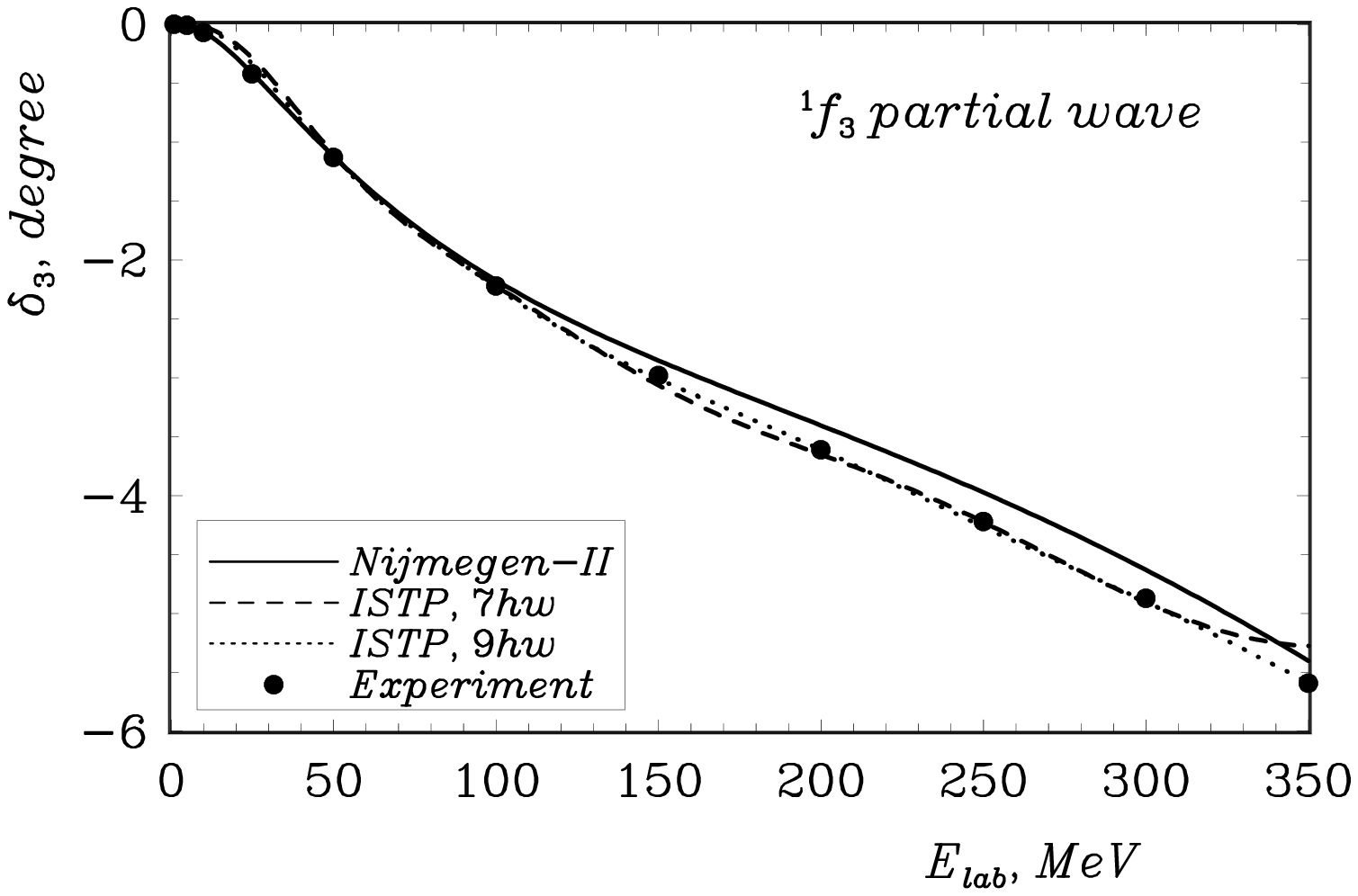,width=0.5\textwidth}}
\caption{$^1f_3$  $np$ scattering phase shifts.
See Fig.~\ref{ph1p1} for details. }
\label{ph1f3}
\vspace{3ex}
\centerline{\psfig{figure=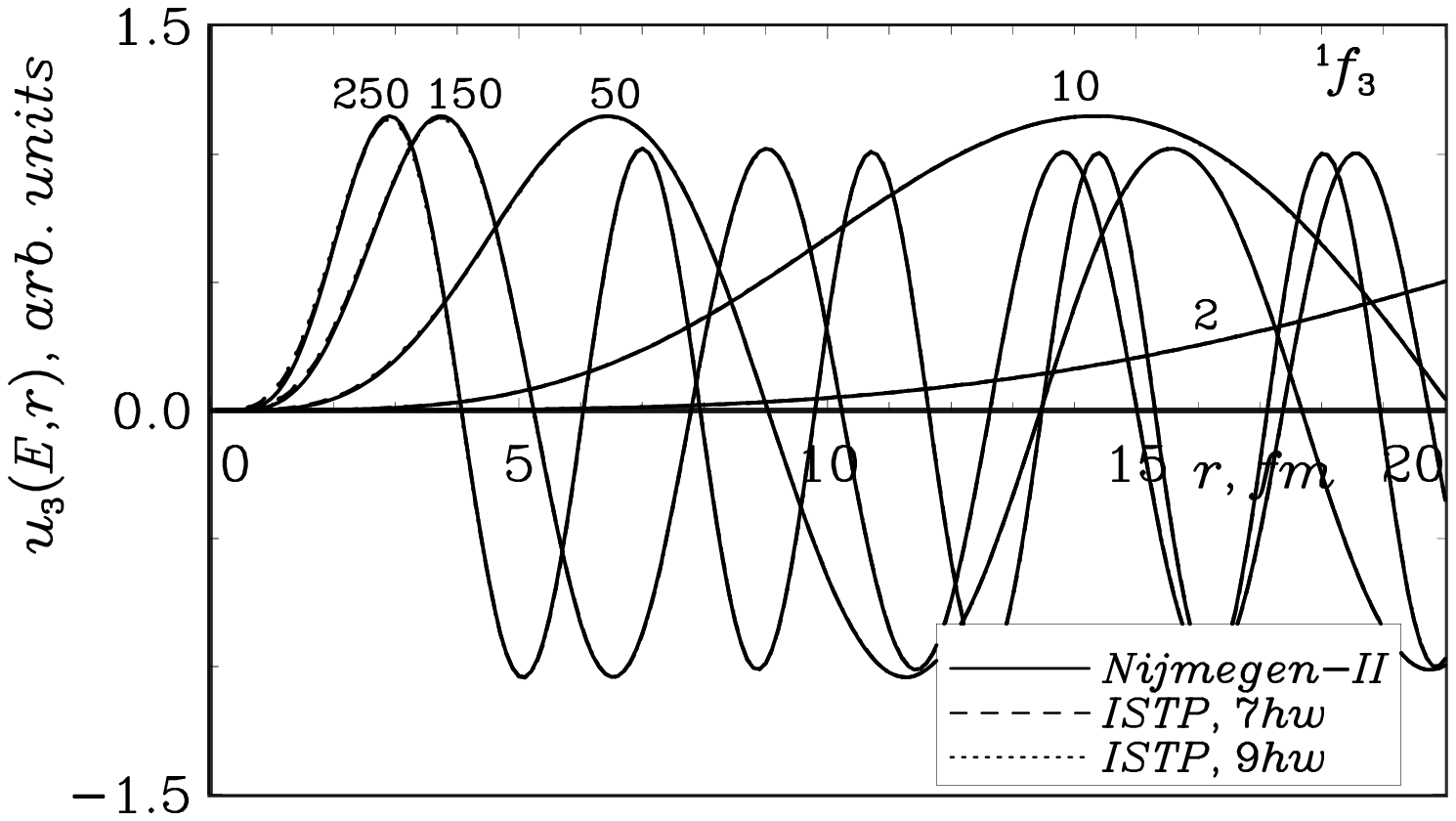,width=0.5\textwidth}}
\caption{$^1f_3$ $np$ scattering wave functions
at the laboratory energies
$E_{\rm lab}= 2$, 10, 50, 150, 250~MeV. See Fig.~\ref{wf1p1} for details.}
\label{wf1f3}
\end{figure}

\begin{figure}
\centerline{\psfig{figure=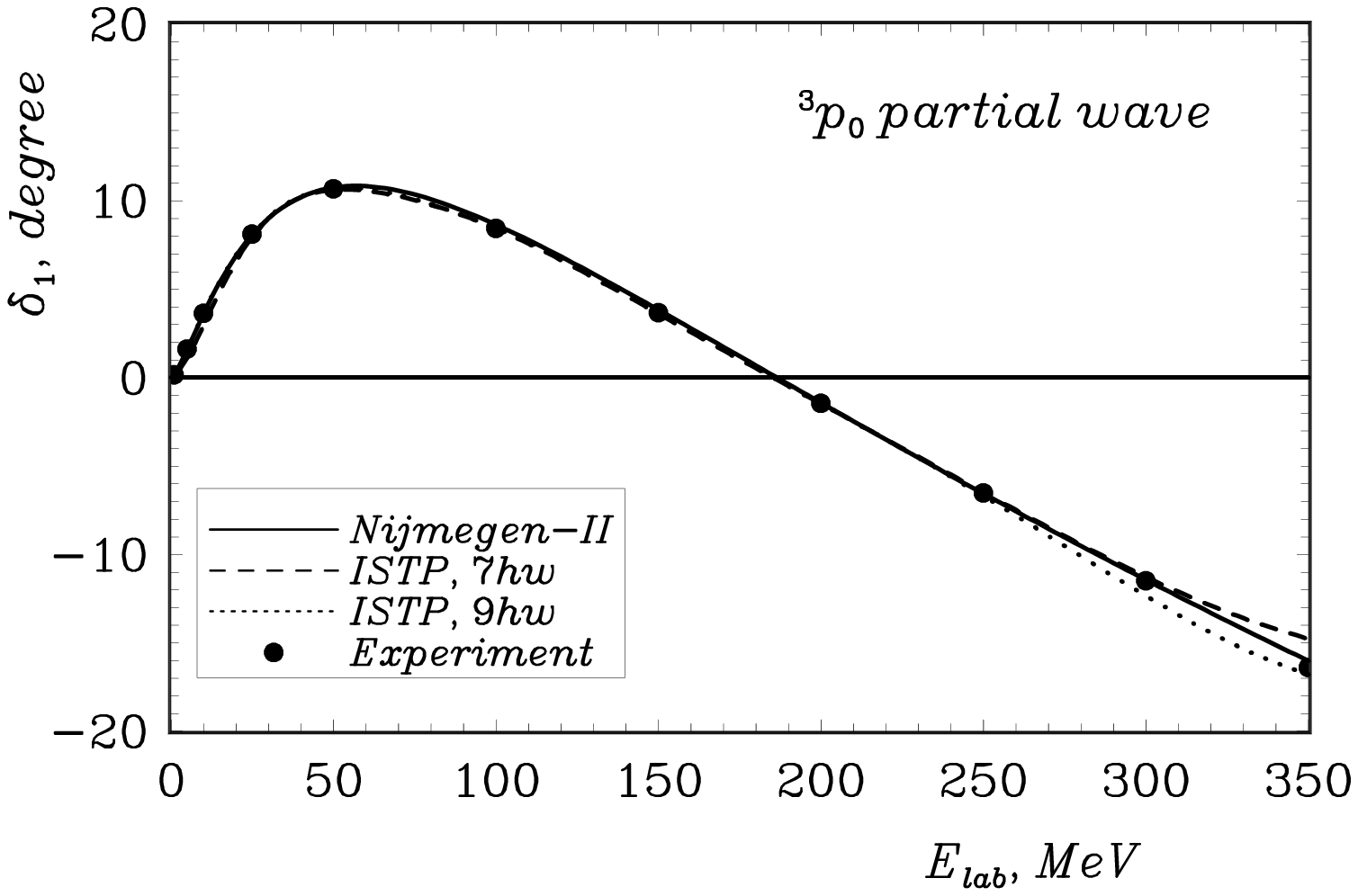,width=0.5\textwidth}}
\caption{$^3p_0$  $np$ scattering phase shifts.
See Fig.~\ref{ph1p1} for details. }
 \label{ph3p0}
\vspace{3ex}
\centerline{\psfig{figure=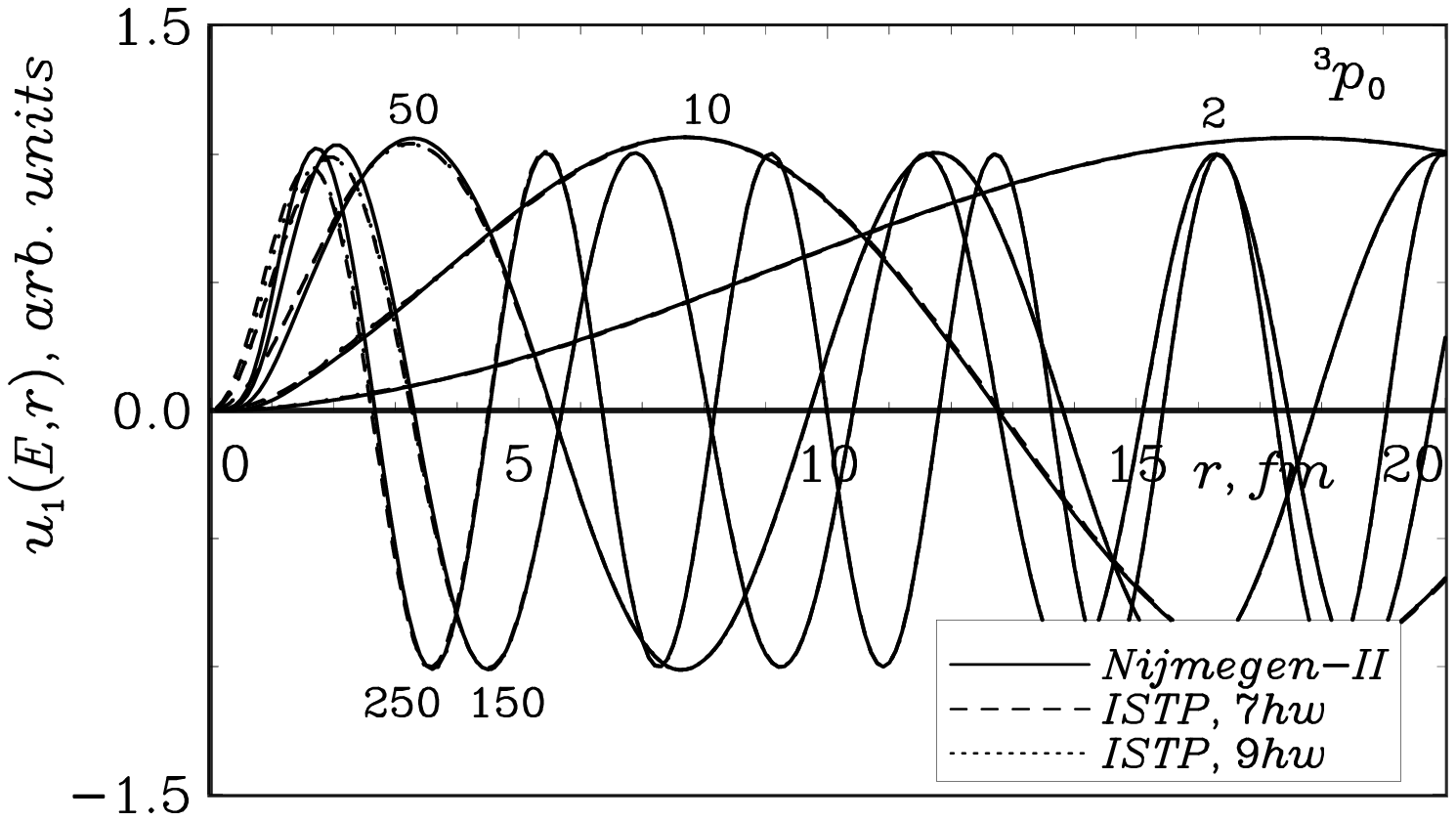,width=0.5\textwidth}}
\caption{$^3p_0$ $np$ scattering wave functions
at  the laboratory energies
$E_{\rm lab}= 2$, 10, 50, 150, 250~MeV. See Fig.~\ref{wf1p1} for details.}
\label{wf3p0}
\end{figure}

\begin{figure}
\centerline{\psfig{figure=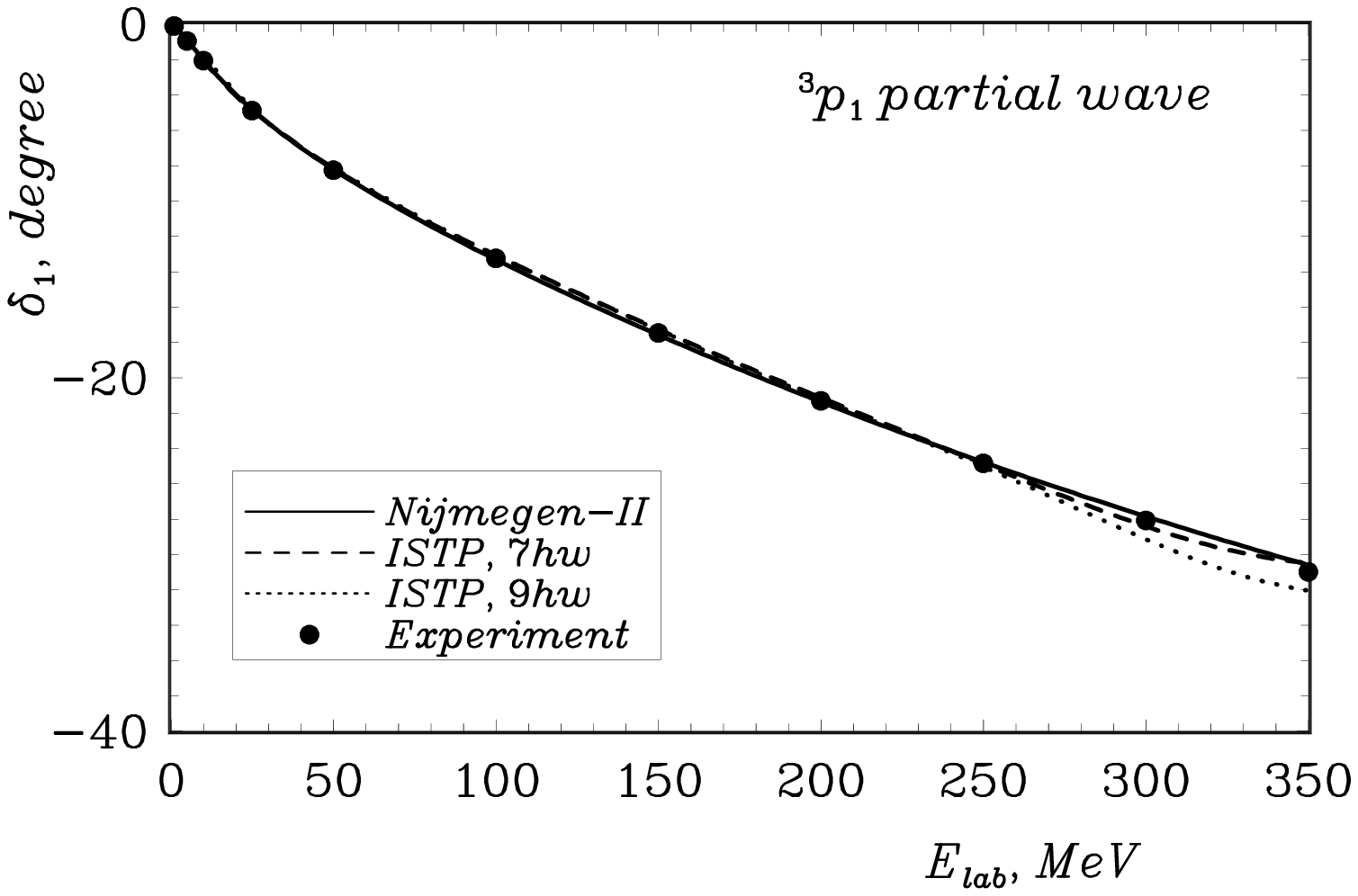,width=0.5\textwidth}}
\caption{$^3p_1$  $np$ scattering phase shifts.
See Fig.~\ref{ph1p1} for details. }
\label{ph3p1}
\vspace{3ex}
\centerline{\psfig{figure=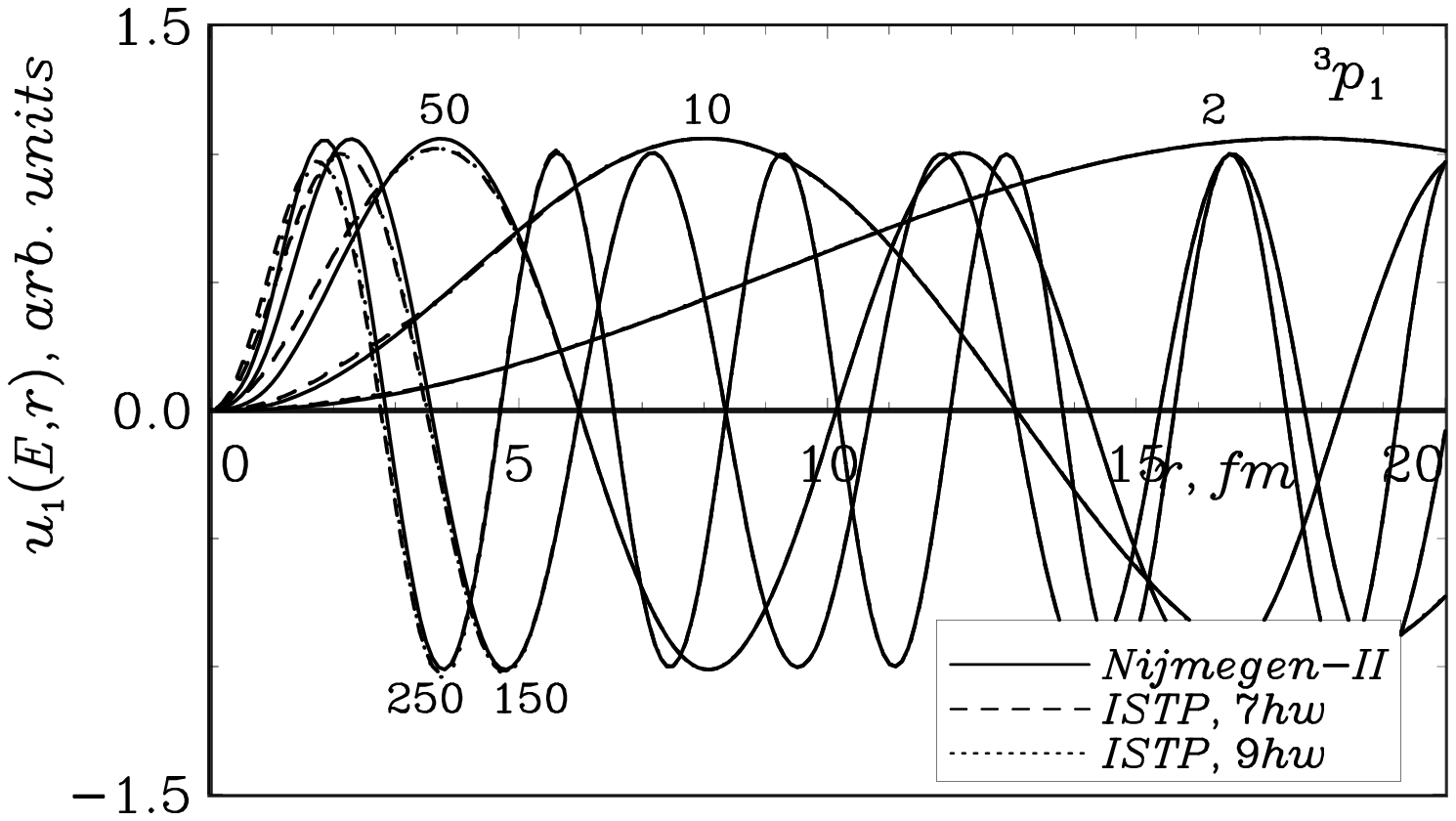,width=0.5\textwidth}}
\caption{$^3p_1$ $np$ scattering wave functions
at the laboratory energies
$E_{\rm lab}= 2$, 10, 50, 150, 250~MeV. See Fig.~\ref{wf1p1} for details.}
\label{wf3p1}
\end{figure}

\begin{figure}
\centerline{\psfig{figure=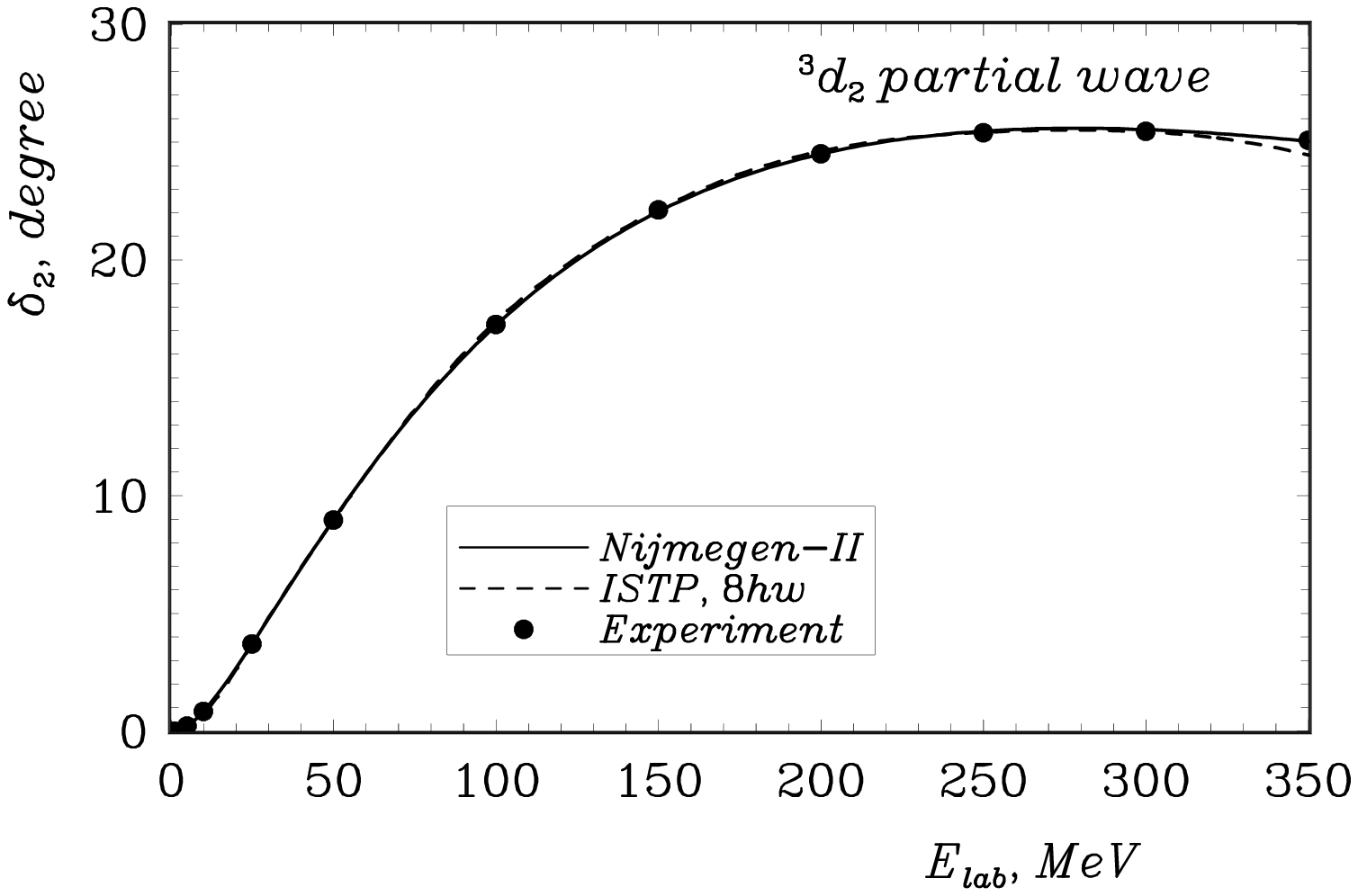,width=0.5\textwidth}}
\caption{$^3d_2$  $np$ scattering phase shifts.
See Fig.~\ref{ph1s0} for details.}
\label{ph3d2}
\vspace{3ex}
\centerline{\psfig{figure=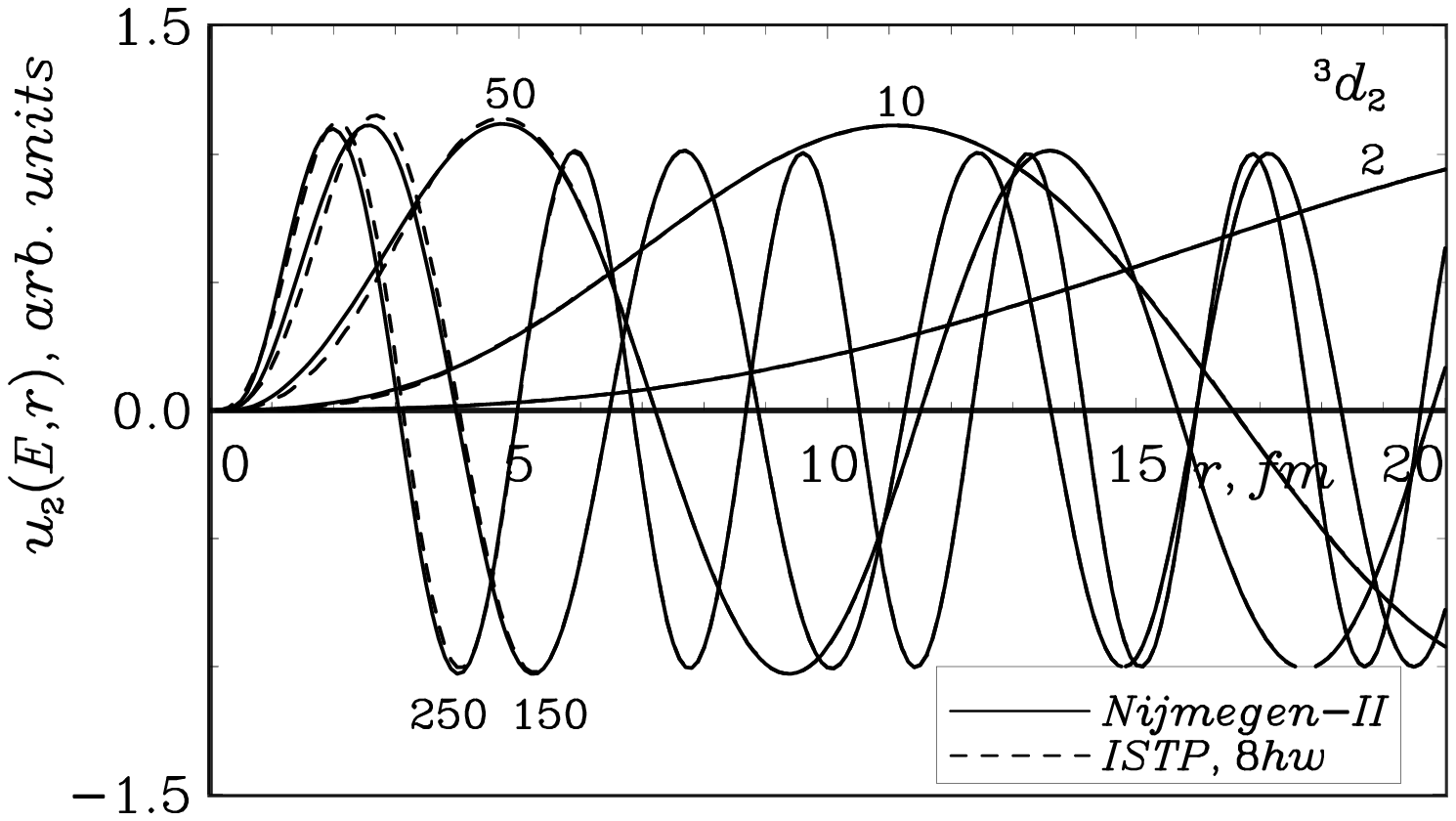,width=0.5\textwidth}}
\caption{$^3d_2$ $np$ scattering wave functions
at the laboratory energies
$E_{\rm lab}= 2$, 10, 50, 150, 250~MeV. See Fig.~\ref{wf1s0} for details.}
\label{wf3d2}
\end{figure}

\begin{figure}
\centerline{\psfig{figure=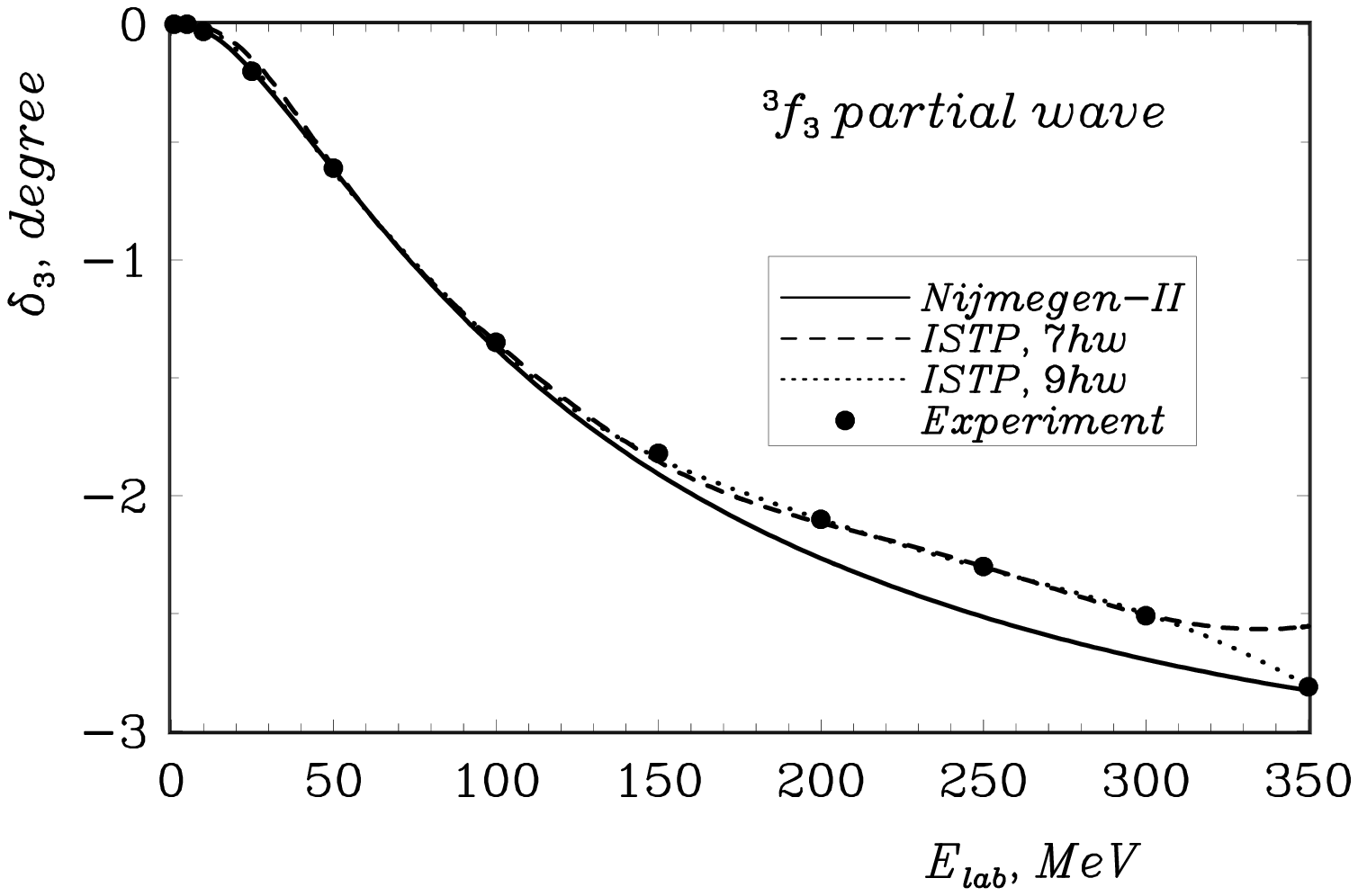,width=0.5\textwidth}}
\caption{$^3f_3$  $np$ scattering phase shifts.
See Fig.~\ref{ph1p1} for details.}
\label{ph3f3}
\vspace{3ex}
\centerline{\psfig{figure=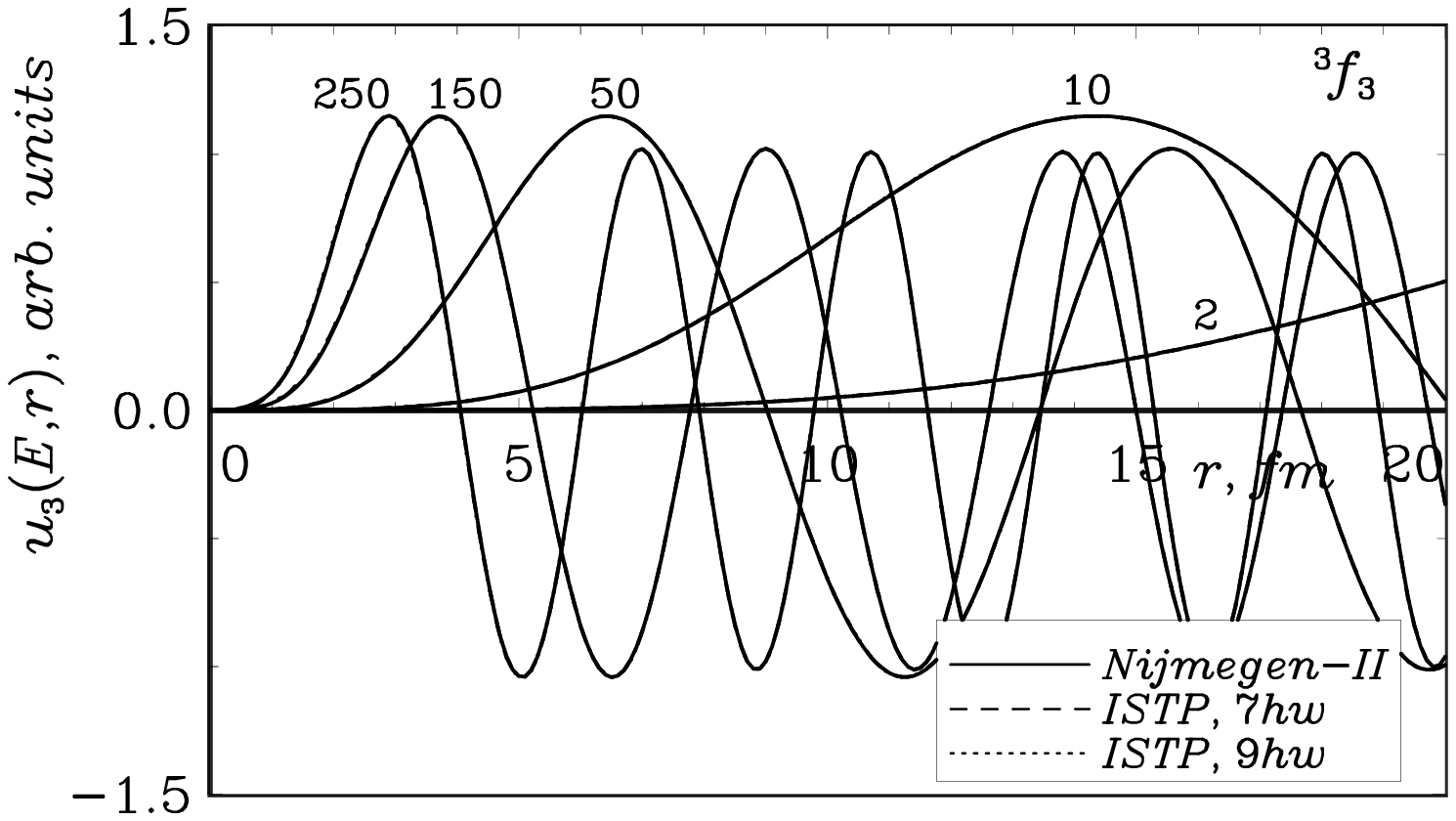,width=0.5\textwidth}}
\caption{$^3f_3$ $np$ scattering wave functions
at the laboratory energies
$E_{\rm lab}= 2$, 10, 50, 150, 250~MeV. See Fig.~\ref{wf1p1} for details.}
\label{wf3f3}
\end{figure}

In Figs.~\ref{ph1s0}--\ref{wf3f3} we present the results of the phase
shift and scattering wave function calculations with our ISTP in
the uncoupled partial waves. The phase shifts are seen to be better
reproduced by ISTP  up to the laboratory energy $E_{\rm lab}=350$~MeV
 than by one of the best realistic meson exchange potentials
Nijmegen-II. Some discrepancies are seen only at large
energies.
These  discrepancies can be eliminated by using larger $N$
values. This is illustrated in
phase shifts of
odd partial waves
presented
in Figs.~\ref{ph1p1}, \ref{ph1f3}, \ref{ph3p0},
\ref{ph3p1}, \ref{ph3f3}.
These are
the results of the phase shift calculations
with the $9\hbar\omega$ ISTP
in addition to the $7\hbar\omega$ ISTP phase
shifts. It is interesting that the differences between the
$7\hbar\omega$ ISTP and  $9\hbar\omega$ ISTP wave functions in odd
partial waves are too small to be seen in Figs.~\ref{wf1p1},
\ref{wf1f3}, \ref{wf3p0}, \ref{wf3p1}, \ref{wf3f3} even at large
energies. We note also that the use of  $7\hbar\omega$ ISTP instead of
$9\hbar\omega$ ISTP in the $^3$H and $^4$He calculations,
%
result in
negligible differences of the  binding energies, wave functions,
etc. The ISTP $np$ scattering wave functions at different energies are
very close to the Nijmegen-II wave functions both in odd and even
partial waves. In other words,
these
ISTP wave functions can be regarded
as realistic.

\section{Two-channel $J$-matrix inverse scattering approach and
ISTP in coupled $NN$ partial waves}

In the case of the
nucleon-nucleon scattering, the spins of two nucleons can couple to
the total spin $S=0$ (singlet spin state) or to the total spin $S=1$
(triplet  spin state). In the case of the singlet spin state, we have
only uncoupled partial waves in the nucleon-nucleon scattering.  In
the case of the triplet  spin state, the total angular momentum
$j=l+1$ can be obtained by the coupling of the total
spin $S=1$ with the orbital angular momentum $l$. On the other hand,
the higher triplet-spin partial wave of the same parity with the
orbital angular momentum $l'=l+2$, can have the same total angular momentum
$j=l+1=l'-1$. Such partial waves are coupled due to the
non-central nature of the $NN$ interaction. The $sd$ coupled partial
waves (the coupling of the $^3s_1$ and $^3d_1$ partial waves) and $pf$
coupled partial waves  (the coupling of the $^3p_2$ and ${^3f}_2$
partial waves) are of
special
interest for applications. The  case of the $sd$
coupled partial waves is of primary importance
due to the existence of
the only $np$ bound state (the
deuteron).
The coupled equations describing  the
$NN$ system in the coupled partial waves, are of the same structure with
the coupled equations describing the two-channel system. In other words, the
description of the coupled waves in the $NN$ scattering is formally
equivalent with the description of the two-channel scattering.

The wave function in the coupled waves case is
\begin{equation}
\Psi=\sum_{\Gamma} 
\frac1r\,
u_{\Gamma\vphantom{'}}(E,r)\,|\Gamma\rangle ,
\label{coup-wf}
\end{equation}
%
%
where $|\Gamma\rangle$ is the spin-angle wave function which includes
the spin variables of two nucleons coupled to the total spin $S=1$,
the spherical function $Y_{l_\Gamma m}(\hat{\vecb{r}})$, and the
coupling of the channel orbital momentum $l_\Gamma$ with the total
spin $S$ into the total angular momentum $j$;
$ u_{\Gamma\vphantom{'}}(E,r)$ is the
radial wave function in the given formal channel $\Gamma=\{l_\Gamma,j\}$.
Generally there are two independent solutions for each  radial
wave function  $ u_{\Gamma\vphantom{'}}(E,r)$.
To distinguish these  
solutions
it is convenient to employ the $K$-matrix formalism associated with
the standing wave asymptotics of the wave
function:\vspace{-1ex}
\begin{gather}
  u_{\Gamma(\Gamma_i)\vphantom{'}}(E,r)
\mathop{\longrightarrow}\limits_{r\to\infty}
\frac{qr}{r_0}\left(\delta_{\Gamma\Gamma_i}\:
 j_{l_\Gamma\vphantom{'}}\!\Big(\frac{qr}{r_0}\Big)
 - K_{\Gamma\Gamma_i}(E)\:
n_{l_\Gamma\vphantom{'}}\!\Big(\frac{qr}{r_0}\Big)\right)  .
  \label{Krow2}
  \end{gather}
%
\noindent Here the index $\Gamma_i$ distinguishes independent radial functions
$u_{\Gamma(\Gamma_i)\vphantom{'}}(E,r)$ in the channel~$\Gamma$,
$K_{\Gamma\Gamma_i}(E)$ is the  $K$-matrix, and ${j}_{\vphantom{'}l}(x)$
and ${n}_{l\vphantom{'}}(x)$ are spherical Bessel and Neumann
functions. The advantage of the  $K$-matrix formalism is that the
radial functions $ u_{\Gamma(\Gamma_i)\vphantom{'}}(E,r) $
 defined according to their  standing wave asymptotics
(\ref{Krow2}) are real contrary to the more conventional $S$-matrix formalism
with complex radial wave functions which are asymptotically a superposition
of ingoing and outgoing spherical waves. The  $K$-matrix
$K_{\Gamma\Gamma_i}(E)$, of course,  can be expressed through the
$S$-matrix. However it is not the $S$-matrix but the so-called
phase shifts $\delta_\Gamma$ and $\delta_{\Gamma_i}$ in each of
the coupled partial waves $\Gamma$ and $\Gamma_i$ and the mixing
parameter $\varepsilon$ that are usually published as functions
of the energy $E$ in the
experimental and theoretical investigations.
 The
$S$-matrix can be parametrized in terms of    $\delta_\Gamma$,
$\delta_{\Gamma_i}$ and $\varepsilon$. 
However for the present
application it is more convenient to  express the
$K$-matrix elements directly through  $\delta_\Gamma$,
$\delta_{\Gamma_i}$
and $\varepsilon$ (see Refs.~\cite{Babikov,Stapp}):
\begin{subequations}
\label{Kss-dd}
\begin{align}
\label{Kss}
K_{ss}(E)&= \frac{\displaystyle \tan \delta_s +
\tan^2\varepsilon\cdot\tan\delta_d}{\displaystyle 1-
\tan^2\varepsilon\cdot\tan \delta_s\cdot\tan\delta_d}, \displaybreak[2]\\[1ex]
\label{Kdd}
K_{dd}(E)&= \frac{\displaystyle \tan \delta_d +
\tan^2\varepsilon\cdot\tan\delta_s}{\displaystyle 1-
\tan^2\varepsilon\cdot\tan \delta_s\cdot\tan\delta_d},
 \displaybreak[2]\\[1ex]
\label{Ksd}
K_{sd}(E)&= K_{ds}(E) \notag\\
&=\frac{\displaystyle
\tan\varepsilon}{\displaystyle \cos
\delta_s\cdot\cos\delta_d\cdot(1- \tan^2\varepsilon\cdot\tan
\delta_s\cdot\tan\delta_d)}.
 \end{align}
\end{subequations}
To be specific, we
have specified
the case of the coupled $sd$
waves
where
the channel indexes $\Gamma$ and
$\Gamma_i$ take the values $s$ or $d$. In the case of the coupled $pf$
waves, one
substitutes
the indexes $s$ and $d$ by the
indexes $p$ and $f$
in the above expressions and in other formulas in this
section.

Within the inverse scattering $J$-matrix approach, the potential
in the coupled partial waves is
fitted with
the form:
\begin{equation}
{V}= \sum_{\Gamma, \, \Gamma '} \;
\sum_{n=0}^{N_{\Gamma}}\;\sum_{n'=0}^{N_{\Gamma '}}\,
| n \Gamma \rangle\: V_{n n'}^{\Gamma \Gamma '}\: \langle n'\Gamma ' | .
\label{Pot}
\end{equation}
Here $V_{n n'}^{\Gamma \Gamma'}\equiv\langle n\Gamma|V|n'\Gamma'\rangle$
is the potential energy matrix element in the oscillator basis%
%
%
\begin{equation}
|n\Gamma \rangle=R_{nl_\Gamma}(r)\,|\Gamma\rangle ,
\label{Gam-bas}
\end{equation}
where the radial oscillator function $R_{nl_\Gamma}(r)$ is given
by Eq.~(\ref{e24}) and $|\Gamma\rangle$ is the spin-angle
function. Different truncation boundaries $N_{\Gamma}$ can be used
in different partial waves $\Gamma$.

The multi-channel $J$-matrix formalism is well known
(see, e.~g., \cite{Yamani,Bang}) and we
will not
discuss it here in detail. The formalism provides
exact solutions for the continuum spectrum wave functions in the
case when the finite-rank potential $V$ of the type (\ref{Pot}) is employed.
In the case of the discrete spectrum states, the exact solutions
are obtained by the calculation of the corresponding $S$-matrix poles as
is discussed in Refs.~\cite{SmSh,Halo-b,Halo-c}. In particular, the
deuteron ground state energy $E_d$ should be associated with the $S$-matrix
pole and its wave function is calculated by means of the
$J$-matrix formalism applied to the negative energy $E=E_d$.

Within the  $J$-matrix formalism, the radial wave function
$u_{\Gamma(\Gamma_i)}(E,r)$ is expanded in the oscillator function  series
  \begin{equation}
 u_{\Gamma(\Gamma_i)\vphantom{'}}(E,r) = \sum_{n=0}^{\infty}
 a_{n\,\Gamma(\Gamma_i)}(E)\, R_{nl_\Gamma}(r).
  \label{eq:m-row}
  \end{equation}
In the  external part of the model space spanned by the functions
(\ref{Gam-bas}) with $n\geq N_\Gamma$, the oscillator representation
wave function $a_{n\,\Gamma(\Gamma_i)}(E)$ fits the
three-term recurrence relation~(\ref{eq:TRS}). Its solutions
corresponding to the asymptotics~(\ref{Krow2})  are
\begin{gather}
\label{an1-S}
a_{n\Gamma(\Gamma_i)}^{\vphantom{a}}(E)
=\delta_{\Gamma\Gamma_i}\,S_{nl_\Gamma}(E)
 + K_{\Gamma\Gamma_i}(E)\,C_{nl_\Gamma}(E) .
\end{gather}
Equation (\ref{an1-S}) can be used for the calculation of
$a_{n\Gamma(\Gamma_i)}^{\vphantom{a}}(E)$ with $n\geq N_\Gamma$ if
the coupled wave phase shifts $\delta_\Gamma$ and
$\delta_{\Gamma_i}$ and the mixing parameter $\varepsilon$ are
known.

The oscillator representation
wave function $a_{n\,\Gamma(\Gamma_i)}(E)$
in the {internal part of the model space} spanned by the functions
(\ref{Gam-bas}) with $n\leq N_\Gamma$, can be expressed through the external
 oscillator representation
wave functions $a_{N_\Gamma\vphantom{'}+1,\Gamma(\Gamma_i)}(E)$ as
  \begin{equation}
a_{n\,\Gamma(\Gamma_i)}(E) = \sum_{\Gamma'}\,
{\mrs G}_{nN_{\Gamma'}}^{\Gamma\Gamma'}\,
T^{l_{\Gamma'}}_{N_{\Gamma'},\,N_{\Gamma'}+1}\:
a_{N_{\Gamma'}+1,\Gamma'(\Gamma_i)}(E) .
  \label{cnin-m}
  \end{equation}
The matrix elements,
  \begin{equation}
  {\mrs G}_{nn'}^{\Gamma\Gamma'} =
  -\sum_{\lambda ^{\prime}=0}^{N}
  \frac{ \langle n\Gamma|\lambda ^{\prime}\rangle\,
    \langle\lambda ^{\prime} |n'\Gamma'\rangle }
  { E_{\lambda ^{\prime}}-E } ,
  \label{oscrm-m}
  \end{equation}
where $N=N_\Gamma+N_{\Gamma'}+1$,
are expressed within the direct $J$-matrix formalism
through the eigenvalues $E_\lambda$ and eigenvectors
$\langle n\Gamma|\lambda\rangle$ of the truncated Hamiltonian matrix, i.~e.
$E_\lambda$ and $\langle n\Gamma|\lambda\rangle$ are obtained by solving the
algebraic problem
\begin{equation}
\sum_{\Gamma'}\sum_{n'=0}^{N_{\Gamma'}} H_{nn'}^{\Gamma\Gamma'}\,
\langle n'\Gamma'|\lambda\rangle
= E_\lambda\,\langle n\Gamma|\lambda\rangle,
\qquad n\leq N_\Gamma .
\label{Alge-m}
\end{equation}
Here $H_{nn'}^{\Gamma\Gamma'}\equiv\langle n\Gamma|H|n'\Gamma'\rangle$
are the Hamiltonian matrix elements.

Within the inverse $J$-matrix approach, we start with assigning some
values to the potential truncation boundaries $N_\Gamma$ [see
Eq.~(\ref{Pot})] in each of the partial waves~$\Gamma$.
As a next step, we
calculate the sets of eigenvalues $E_\lambda$ and respective
eigenvector components $\langle N_\Gamma\Gamma|\lambda\rangle$. This
can be done using the set of the $J$-matrix matching conditions
which are obtained from Eq.~(\ref{cnin-m}) supposing $n=N_\Gamma$.
In more detail,
these matching conditions are (to be specific, we again
take
the case of the coupled $sd$
waves
so
the channel indexes $\Gamma$ and
$\Gamma_i$ take the values $s$ or $d$)
\begin{subequations}
\label{Match}
\begin{gather}
\label{M1} a_{N_{s}s(s)}^{\vphantom{a}}(E)= \sum_{\Gamma'=s,d} \,
{\cal G}_{s {\Gamma '}}\,
 T^{\Gamma'}_{N_{\Gamma'},N_{\Gamma'}+1}\;a_{N_{\Gamma'}+1,\Gamma' (s)}(E),
\displaybreak[3]  \\
\label{M2}
a_{N_{d}d(s)}^{\vphantom{a}}(E)= \sum_{\Gamma'=s,d}\,
{\cal G}_{d {\Gamma '}}\,
 T^{\Gamma'}_{N_{\Gamma'},N_{\Gamma'}+1}\;a_{N_{\Gamma'}+1,\Gamma' (s)}(E),
\displaybreak[3] \\
\label{M3}
a_{N_{s}s(d)}^{\vphantom{a}}(E)= \sum_{\Gamma'=s,d}\,
{\cal G}_{s {\Gamma '}}\,
 T^{\Gamma'}_{N_{\Gamma'},N_{\Gamma'}+1}\;a_{N_{\Gamma'}+1,\Gamma'(d)}(E),
\intertext{and}
\label{M4} a_{N_{d}d(d)}^{\vphantom{a}}(E)= \sum_{\Gamma'=s,d}\,
{\cal G}_{d {\Gamma '}}\,
 T^{\Gamma'}_{N_{\Gamma'},N_{\Gamma'}+1}\;a_{N_{\Gamma'}+1,\Gamma' (d)}(E),
\end{gather}
\end{subequations}
where we introduced the shortened notation
\begin{equation}
{\cal G}_{\Gamma\Gamma'}\equiv{\mrs G}_{N_{\Gamma}N_{\Gamma'}}^{\Gamma\Gamma'}
 =   -\sum_{\lambda ^{\prime}=0}^{N}
  \frac{ \langle N_\Gamma\Gamma|\lambda ^{\prime}\rangle\,
    \langle\lambda ^{\prime} |N_{\Gamma'}\Gamma'\rangle }
  { E_{\lambda ^{\prime}}-E } .
\label{G-sh}
\end{equation}
To calculate $a_{N_{\Gamma\vphantom{'}}\Gamma(\Gamma_i)}^{\vphantom{a}}(E)$
and  $a_{N_{\Gamma\vphantom{'}}+1,\,\Gamma(\Gamma_i)}^{\vphantom{a}}(E)$
entering Eqs.~(\ref{Match}), we can
use Eq.~(\ref{an1-S}) with the $K$-matrix elements expressed through
the experimental data by Eqs.~(\ref{Kss-dd}). Therefore
${\cal G}_{ss}$, ${\cal G}_{sd}$, ${\cal G}_{ds}$ and ${\cal G}_{dd}$
are the only unknown  quantities in Eqs.~(\ref{Match}) and they can be
obtained as the solutions of the algebraic problem~(\ref{Match})
at any positive energy $E$.

These solutions may be expressed as
\begin{subequations}
\label{G-sol}
\begin{align}
 \label{Gsst}
{\cal G}_{ss} & = \frac{\Delta_{ss}(E)}{T^s_{N_s, N_s+1}\, \Delta(E)} ,
\\[1.5ex]
 \label{Gddt}
{\cal G}_{dd} & = \frac{\Delta_{dd}(E)}{T^d_{N_d,N_d+1}\, \Delta(E)} ,
\intertext{and}
 \label{Gsdt}
{\cal G}_{sd}= {\cal G}_{ds}& =
 -\frac{r_{0\vphantom{'}}\sqrt{2 E} \,K_{sd} }
 { 2\, T^s_{N_s, N_s+1}\, T^d_{N_d,N_d+1}\, \Delta (E) },
\end{align}
\end{subequations}
where
\begin{widetext}
\begin{subequations}
\label{D-sol}
\begin{align}
&\Delta_{ss}(E) = \Big(S_{N_ss}(E) + K_{ss}(E)\,C_{N_ss}(E) \Big)
\Big(S_{N_d+1,d}(E) + K_{dd}(E)\, C_{N_d+1,d}(E)\Big)
- K^2_{sd}(E)\,C_{N_ss}(E)\,C_{N_d+1,d}(E) ,
\label{Dss}
\\ 
&\Delta_{dd}(E) =
\Big(S_{N_s+1,s}(E) + K_{ss}(E)\, C_{N_s+1,s}(E) \Big)
 \Big(S_{N_dd}(E) + K_{dd}(E)\, C_{N_dd}(E)\Big)
- K^2_{sd}(E)\,C_{N_s+1,s}(E)\,C_{N_dd}(E) ,
\label{Ddd}
\displaybreak[3]\\ 
\intertext{and}
&\Delta(E) =
\Big(S_{N_s+1,s}(E) + K_{ss}(E)\,C_{N_s+1,s}(E) \Big)
\Big(S_{N_d+1,d}(E) + K_{dd}(E)\, C_{N_d+1,d}(E)\Big)
- K^2_{sd}(E)\,C_{N_s+1,s}(E)\,C_{N_d+1,d}(E) .
 \label{De} 
\end{align}
\end{subequations}
To derive Eq.~(\ref{Gsdt}), we used the following expression for the
Casoratian determinant \cite{SmSh,Bang}:
\begin{equation}
{\mrs K}^l_n(C,S)\equiv C_{n+1,l}(E)\,S_{nl}(E)-S_{n+1,l}(E)\,C_{nl}(E)
=\frac{r_0 \, \sqrt{2E}}{2\,T^l_{n,n+1}}.
\label{Casorati}
\end{equation}
\end{widetext}

It is obvious from Eqs.~(\ref{G-sh}) and (\ref{G-sol})
that the eigenvalues $E_\lambda$ can be found
by solving the following equation:
\begin{gather}
 \label{DZero}
 \Delta (E_\lambda) = 0.
\end{gather}
The eigenvector components $\langle N_\Gamma\Gamma|\lambda\rangle$ can
be obtained from Eqs.~(\ref{Gsst})--(\ref{Gddt}) in the limit
$E\to E_\lambda$ in the same manner as Eq.~(\ref{Nlabda}) in the
single-channel case:
\begin{gather}
 \label{Gams}
|\langle N_ss | \lambda \rangle |^2  =
\frac{\Delta_{ss}(E_{\lambda})}{T^s_{N_s, N_s+1}\:\Delta^{\lambda}} 
\intertext{and}
 \label{Gamd}
|\langle N_dd | \lambda \rangle |^2  =
\frac{\Delta_{dd}(E_{\lambda})}{T^d_{N_d, N_d+1}\: \Delta^{\lambda}} ,
\intertext{where}
 \label{Delpr}
\Delta^{\lambda} =
\left. \frac{d\Delta (E)}{dE} \right| _{E=E_{\lambda}} .
\end{gather}
Equations   (\ref{Gams})--(\ref{Gamd}) make it possible to calculate
the absolute values of $\langle N_ss | \lambda \rangle $ and $\langle
N_dd | \lambda \rangle$ only. However the relative sign of these
eigenvector components is important. This relative sign can be
established using the relation
\begin{multline}
 \label{Rel}
\frac{\langle N_ss | \lambda \rangle\: T^s_{N_s, N_s+1}}
 {\displaystyle \langle N_dd | \lambda \rangle\: T^d_{N_d, N_d+1}} =
 - \frac{a_{N_d+1,d(s)}(E_{\lambda})}{a_{N_s+1,s(s)}(E_{\lambda})}\\
=  - \frac{a_{N_d+1,d(d)}(E_{\lambda})}{a_{N_s+1,s(d)}(E_{\lambda})}
\end{multline}
that can be easily obtained from Eqs.~(\ref{Match}).

Using Eqs. (\ref{DZero})--(\ref{Rel}) we obtain all eigenvalues
$E_\lambda>0$ and corresponding eigenvector components
$\langle N_\Gamma\Gamma|\lambda\rangle$.
For example, in
the case of
the
coupled $pf$ waves when the $NN$ system does not have a bound state,
all eigenvalues $E_\lambda$ are positive and by means of
Eqs.~(\ref{DZero})--(\ref{Rel}) we obtain a complete set of eigenvalues
$E_\lambda=0$, 1,~...~, $N$ and the complete set of the
eigenvector's
last
components $\langle N_\Gamma\Gamma|\lambda\rangle$ providing the best
description of the `experimental' (obtained by means of phase shift
analysis) phase shifts $\delta_1(E)$ and $\delta_3(E)$ and mixing
parameter $\varepsilon$.  However, as in the case of the uncoupled
waves, we should take care of fitting the completeness relation 
for the eigenvectors $\langle n\Gamma|\lambda\rangle$ that  in the
coupled wave  case takes the form 
\begin{gather}
\label{mcomplete}
 \sum_{\lambda=0 }^N
\langle n\Gamma|\lambda\rangle\langle\lambda| n'\Gamma'\rangle
= \delta _{nn'}  \delta_{\Gamma \Gamma '}.
\end{gather}
Due to Eq.~(\ref{mcomplete}), in the two-channel case, we should
perform variation of 
the components $\langle N_\Gamma\Gamma|\lambda\rangle$ 
associated with the two largest eigenenergies $E_{\lambda=N}$ and
$E_{\lambda=N-1}$ to
fit three relations
\begin{subequations}
\label{pfmcomplete}
\begin{align}
\sum_{\lambda=0 }^N
\langle N_{\Gamma_1}\Gamma_1|\lambda\rangle
    \langle\lambda| N_{\Gamma_1}\Gamma_1\rangle
&= 1, 
\label{ppcomplete} \\[1ex]
\sum_{\lambda=0 }^N
\langle N_{\Gamma_1}\Gamma_1|\lambda\rangle
    \langle\lambda| N_{\Gamma_2}\Gamma_2\rangle
&= 0, 
\label{pfcomplete} 
\intertext{and}
\sum_{\lambda=0 }^N
\langle N_{\Gamma_2}\Gamma_2|\lambda\rangle
    \langle\lambda| N_{\Gamma_2}\Gamma_2\rangle
&=1.
\label{ffcomplete}
\end{align}
\end{subequations}
This immediately spoils the description of the scattering data that
can be restored by the additional variation of the  eigenenergies
$E_{\lambda=N}$ and $E_{\lambda=N-1}$. As a result, in the case of
the coupled $pf$ waves, we perform a standard fit to the data by
minimizing $\chi^2$ by the variation of 
${\langle N_pp|\lambda=N\rangle}$, ${\langle N_pp|\lambda=N-1\rangle}$, 
${\langle N_ff|\lambda=N\rangle}$, ${\langle N_ff|\lambda=N-1\rangle}$,
$E_{\lambda=N}$ and $E_{\lambda=N-1}$. These 6 parameters should fit
three relations (\ref{pfmcomplete}), hence we face a simple problem of
a three-parameter fit.

In the
case of the coupled $sd$ waves, the $np$ system has a bound state (the
deuteron) at the energy $E_d$ ($E_d<0$) and one of the eigenvalues
$E_\lambda$ is negative:
$E_0<0$. We should extend the above theory to the case of a system
with bound states. For the coupled $sd$ waves case when the $np$
system has only one bound state, we need three additional equations to
calculate $E_0$ and
the
components $\langle N_ss|\lambda=0\rangle$ and
$\langle N_dd|\lambda=0\rangle$.

The deuteron energy $E_d$ should be associated with the $S$-matrix
pole. As it was already noted, the technique of the $S$-matrix pole
calculation within the $J$-matrix formalism
is discussed  together with some applications
in Refs.~\cite{Halo-b,Halo-c}. In the case of the finite-rank potentials of
the type~(\ref{Pot}),  one can  obtain
the exact value of the bound state energy $E_d$ and the exact  bound
state wave function by the $S$-matrix pole calculation within the
$J$-matrix formalism. 
To calculate the $S$-matrix, we
use the standard
outgoing-ingoing spherical wave asymptotics and the respective
expression for the $J$-matrix oscillator space wave function in
the external part of the model space discussed, e.~g.,  in
Refs.~\cite{SmSh,Bang,Halo-b,Halo-c}
instead of the standing wave asymptotics~(\ref{Krow2}) and
respectively modified expression~(\ref{an1-S}) for the $J$-matrix oscillator
space wave function.
Using the expressions for the multi-channel $S$-matrix within the
$J$-matrix formalism
presented in Refs.~\cite{SmSh,Bang,Halo-b,Halo-c},  it is easy to
obtain the following
expressions~\cite{Zret} for the  two-channel $S$-matrix elements:
\begin{widetext}
\begin{subequations}
\label{S-matr}
\begin{multline}
S_{ss} = \frac{1}{D(E)}
  \left \{ \left(C_{N_{s}s}^{(-)}(E)-
{\cal G}_{ss}\:T_{N_s,N_s+1}^s\, C_{N_{s}+1,s}^{(-)}(E) \right)
 \right.  
\left(C_{N_{d}d}^{(+)}(E)-
 {\cal G}_{dd}\:T_{N_d,N_d+1}^d\, C_{N_{d}+1,d}^{(+)}(E) \right)  \\[1ex]
\left. -{\cal G}_{sd}^{\, 2}\:
 T_{N_s,N_s+1}^s\, T_{N_d,N_d+1}^d\,
           C_{N_{s}+1,s}^{(-)}(E)\,C_{N_{d}+1,d}^{(+)}(E)
   \right\},
\label{S11}
\end{multline}
%
\begin{multline}
S_{dd} = \frac{1}{D(E)}
  \left \{ \left(C_{N_{s}s}^{(+)}(E)-
{\cal G}_{ss}\:T_{N_s,N_s+1}^s\, C_{N_{s}+1,s}^{(+)}(E) \right)
 \right. 
\left(C_{N_{d}d}^{(-)}(E)-
{\cal G}_{dd}\:T_{N_d,N_d+1}^d\, C_{N_{d}+1,d}^{(-)}(E) \right)\\[1ex]
\left. -{\cal G}_{sd}^{\, 2}\:
 T_{N_s,N_s+1}^s\, T_{N_d,N_d+1}^d\,
           C_{N_{s}+1,s}^{(+)}(E)\,C_{N_{d}+1,d}^{(-)}(E)
   \right\},
\label{S22}
\end{multline}
and
\begin{equation}
S_{sd}=S_{ds}=- \frac{i r_{0\vphantom{'}} \sqrt{2E}\,
{\cal G}_{sd}} {D(E)}, \label{S12}
\end{equation}
\end{subequations}
where
\begin{multline}
  D(E) =\left( C_{N_ss}^{(+)}(E)-
  \mathcal{G}_{ss}\:T_{N_s,N_s+1}^s\, C_{N_s+1,s}^{(+)}(E)\right) 
\left( C_{N_dd}^{(+)}(E)-
  \mathcal{G}_{dd}\:T_{N_d,N_d+1}^d\, C_{N_d+1,d}^{(+)}(E)\right)
\\[1ex]
-\mathcal{G}_{sd}^{\,2}\:T_{N_s,N_s+1}^s\, T_{N_d,N_d+1}^d\,
  C_{N_s+1,s}^{(+)}(E) \, C_{N_d+1,d}^{(+)}(E)
 \label{Dt}
\end{multline}
\end{widetext}
and
\begin{equation}
C_{nl}^{(\pm)}(E)= C_{nl}(E)\pm {i}S_{nl}(E).
\label{Cpm}
\end{equation}
We need to calculate $C_{nl}^{(\pm)}(E)$ at negative energy $E=E_d$
which
can be
done using Eqs.~(\ref{Cpm}), (\ref{eq:Snl}) and (\ref{eq:Cnl})
where imaginary  values of $q=q_d=i\sqrt{2|E_d|}$
are employed.
Extension of these expressions to the complex  $q$ plane is
discussed in Ref.~\cite{SmSh}.

Since we associate the deuteron energy $E_d$ with the $S$-matrix pole,
from Eqs.~(\ref{S-matr}) we have
\begin{equation}
  D(E_d)=0.
\label{ZDt}
\end{equation}
Assigning the experimental deuteron ground state energy to $E_d$  in
Eq.~(\ref{ZDt}) and substituting  $ D(E_d)$ in this
formula by its expression~(\ref{Dt}), we obtain one of the equations
needed to calculate $E_0$, $\langle N_ss|\lambda=0\rangle$ and
$\langle N_dd|\lambda=0\rangle$.

Two other equations utilize information about the asymptotic
normalization constants of the deuteron bound state ${\mrs A}_s$ and
${\mrs A}_d$. If the $S$-matrix is treated as a function of the
complex momentum $q$, then its residue can be expressed
through~${\mrs A}_s$ and ${\mrs A}_d$~\cite{BZP,BBD}:
\begin{equation}
i \mathop{\rm Res}_{q=iq_d}S_{l_{\Gamma\vphantom{'}} l_{\Gamma'}} 
=r_0\,e^{i\frac{\pi}{2}\left(l_{\Gamma\vphantom{'}}+l_{\Gamma'}\right)}
{\mrs A}_{l_{\Gamma\vphantom{'}}} {\mrs A}_{l_{\Gamma'}}.
\label{resS}
\end{equation}
(the factor $r_0$ in the  right-hand-side originates from the
use of the dimensionless momentum $q$). ${\mrs A}_s$ and
$\eta =\displaystyle\frac{{\mrs A}_d}{{\mrs A}_s}$ are determined
experimentally. Therefore it is useful to
rewrite equations (\ref{resS}) as 
\begin{subequations}
\label{asC}
\begin{align}
&{i}\lim _{q\to iq_{d\vphantom{'}}}(q-{i}q_{d\vphantom{'}})S_{ss}
= r_0 {\mrs A}_{s}^2   
\label{asC1} 
\intertext{and}
&i\lim_{q \to iq_{d\vphantom{'}}}(q-iq_{d\vphantom{'}})S_{sd}=
-r_0  \eta {\mrs A}_{s}^2 .
\label{asC2}
\end{align}
\end{subequations}
Substituting $S_{ss}$ and $S_{sd}$ by its expressions
(\ref{S-matr})--(\ref{Dt}), we obtain 
two additional equations for the calculation of $E_0$, 
$\langle N_ss|\lambda=0\rangle$ and 
$\langle N_dd|\lambda=0\rangle$.

Clearly, in the case of coupled $sd$ waves, we should also 
fit the  completeness relation~(\ref{mcomplete}). 
We employ the following method of calculation of the sets of the
eigenvalues $E_\lambda$ and the components $\langle N_ss|\lambda\rangle$ and 
$\langle N_dd|\lambda\rangle$. The  $E_\lambda$ values with
$\lambda=1$, 2,~...~, ${N-2}$ are obtained by solving
Eq.~(\ref{DZero}) while the respective eigenvector's last components
$\langle N_ss|\lambda\rangle$ and  
$\langle N_dd|\lambda\rangle$  are calculated using
Eqs.~(\ref{Gams})--(\ref{Rel}). 
Next we perform a $\chi^2$ fit to
the scattering data of the parameters $E_0$, $E_{\lambda=N-1}$,
$E_{\lambda=N}$, $\langle N_ss|\lambda=0\rangle$,
 $\langle N_ss|\lambda=N-1\rangle$, $\langle N_dd|\lambda=N\rangle$,
$\langle N_dd|\lambda=0\rangle$, $\langle N_dd|\lambda=N-1\rangle$, and
$\langle N_dd|\lambda=N\rangle$. These 9 parameters fit 6 relations 
(\ref{ppcomplete}), (\ref{pfcomplete}), (\ref{ffcomplete}),
(\ref{ZDt}), (\ref{asC1}) and (\ref{asC2}), i.~e. we should perform a
three-parameter fit as in the case of coupled $pf$ waves.


Now we turn to the calculation of the remaining eigenvector components
$\langle n\Gamma|\lambda\rangle$ with $n<N_\Gamma$ and  the Hamiltonian
matrix elements $H_{nn'}^{\Gamma\Gamma'}$ with
$n\leq N_{\Gamma}$ and $n'\leq N_{\Gamma'}$ entering
Eq.~(\ref{Alge-m}). The coupled waves Hamiltonian matrix obtained by
the general $J$-matrix inverse scattering method is ambiguous; the
ambiguity originates from the multi-channel generalization of the
phase equivalent transformation mentioned
in the single channel case.
As in the single channel case, we
eliminate
the ambiguity by adopting a
particular form of the potential energy matrix.

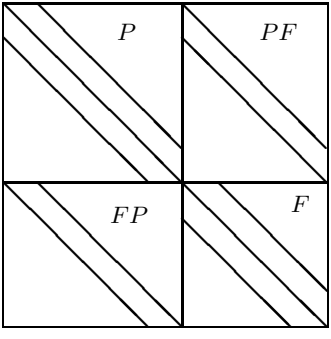
\begin{figure} 
\setlength{\unitlength}{747sp}%
\begin{picture}(10791,10791)(43,-9994)
\put( 76,-5161){\framebox(5925,5925){}}
\put(6001,-9961){\framebox(4800,4800){}} \thinlines
\put(6001,-5161){\framebox(4800,5925){}} \put(
76,-9961){\framebox(5925,4800){}} \thicklines \put( 76,764){\line(
1,-1){5925}} \put(6001,-5161){\line( 1,-1){4800}}
\put(6001,764){\line( 1,-1){4762.500}} \put(10801,-5161){\line(-1,
1){4800}} \put(1201,764){\line( 1,-1){4762.500}} \put(
76,-361){\line( 1,-1){4762.500}} \put(6001,-6361){\line(
1,-1){3600}} \put(7201,-5161){\line( 1,-1){3600}} \put(
76,-5161){\line( 1,-1){4762.500}} \put(1201,-5161){\line(
1,-1){4800}} \put(3826,-436){\makebox(0,0)[lb]{\smash{$P$}}}
\put(9601,-6136){\makebox(0,0)[lb]{\smash{$F$}}}
\put(8551,-436){\makebox(0,0)[lb]{\smash{$PF$}}}
\put(3601,-6511){\makebox(0,0)[lb]{\smash{$FP$}}}
\end{picture}
\caption{Structure of the ISTP matrix in the coupled $pf$ waves and of
the Version~0 ISTP in the coupled $sd$ waves. The location of
non-zero matrix is schematically illustrated by
solid lines. }\label{pf43pot} 
\end{figure}

As in the case of uncoupled partial waves, we construct $8\hbar\omega$
ISTP in the coupled $sd$ waves. Therefore
$2N_\Gamma+l_{\Gamma\vphantom{'}}=8$, or $2N_s+0=8$ and $2N_d+2=8$;
hence $N_s=N_d+1$. In the coupled $pf$ waves,  we construct
$7\hbar\omega$ and $9\hbar\omega$ ISTP; clearly we again have
$N_p=N_f+1$.  Thus the potential matrix $V_{nn'}^{\Gamma\Gamma'}$
has the
following structure: the submatrices $V_{nn'}^{\Gamma\Gamma}$ coupling
the oscillator components of the same partial wave are
quadratic [e.~g., $(N_p+1)\times(N_p+1)$ submatrix
$V_{nn'}^{pp}$ in the $^3p_2$ wave] while the submatrices
$V_{nn'}^{\Gamma\Gamma'}$ with $\Gamma\ne\Gamma'$ coupling
the oscillator components of different partial waves are
$(N_\Gamma+1)\times N_\Gamma$ or $N_\Gamma\times(N_\Gamma+1)$ matrices
 [e.~g., $(N_p+1)\times(N_p)$ submatrix $V_{nn'}^{pf}$ coupling   the
$^3p_2$ and $^3f_2$ waves]. Our assumptions are: we adopt  (i)~the
tridiagonal form of the quadratic submatrices $V_{nn'}^{\Gamma\Gamma}$
and (ii)~the simplest two-diagonal form of the non-quadratic
submatrices~$V_{nn'}^{\Gamma\Gamma'}$ with $\Gamma\ne\Gamma'$ coupling
the oscillator components of different partial waves. The structure of
the ISTP matrices in coupled partial waves is illustrated by
Fig.~\ref{pf43pot}.

Due to these assumptions, the algebraic problem (\ref{Alge-m}) takes
the following form:
\begin{widetext}
\begin{subequations}
\label{Systm}
\begin{align}
&H^{ss}_{00}  
\langle 0\,s|\lambda\rangle
 +H^{ss}_{01} 
\langle 1s|\lambda\rangle
 + H^{sd}_{00} 
\langle 0d|\lambda\rangle
 =E_\lambda\langle0s|\lambda\rangle,
 \label{Systm1a}  \\[1ex]
& H^{ds}_{00} 
\langle 0s|\lambda\rangle
 +H^{ds}_{01} 
\langle 1s|\lambda\rangle
 + H^{dd}_{00} 
\langle 0d|\lambda\rangle
 +H^{dd}_{01} 
\langle 1d|\lambda\rangle
  =E_\lambda\langle0d|\lambda\rangle ,
 \label{Systm1}  \\[1ex]
%
&H^{ss}_{n,n-1} 
\langle n-1,s|\lambda\rangle
 +H^{ss}_{nn} 
\langle ns|\lambda\rangle
 +H^{ss}_{n,n+1} 
\langle n+1,s|\lambda\rangle  
+H^{sd}_{n,n-1} 
\langle n-1,d|\lambda\rangle  
  +H^{sd}_{nn} 
\langle nd|\lambda\rangle
  =E_\lambda\langle ns|\lambda\rangle
\notag\\
&\hspace{356pt}
\qquad (n=1,\,2,\,...\,,N_s-1),
 \label{Systm2a} \displaybreak[2]  \\[1ex]
& H^{ds}_{nn} 
\langle ns|\lambda\rangle
 + H^{ds}_{n,n+1} 
\langle n+1,s|\lambda\rangle
 + H^{dd}_{n,n-1} 
\langle n-1,d|\lambda\rangle
 +H^{dd}_{nn} 
\langle nd|\lambda\rangle  +H^{dd}_{n,n+1} 
\langle n+1,d|\lambda\rangle
  =E_\lambda\langle i\,d|\lambda\rangle \notag\\
&\hspace{356pt}
\qquad (n=1,\,2,\,...\,,N_d-1),
 \label{Systm2} \displaybreak[3] \\[1ex]
%
&H^{ss}_{N_s,N_s-1} 
\langle N_s-1,s|\lambda\rangle
 +H^{ss}_{N_sN_s} 
\langle N_ss|\lambda\rangle
 +H^{sd}_{N_sN_d} 
\langle N_dd|\lambda\rangle
  =E_\lambda\langle N_ss|\lambda\rangle,
 \label{Systm3a} 
\intertext{and}
& H^{ds}_{N_d,N_s-1} 
\langle N_s-1,s|\lambda\rangle
 + H^{ds}_{N_dN_s} 
\langle N_ss|\lambda\rangle
+ H^{dd}_{N_d,N_d-1} 
\langle N_d-1,d|\lambda\rangle   
+ H^{dd}_{N_dN_d} 
\langle N_dd|\lambda\rangle
  =E_\lambda\langle N_dd|\lambda\rangle .
 \label{Systm3}
\end{align}
\end{subequations}
Even though
this set of equations 
is more complicated than the set (\ref{Syst1}) discussed in the
uncoupled waves case,
%
it can be solved in the same manner.

Multiplying Eqs.~(\ref{Systm3a})--(\ref{Systm3}) by
$\langle N_ss|\lambda\rangle$ and $\langle N_dd|\lambda\rangle$,
summing the results over $\lambda$ and using the completeness
relation~(\ref{mcomplete}) we obtain 
\begin{subequations}
\label{HNN}
\begin{align}
 \label{hNsNs}
&H^{ss}_{N_sN_s} 
= \sum_{\lambda=0}^N
 E_{\lambda} \langle N_ss|\lambda\rangle ^2 , \\[1ex]
 \label{hNdNd}
& H^{dd}_{N_dN_d} 
= \sum_{\lambda=0}^N
 E_{\lambda} \langle N_dd|\lambda\rangle ^2 , 
\intertext{and}
 \label{hNsNd}
& H^{sd}_{N_sN_d} 
= \sum_{\lambda=0}^N
 E_{\lambda} \langle N_ss|\lambda\rangle
\langle \lambda |  N_d\,d \rangle .
\end{align}
\end{subequations}
Now we multiply each of the equations~(\ref{Systm3a})--(\ref{Systm3})
by its Hermitian conjugate and one of these equations
by the  Hermitian conjugate
of the other, sum the results over $\lambda$ and use
(\ref{mcomplete}) to obtain
\begin{subequations}
\label{HN1N}
\begin{align}
\label{hNs1Ns}
&H^{ss}_{N_s,N_s-1} 
= -\,\sqrt{\sum_{\lambda=0}^N E_{\lambda}^2\,{\langle N_ss|\lambda\rangle}^2
- \left(H^{ss}_{N_sN_s}\right)^2 
- \left(H^{sd}_{N_sN_d}\right)^2 
           }, \\[1ex]
 \label{hNs1Nd}
& H^{ds}_{N_d,N_s-1}    
=\frac{1}{H^{ss}_{N_s,N_s-1}}
\left[ \sum_{\lambda=0}^N E_{\lambda}^2\,\langle N_ss|\lambda\rangle
\, \langle\lambda | N_d\,d\rangle
- H^{sd}_{N_sN_d} 
\Bigl(  H^{ss}_{N_sN_s} 
 +  H^{dd}_{N_dN_d} 
 \Bigr) \right], 
\intertext{and}
\label{hNd1Nd}
& H^{dd}_{N_d,N_d-1} 
= -\,\sqrt{\sum_{\lambda=0}^N E_{\lambda}^2\,{\langle N_d\,d|\lambda\rangle}^2
- \left(H^{dd}_{N_dN_d}\right)^2 
- \left(H^{sd}_{N_sN_d}\right)^2 
-\big(H^{ds}_{N_d,N_s-1}\big)^2 
           }.
\end{align}
\end{subequations}
As in the case of uncoupled waves, we
take
the off-diagonal
matrix elements\- $H^{ss}_{N_s,N_s\pm 1}$ and $H^{dd}_{N_d,N_d\pm 1}$
to be
dominated by the respective kinetic energy\- matrix elements
$T^{s}_{N_s,N_s\pm 1}$ and $T^{d}_{N_d,N_d\pm 1}$ and therefore choose
the minus sign in the right-hand-sides of Eqs.~(\ref{hNs1Ns})
and~(\ref{hNd1Nd}).

By means of Eqs.~(\ref{HNN})--(\ref{HN1N}) we obtain all matrix
elements $H^{\Gamma\Gamma'}_{nn'}$ entering
Eqs.~(\ref{Systm3a})--(\ref{Systm3}). Using this information,  the eigenvector
components $\langle N_s-1,s|\lambda\rangle$
and $\langle N_d-1,d|\lambda\rangle$ can be extracted directly from
Eqs.~(\ref{Systm3a})--(\ref{Systm3}):
\begin{subequations}
\begin{align}
\label{gN1sNs}
& \langle N_s-1,s|\lambda\rangle =
\frac{1}{H^{ss}_{N_s,N_s-1}}
\Bigl( E_{\lambda}\, \langle N_ss|\lambda\rangle
- H^{ss}_{N_sN_s} 
\langle N_ss|\lambda\rangle
- H^{sd}_{N_sN_d} 
\langle N_dd|\lambda\rangle \Bigr)  
\intertext{and}
%
\label{gN1dNd}
& \langle N_d-1,d|\lambda\rangle =
\frac{1}{H^{dd}_{N_d,N_d-1}}
\Bigl( E_{\lambda}\, \langle N_d\,d|\lambda\rangle
 -H^{dd}_{N_dN_d} 
\langle N_dd|\lambda\rangle
- H^{ds}_{N_dN_s} 
\langle N_ss|\lambda\rangle 
 - H^{ds}_{N_d,N_s-1} 
\langle N_s-1,s|\lambda\rangle
  \Bigr) .
\end{align}
\end{subequations}

Now we can perform the same manipulations with
Eqs.~(\ref{Systm1a})--(\ref{Systm2}). We
take
$n=N_s-1$, $N_s-2$,~...~, 1 in Eq.~(\ref{Systm2a}) and $n=N_d-1$,
$N_d-2$,~...~,~1 in Eq.~(\ref{Systm2}).
Equations~(\ref{Systm2a})--(\ref{Systm2})  are a bit more
complicated than  Eqs.~(\ref{Systm3a})--(\ref{Systm3}), however the
additional terms in Eqs.~(\ref{Systm2a})--(\ref{Systm2}) include only
the quantities calculated on the previous step.
As a result,
we
obtain the following relations for the calculation of the
 matrix elements $H^{\Gamma\Gamma'}_{nn}$:
\begin{subequations}
\label{HGG}
\begin{align}
 \label{hisis}
& H^{ss}_{nn} 
= \sum_{\lambda=0}^N
 E_{\lambda} \,{\langle ns|\lambda\rangle} ^2 , \\[1ex]
 \label{hidid}
& H^{dd}_{nn} 
= \sum_{\lambda=0}^N
 E_{\lambda} \,{\langle nd|\lambda\rangle }^2 , 
\intertext{and}
 \label{hisid}
&  H^{sd}_{nn} 
= \sum_{\lambda=0}^N
 E_{\lambda} \,\langle n|\lambda\rangle
\langle \lambda |  nd \rangle .
\end{align}
\end{subequations}
Equation~(\ref{hisis}) is valid for $n=N_s,$ $N_s-1$,~...~,~0 while
equations~(\ref{hidid})--(\ref{hisid}) are valid for $n=N_d,$
$N_d-1$,~...~,~0.

For the matrix elements $H^{\Gamma\Gamma'}_{n,n-1}$
we obtain
\begin{subequations}
\label{HGG1}
\begin{align}
\label{hidi1d}
& H^{dd}_{n,n-1} 
=-\,\sqrt{\sum_{\lambda=0}^N E_{\lambda}^2\, {\langle nd|\lambda\rangle}^2
  - \big(H^{ds}_{nn}\big)^2 
- \left(H^{ds}_{n,n+1}\right)^2 
  -\big( H^{dd}_{nn}\big)^2 
-\big( H^{dd}_{n,n+1}\big)^2 
                       }, \\[1ex]
 \label{hisi1d}
& H^{sd}_{n,n-1}  
= \frac{1}{H^{dd}_{n,n-1}}
\left[ \sum_{\lambda=0}^N
E_{\lambda}^2\, \langle ns|\lambda\rangle   \langle\lambda | nd\rangle
- H^{sd}_{nn} 
\Bigl(H^{ss}_{nn}  
 + H^{dd}_{nn} 
       \Bigr)
 - H^{ss}_{n,n+1} 
\,H^{ds}_{n,n+1}  
 \right], 
\intertext{and}
\label{hisi1s}
& H^{ss}_{n,n-1} 
=-\,\sqrt{\sum_{\lambda=0}^N E_{\lambda}^2\,{\langle ns|\lambda\rangle}^2
  -\big(H^{ss}_{nn}\big)^2 
- \big(H^{ss}_{n,n+1}\big)^2 
  -\big(H^{sd}_{n,n-1}\big)^2 
- \big(H^{sd}_{nn}\big)^2 
                 }.
\end{align}
 \end{subequations}
Equation (\ref{hidi1d}) is valid for $n=N_d-1$, $N_d-2$,~...~,~1;
equation~(\ref{hisi1d}) is valid for $n=N_d$, $N_d-1$,~...~,~1,
and equation~(\ref{hisi1s}) is valid for $n=N_s-1$, $N_s-2$,~...~,~1.

The eigenvector components  $\langle n-1,s|\lambda\rangle$ with
$n=N_s-1$, $N_s-2$,~...~,~1 and
$\langle n-1,d|\lambda\rangle $ with $n=N_d-1$, $N_d-2$,~...~,~1
can be calculated using the following expressions:
\begin{subequations}
\label{EigNN1}
\label{gi1ss}
\begin{align}
& \langle n-1,s|\lambda\rangle =
\frac{1}{H^{ss}_{n,n-1}}
 \Bigl( E_{\lambda} \langle ns|\lambda\rangle
- H^{ss}_{nn} 
\langle ns|\lambda\rangle
-  H^{ss}_{n,n+1} 
\langle n+1,s|\lambda\rangle  
-  H^{sd}_{n,n-1} 
\langle n-1,d|\lambda\rangle
- H^{sd}_{nn} 
\langle nd|\lambda\rangle \Bigr) 
\intertext{and}
\label{gi1dd}
& \langle n-1,d|\lambda\rangle =
\frac{1}{H^{dd}_{n,n-1}} 
 \Bigl( E_{\lambda} \langle nd|\lambda\rangle
- H^{ds}_{nn} 
\langle ns|\lambda\rangle
-H^{ds}_{n,n+1} 
\langle n+1,s|\lambda\rangle 
- H^{dd}_{nn} 
\langle nd|\lambda\rangle
- H^{dd}_{n,n+1} 
\langle n+1,d|\lambda\rangle \Bigr) .
\end{align}
\end{subequations}
\end{widetext}

Having calculated the Hamiltonian matrix elements
$H^{\Gamma\Gamma'}_{nn'}$, we obtain the potential energy matrix
elements $V^{\Gamma\Gamma'}_{nn'}$ by subtracting the kinetic energy.

We
recall
here that we arbitrarily assigned the values $s$ and $d$ to
the channel index $\Gamma$
but
the above theory can be
applied to any pair of coupled partial waves. The only equations
specific for the $sd$ coupled partial waves case are Eqs.~
(\ref{ZDt})--(\ref{asC}) that are needed to account for the
experimental information about the bound state which is present in the
$np$ system in the $sd$ coupled partial waves. In
equations~(\ref{Kss-dd}), (\ref{Match})--(\ref{Rel}) and
(\ref{Systm})--(\ref{EigNN1}) one can substitute $s$ and $d$ by $p$
and $f$, respectively, and use them for constructing the ISTP in the
coupled $pf$ waves.

 We construct ISTP in the coupled $NN$ partial waves  using as
input the $np$ scattering phase shifts and mixing parameters
reconstructed from the experimental data by the Nijmegen group
\cite{Stocks}. We start the discussion from the ISTP in the coupled
$pf$ waves.

\begin{figure}
\centerline{\psfig{figure=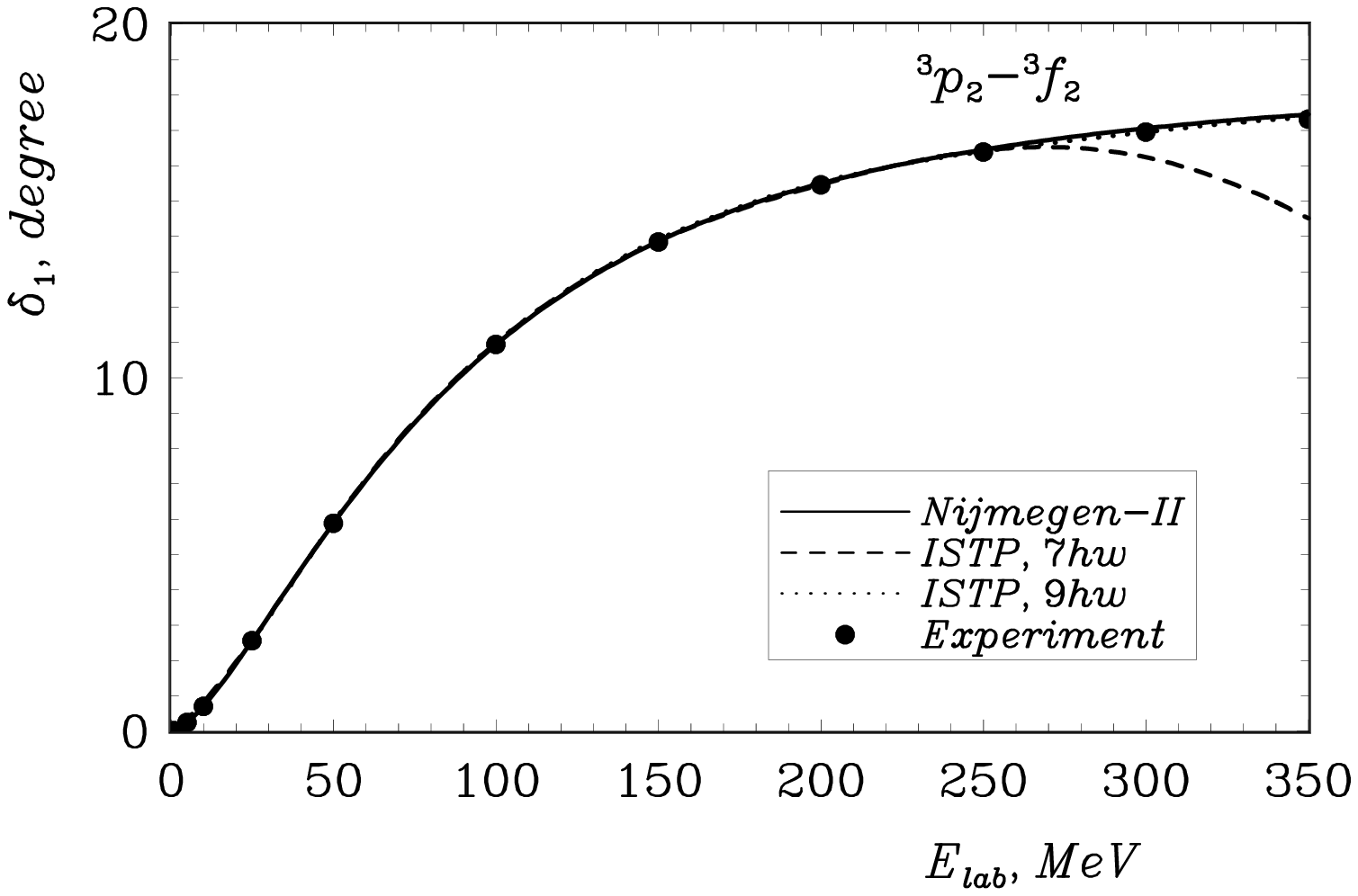,width=0.5\textwidth}}
\caption{${^3p_2}$  $np$ scattering phase shifts $\delta_p$ (coupled $pf$
waves).
Filled circles --- experimental data of Ref.~\cite{Stocks}; solid
line~--- realistic meson exchange Nijmegen-II potential~\cite{Stocks}
phase shifts; dashed line --- $7\hbar\omega$ ISTP phase shifts; dotted
line~--- $9\hbar\omega$ ISTP phase shifts.} \label{php3pf2}
\vspace{3ex}
\centerline{\psfig{figure=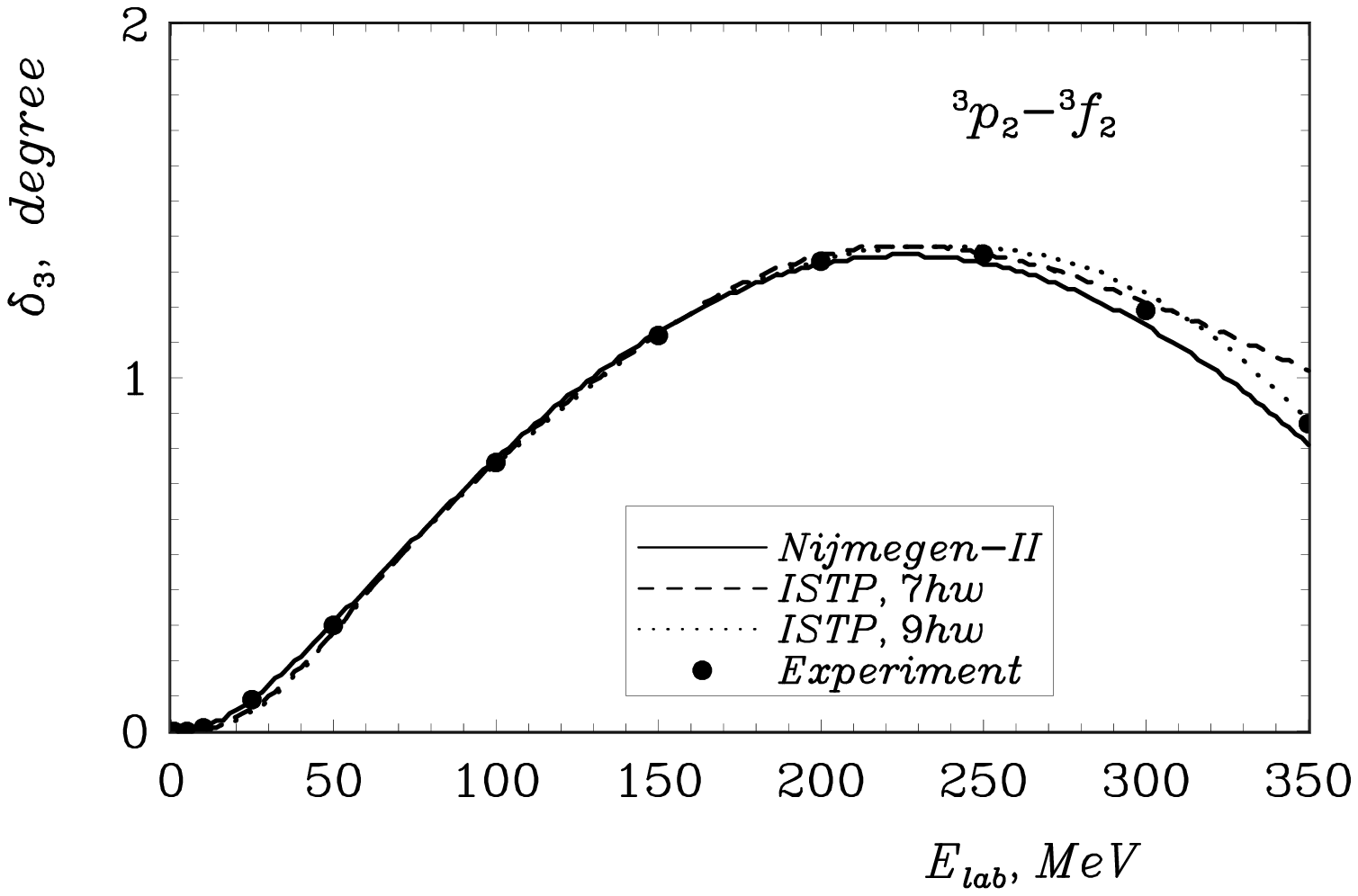,width=0.5\textwidth}}
\caption{${^3f_2}$  $np$ scattering phase shifts $\delta_f$ (coupled $pf$
waves). See
Fig.~\ref{php3pf2} for details.} \label{phf3pf2}
\vspace{3ex}
%
%
\centerline{\psfig{figure=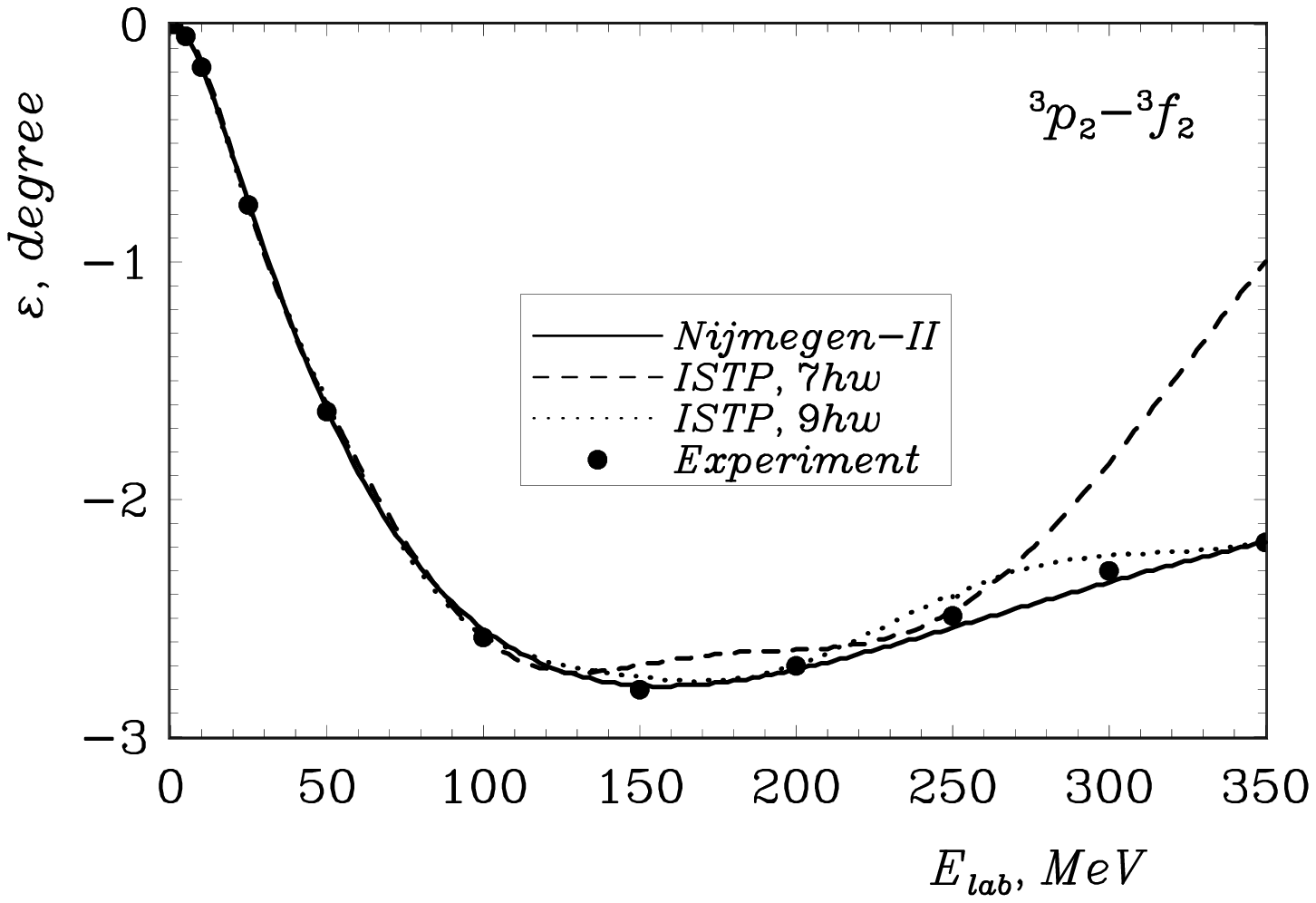,width=0.5\textwidth}}
\caption{ $np$ scattering mixing parameter $\varepsilon$ in the coupled
$pf$ waves. See
Fig.~\ref{php3pf2} for details.} \label{eps3pf2}
\end{figure}

  The non-zero potential energy matrix elements of the
obtained $7\hbar\omega$ $pf$-ISTP  are
given in Table~\ref{pot3pf1}  (in $\hbar\omega=40$~MeV units).
The description of the  phase shifts $\delta_p$ and
$\delta_f$ and of the  mixing parameter $\varepsilon$ is shown in
Figs.~\ref{php3pf2}--\ref{eps3pf2}.  The phenomenological data are
seen to be well reproduced by the $7\hbar\omega$ ISTP up to the
laboratory energy $E_{\rm lab}\approx 270$~MeV; at higher energies
there are discrepancies between the ISTP predictions and the
experimental data that are most pronounced in the ${^3p_2}$ partial
wave (note the very different scales in Fig.~\ref{php3pf2} and
Figs.~\ref{phf3pf2}--\ref{eps3pf2}). These discrepancies are seen to be
eliminated by constructing the  $9\hbar\omega$ $pf$-ISTP.

\begin{figure}
\centerline{\psfig{figure=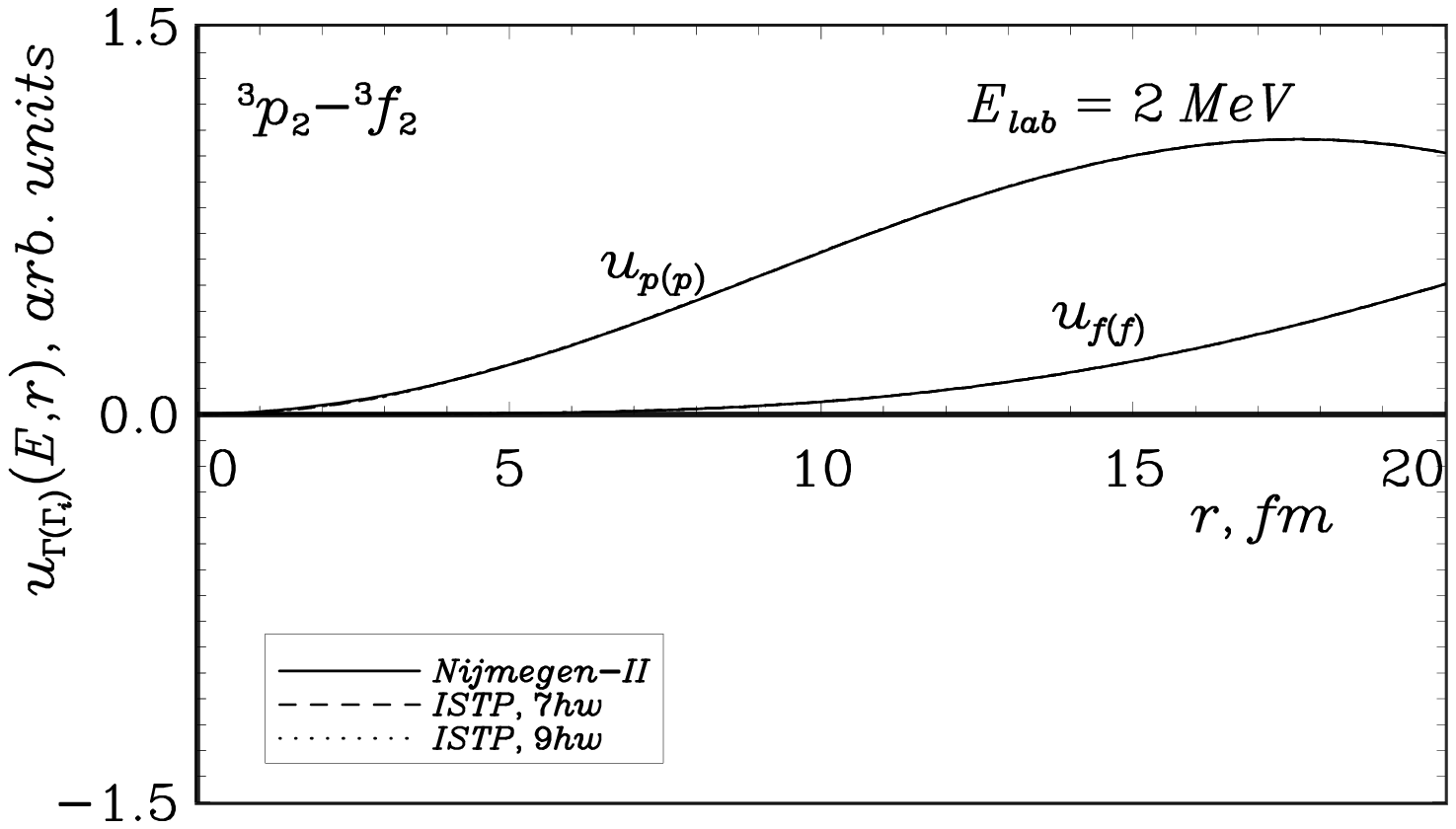,width=0.5\textwidth}}
\caption{Large components $u_{p(p)\protect\vphantom{'}}(E,r)$ and
$u_{f(f)\protect\vphantom{'}}(E,r)$  
of the coupled $pf$ waves
$np$ scattering wave function at the laboratory energy
$E_{\rm lab}=2$~MeV. See
Fig.~\ref{php3pf2} for details.} \label{wf02pf}
\vspace{4ex}
%
\centerline{\psfig{figure=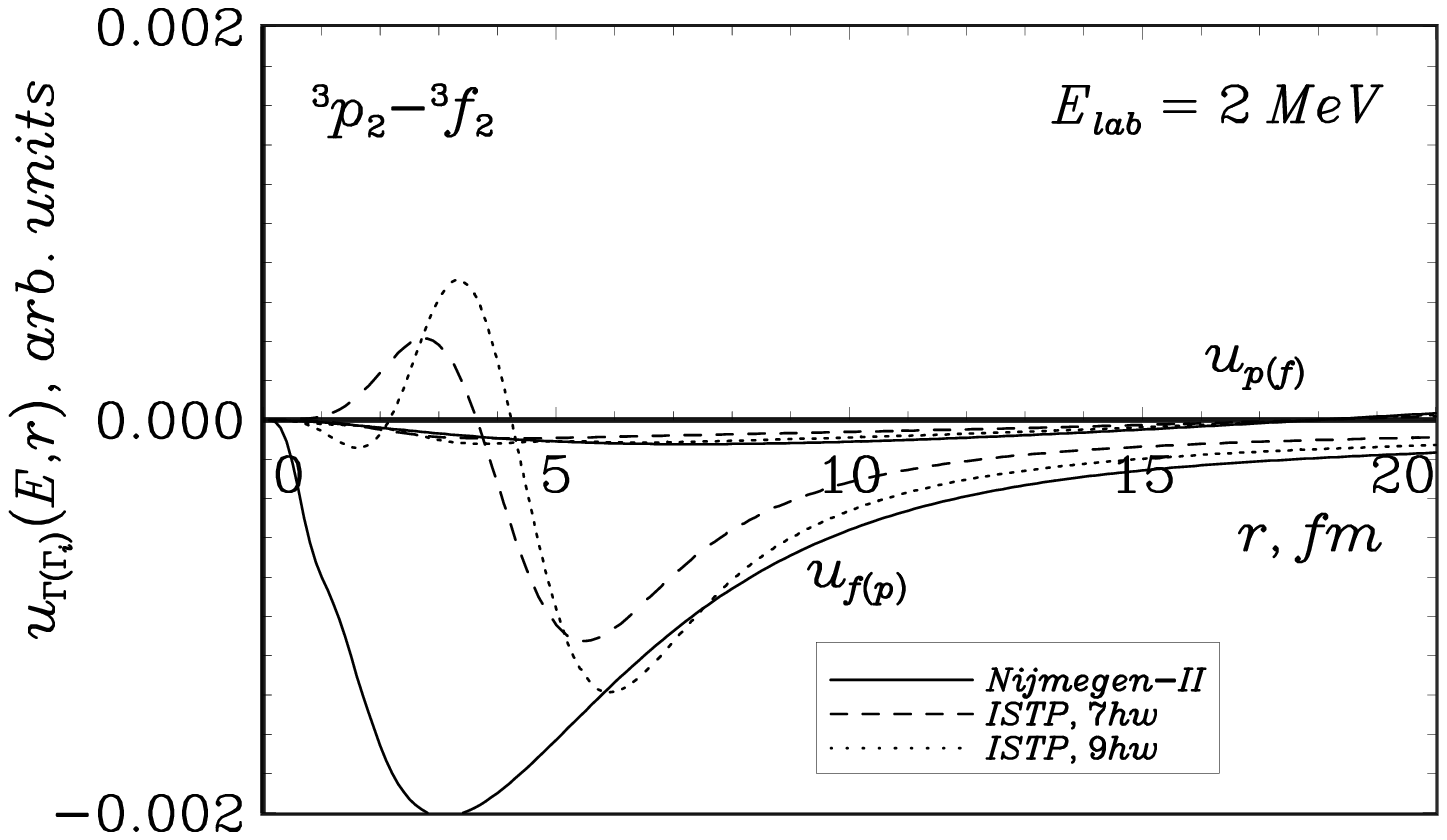,width=0.5\textwidth}}
\caption{Small components $u_{p(f)\protect\vphantom{'}}(E,r)$ and
$u_{f(p)\protect\vphantom{'}}(E,r)$  
of the coupled $pf$ waves
$np$ scattering wave function at the laboratory energy
$E_{\rm lab}=2$~MeV. See
Fig.~\ref{php3pf2} for details.} \label{wf02pfs}
\vspace{4ex}
%
\centerline{\psfig{figure=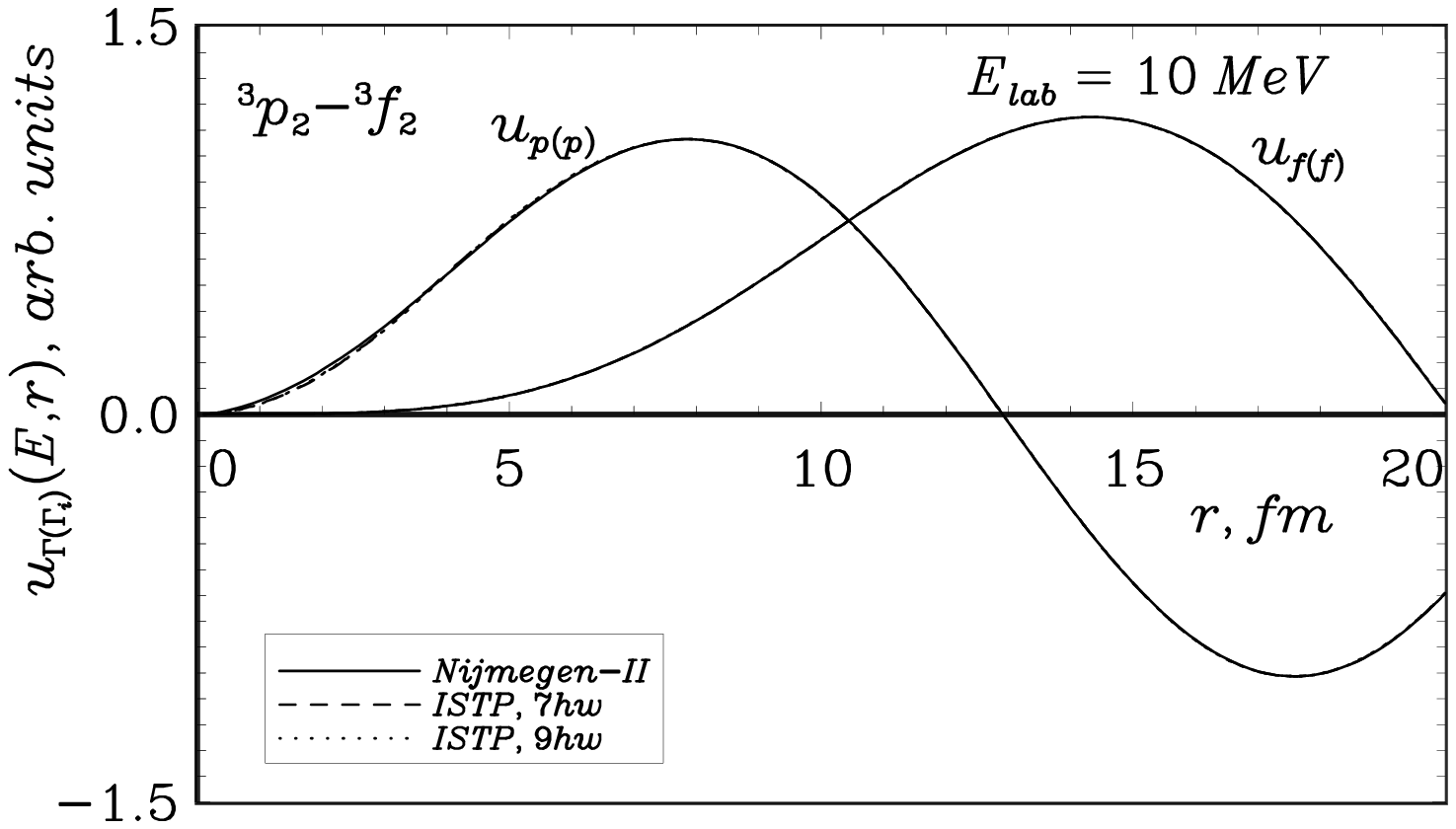,width=0.5\textwidth}}
\caption{Large components $u_{p(p)\protect\vphantom{'}}(E,r)$ and
$u_{f(f)\protect\vphantom{'}}(E,r)$  of the coupled $pf$ waves
$np$ scattering wave function at the laboratory energy
$E_{\rm lab}=10$~MeV. See
Fig.~\ref{php3pf2} for details.}
\label{wf10pf}
\end{figure}

\begin{figure}
\centerline{\psfig{figure=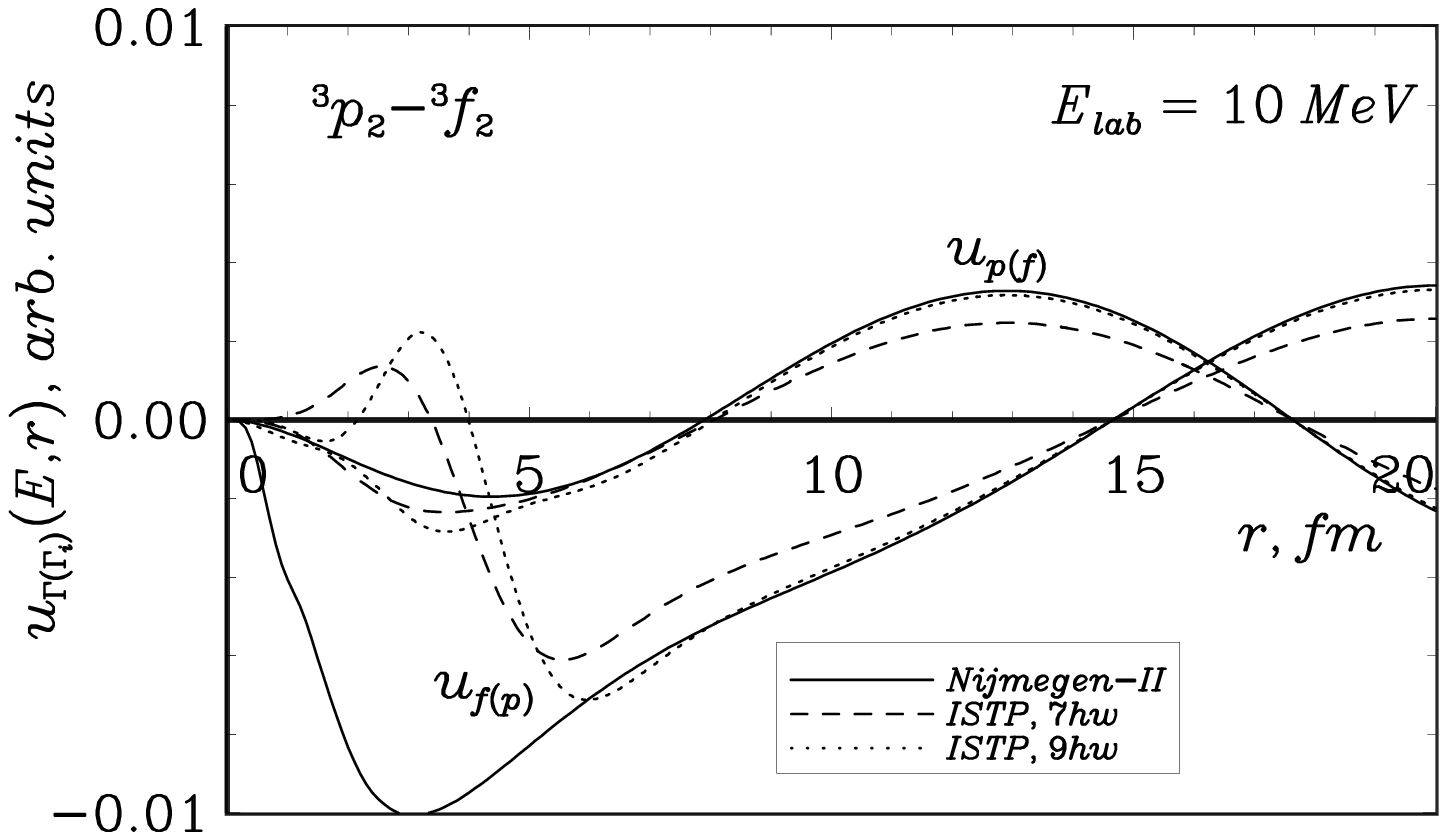,width=0.5\textwidth}}
\caption{Small components $u_{p(f)\protect\vphantom{'}}(E,r)$ and
$u_{f(p)\protect\vphantom{'}}(E,r)$  of the coupled $pf$ waves
$np$ scattering wave function at the laboratory energy
$E_{\rm lab}=10$~MeV. See
Fig.~\ref{php3pf2} for details.}
\label{wf10pfs}
\vspace{4ex}
\centerline{\psfig{figure=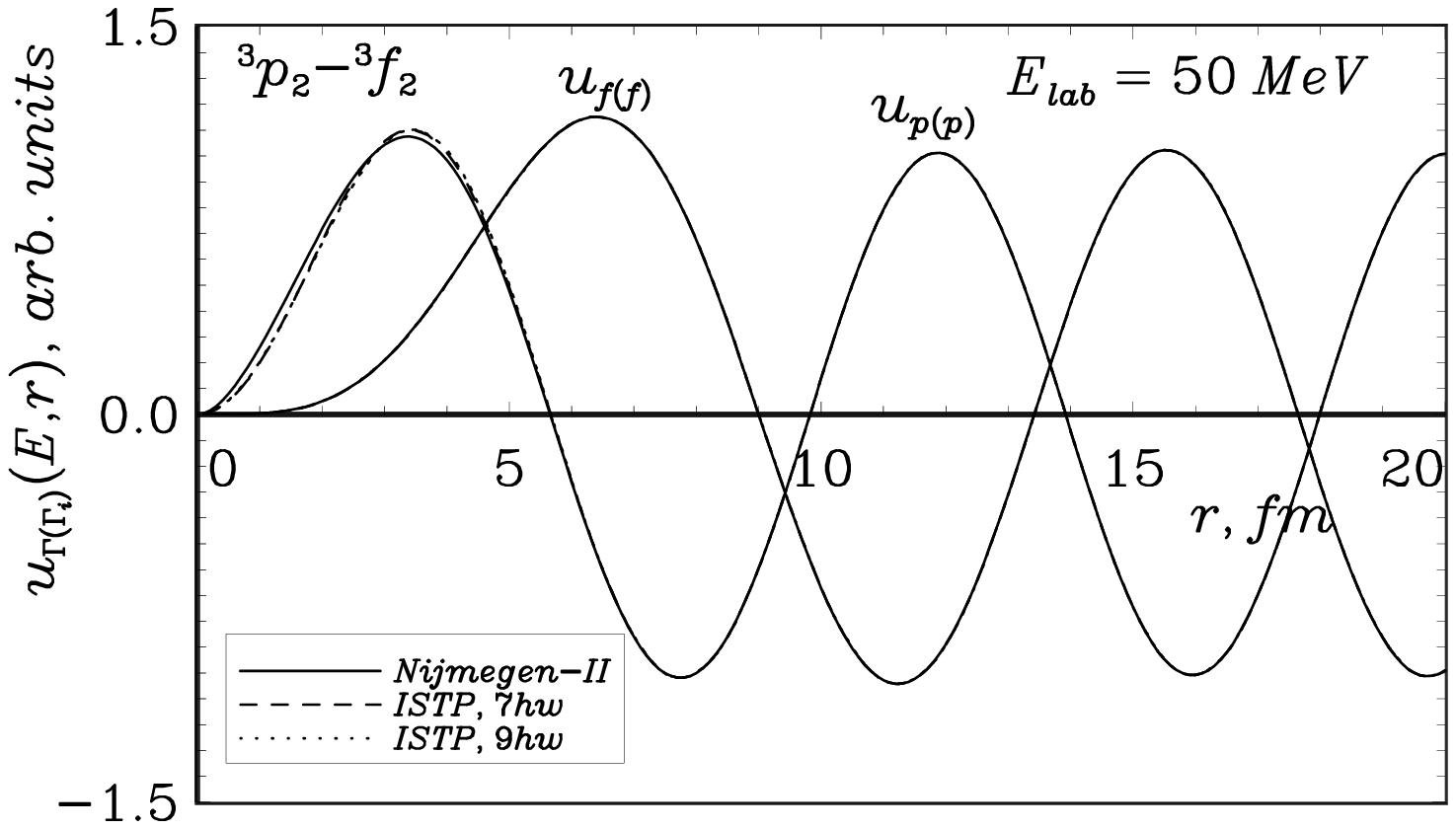,width=0.5\textwidth}}
\caption{Large components $u_{p(p)\protect\vphantom{'}}(E,r)$ and
$u_{f(f)\protect\vphantom{'}}(E,r)$  of the coupled $pf$ waves
$np$ scattering wave function at the laboratory energy
$E_{\rm lab}=50$~MeV. See
Fig.~\ref{php3pf2} for details.}
\label{wf50pf}
\vspace{4ex}
%
\centerline{\psfig{figure=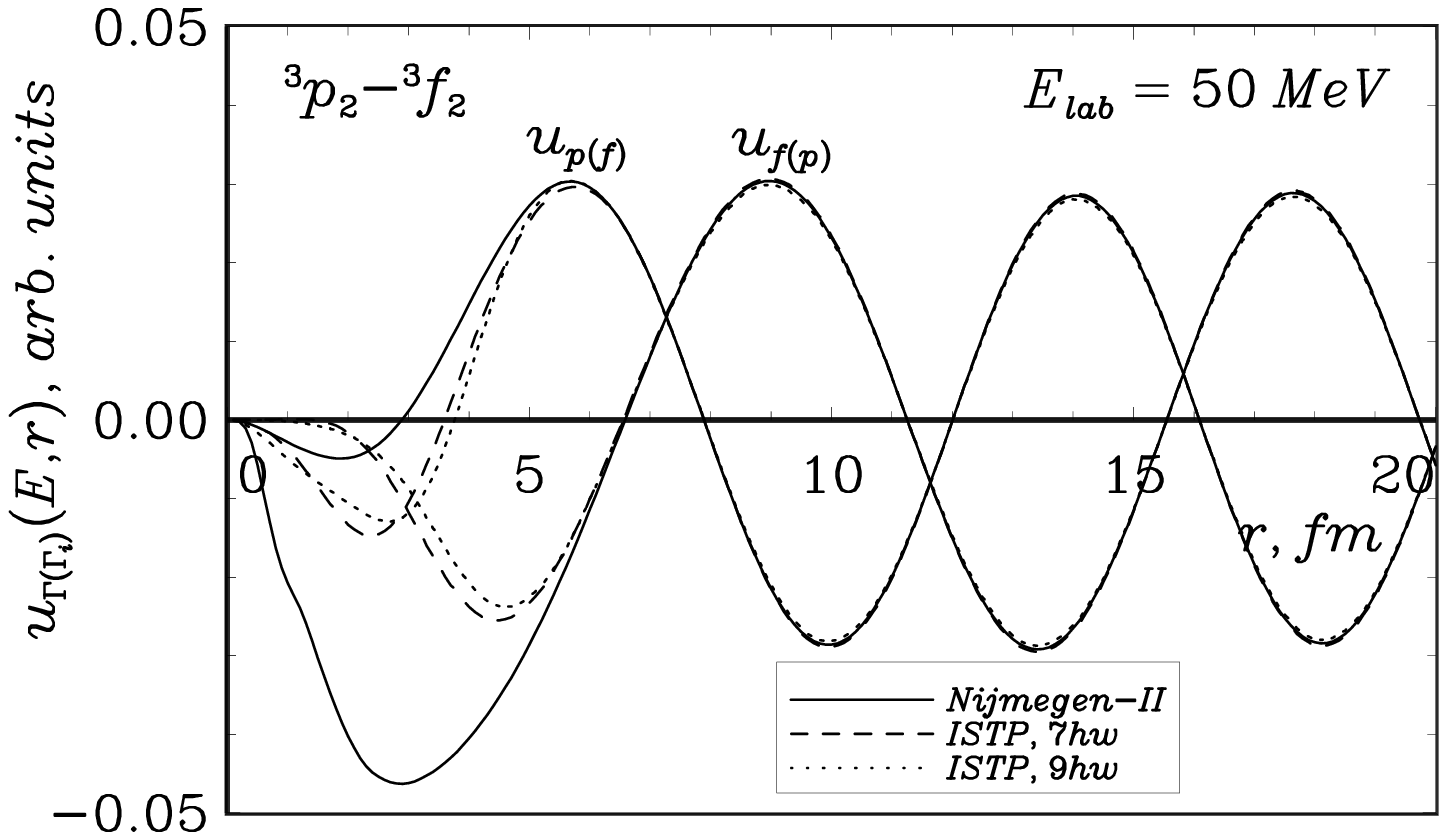,width=0.5\textwidth}}
\caption{Small components $u_{p(f)\protect\vphantom{'}}(E,r)$ and
$u_{f(p)\protect\vphantom{'}}(E,r)$  of the coupled $pf$ waves
$np$ scattering wave function at the laboratory energy
$E_{\rm lab}=50$~MeV. See
Fig.~\ref{php3pf2} for details.}
\label{wf50pfs}
\end{figure}

\begin{figure}
\centerline{\psfig{figure=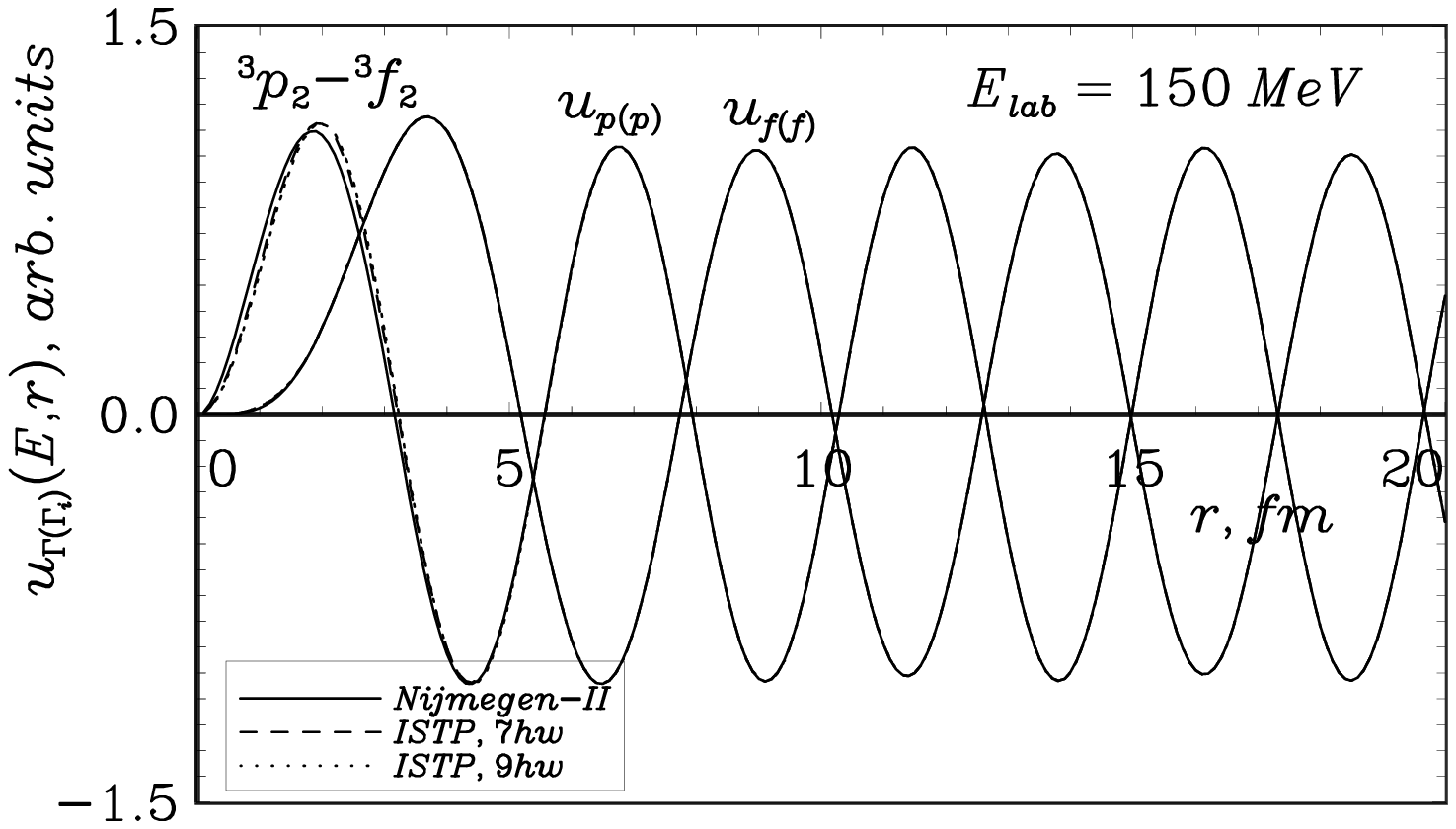,width=0.5\textwidth}}
\caption{Large components $u_{p(p)\protect\vphantom{'}}(E,r)$ and
$u_{f(f)\protect\vphantom{'}}(E,r)$  of the coupled $pf$ waves
$np$ scattering wave function at the laboratory energy
$E_{\rm lab}=150$~MeV. See
Fig.~\ref{php3pf2} for details.}
\label{wf150pf}
\vspace{4ex}
%
\centerline{\psfig{figure=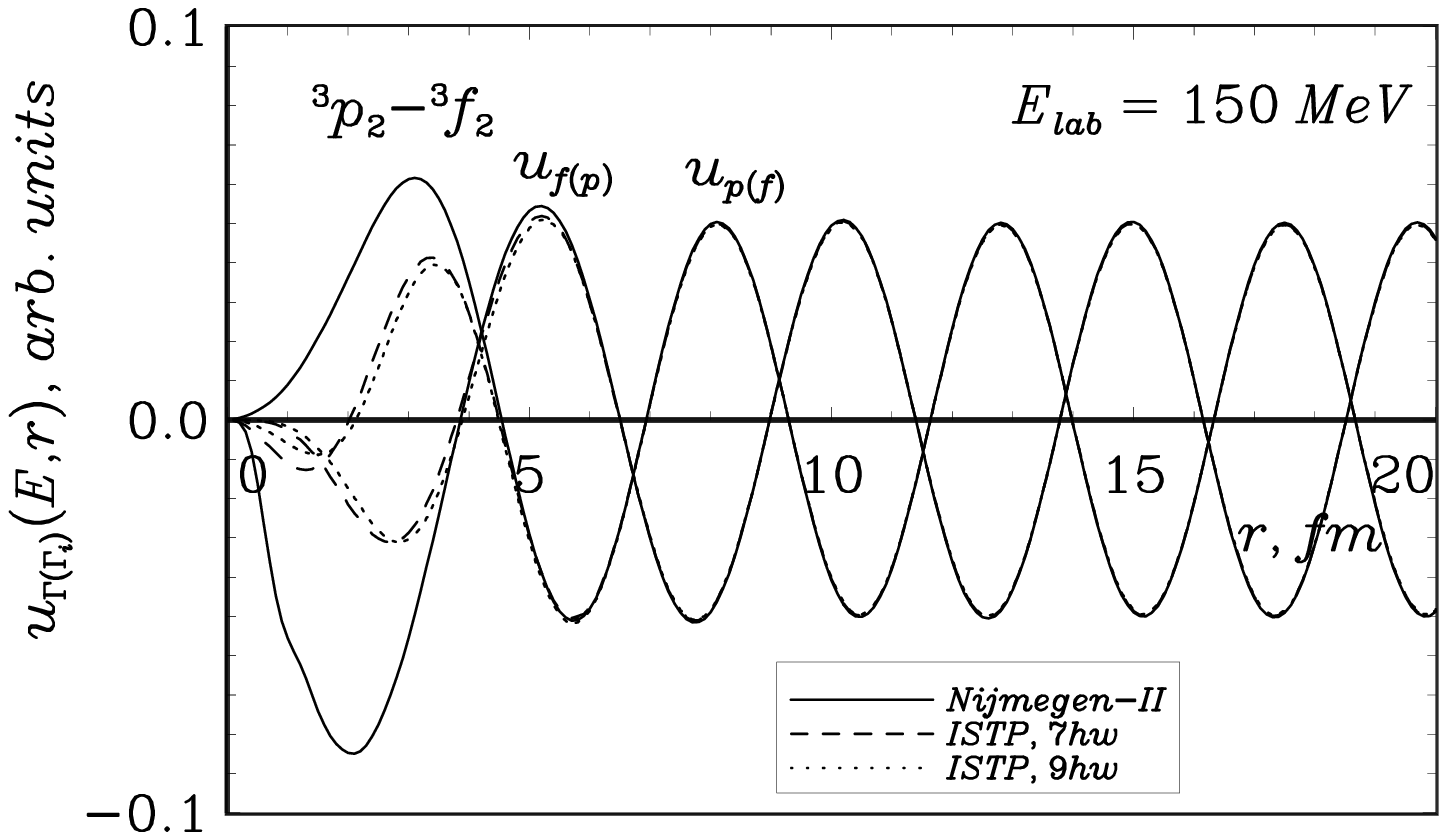,width=0.5\textwidth}}
\caption{Small components $u_{p(f)\protect\vphantom{'}}(E,r)$ and
$u_{f(p)\protect\vphantom{'}}(E,r)$  of the coupled $pf$ waves
$np$ scattering wave function at the laboratory energy
$E_{\rm lab}=150$~MeV. See
Fig.~\ref{php3pf2} for details.}
\label{wf150pfs}
\vspace{4ex}
\centerline{\psfig{figure=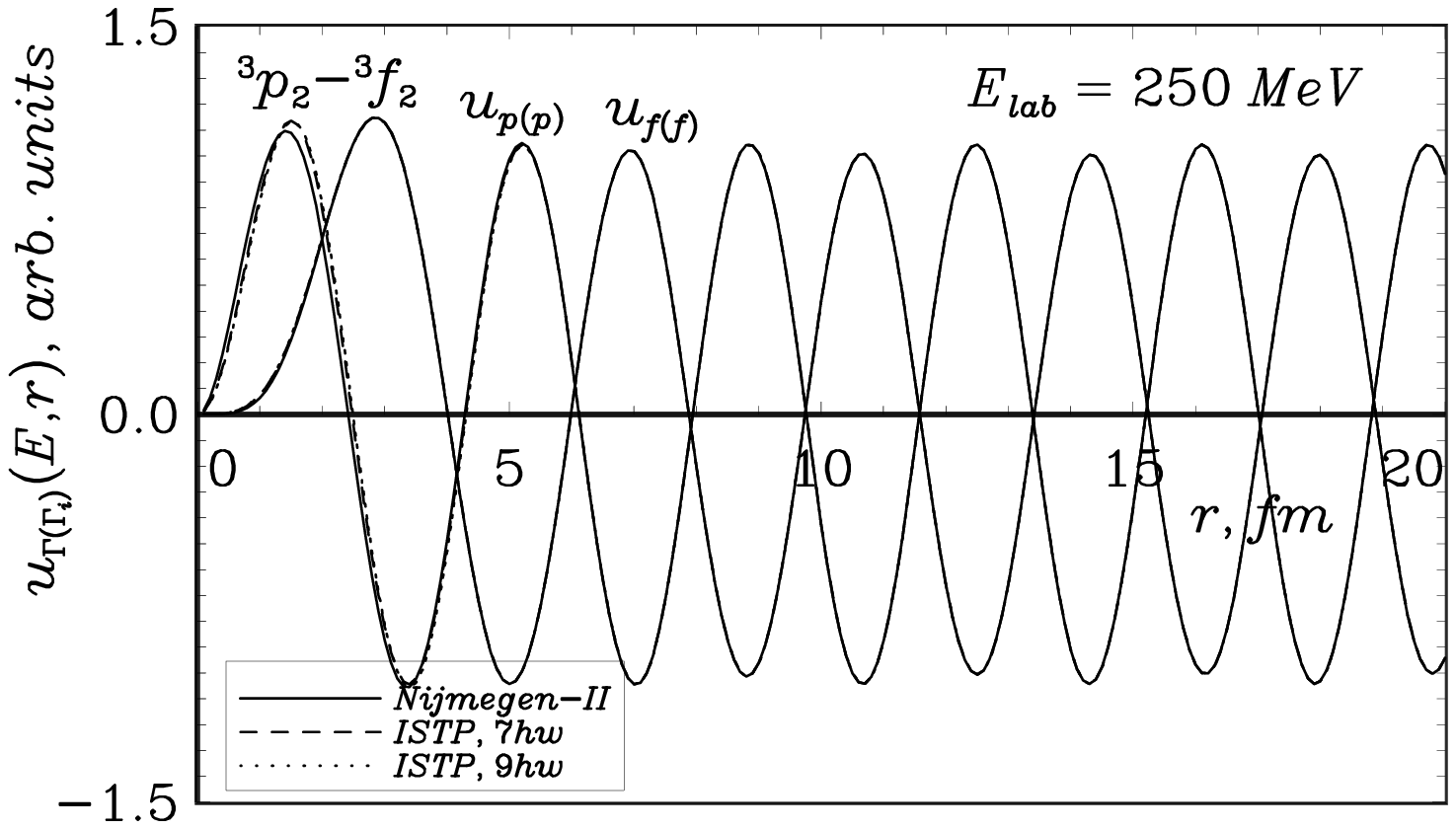,width=0.5\textwidth}}
\caption{Large components $u_{p(p)\protect\vphantom{'}}(E,r)$ and
$u_{f(f)\protect\vphantom{'}}(E,r)$ of the coupled $pf$ waves
$np$ scattering wave function at the laboratory energy
$E_{\rm lab}=250$~MeV. See
Fig.~\ref{php3pf2} for details.}
\label{wf250pf}
\end{figure}

\begin{figure}
\centerline{\psfig{figure=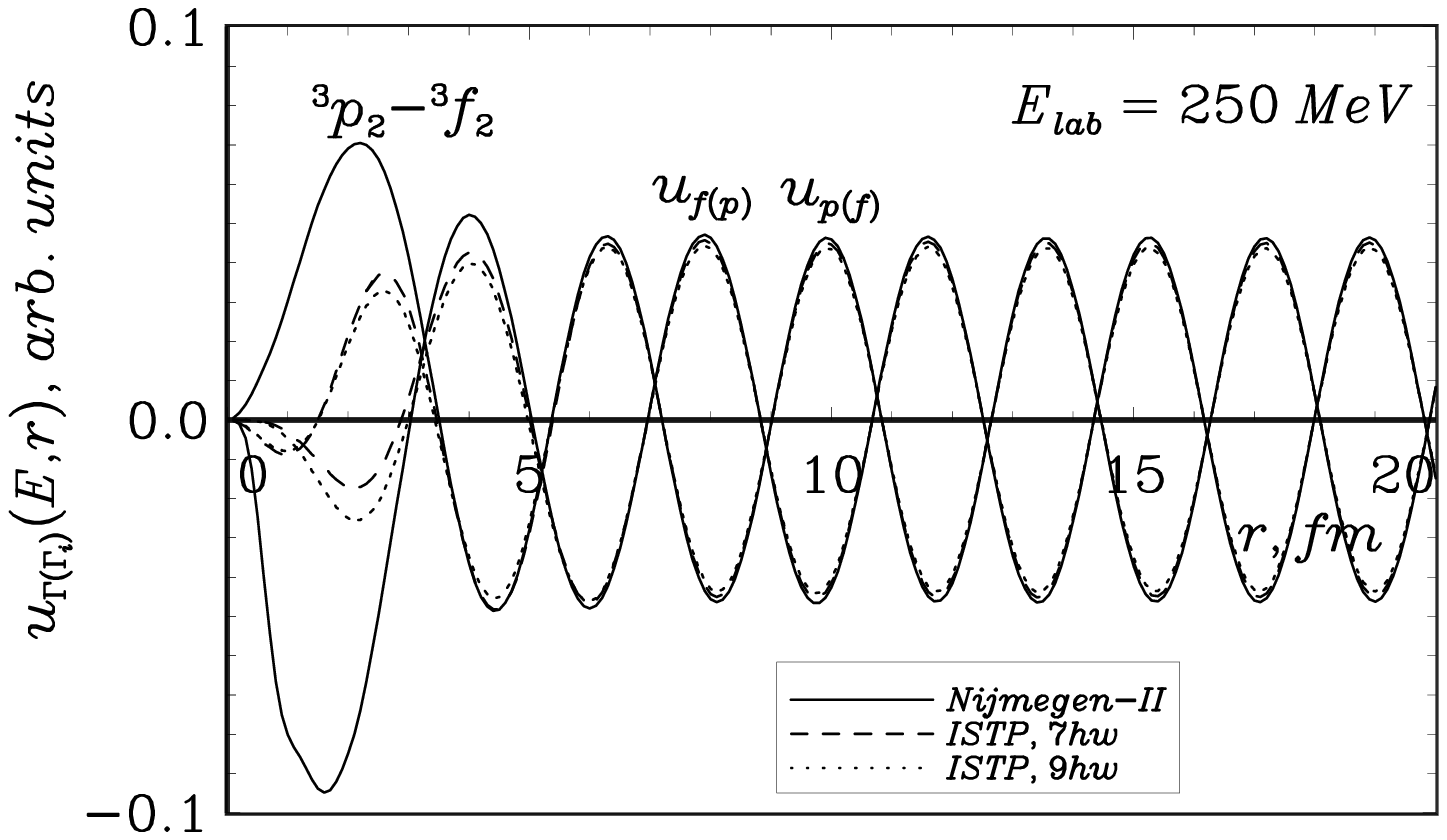,width=0.5\textwidth}}
\caption{Small components $u_{p(f)\protect\vphantom{'}}(E,r)$ and
$u_{f(p)\protect\vphantom{'}}(E,r)$ of the coupled $pf$ waves
$np$ scattering wave function at the laboratory energy
$E_{\rm lab}=250$~MeV. See
Fig.~\ref{php3pf2} for details.}
\label{wf250pfs}
\end{figure}

\begin{table} 
\caption{Non-zero matrix elements in $\hbar\omega$ units of the
$7\hbar\omega$ ISTP matrix in the $pf$ coupled partial waves.}
\label{pot3pf1}
\begin{ruledtabular}
\begin{tabular}{>{$}c<{$}>{$}c<{$}>{$}c<{$}} 
 \multicolumn{3}{c}{$V^{pp}_{nn'\vphantom{,}}$ matrix elements} \\
  n & V_{n n}^{pp}  & V_{n, n+1}^{pp}=V_{n+1,n}^{pp}\\ \hline
 0&  -0.08320586302194144 & 0.06828130087634332 \\
 1&  -0.1733874789751888 &  0.09710466067410284\\
 2&   -0.1630792532684091 &   0.04737005443342412 \\
 3&  -0.02514449050450729 &  \\[1.5ex] 
 \multicolumn{3}{c}{$V^{ff}_{nn'\vphantom{,}}$ matrix elements} \\
n & V_{n n}^{ff} & V_{n, n+1}^{ff}= V_{n+1,n}^{ff} \\ \hline
&    -0.01860731179640451 &  0.008146529480927311\\
&     -0.01230112258503713 &  0.002878668408624830 \\
&  -0.002274165031966646 & \\[1.5ex] 
  \multicolumn{3}{c}{ $V^{pf}_{nn'\vphantom{,}}$ matrix elements } \\
  n & V_{n, n-1}^{pf}= V_{n-1,n}^{fp}  & V_{n n}^{pf}=V^{fp}_{nn}  \\ \hline
 0&     & 0.03113837433227350  \\
 1&  -0.02731096515966312 &  0.02654889981453685  \\
 2&  -0.005320397951202732 & -0.007039900977944484 \\
 3&   0.009906839670436384 &     
\end{tabular}
\end{ruledtabular}
\end{table}

Generally, for the
coupled $pf$ waves,
we have 4 radial wave function components
$u_{p(p)\vphantom{'}}(E,r)$, $u_{p(f)\vphantom{'}}(E,r)$,
$u_{f(p)\vphantom{'}}(E,r)$ and  $u_{f(f)\vphantom{'}}(E,r)$ defined
according to their standing wave asymptotics~(\ref{Krow2}).
We present in Figs. \ref{wf02pf}--\ref{wf250pfs} the plots of these 
components at the laboratory energies $E_{\rm lab}=2$, 10, 50, 150 and
250~MeV obtained with the $7\hbar\omega$ and  $9\hbar\omega$ ISTP in
comparison with the respective Nijmegen-II wave function components. 
%

It is seen from the figures that the $9\hbar\omega$ ISTP
and  Nijmegen-II `large' (diagonal) wave function components 
$u_{p(p)\protect\vphantom{'}}(E,r)$ and
$u_{f(f)\protect\vphantom{'}}(E,r)$
are indistinguishable. The same 
$7\hbar\omega$ ISTP components 
 differ a little from
those of Nijmegen\mbox{-}II at high energies. At the same time, the
`small' (non-diagonal) ISTP wave function components 
$u_{p(f)\protect\vphantom{'}}(E,r)$ and
$u_{f(p)\protect\vphantom{'}}(E,r)$ differ essentially at small
distances  from the
Nijmegen\mbox{-}II ones. It is a clear indication of a very
different nature of the ISTP tensor interaction. 

%

\begin{figure}
\centerline{\psfig{figure=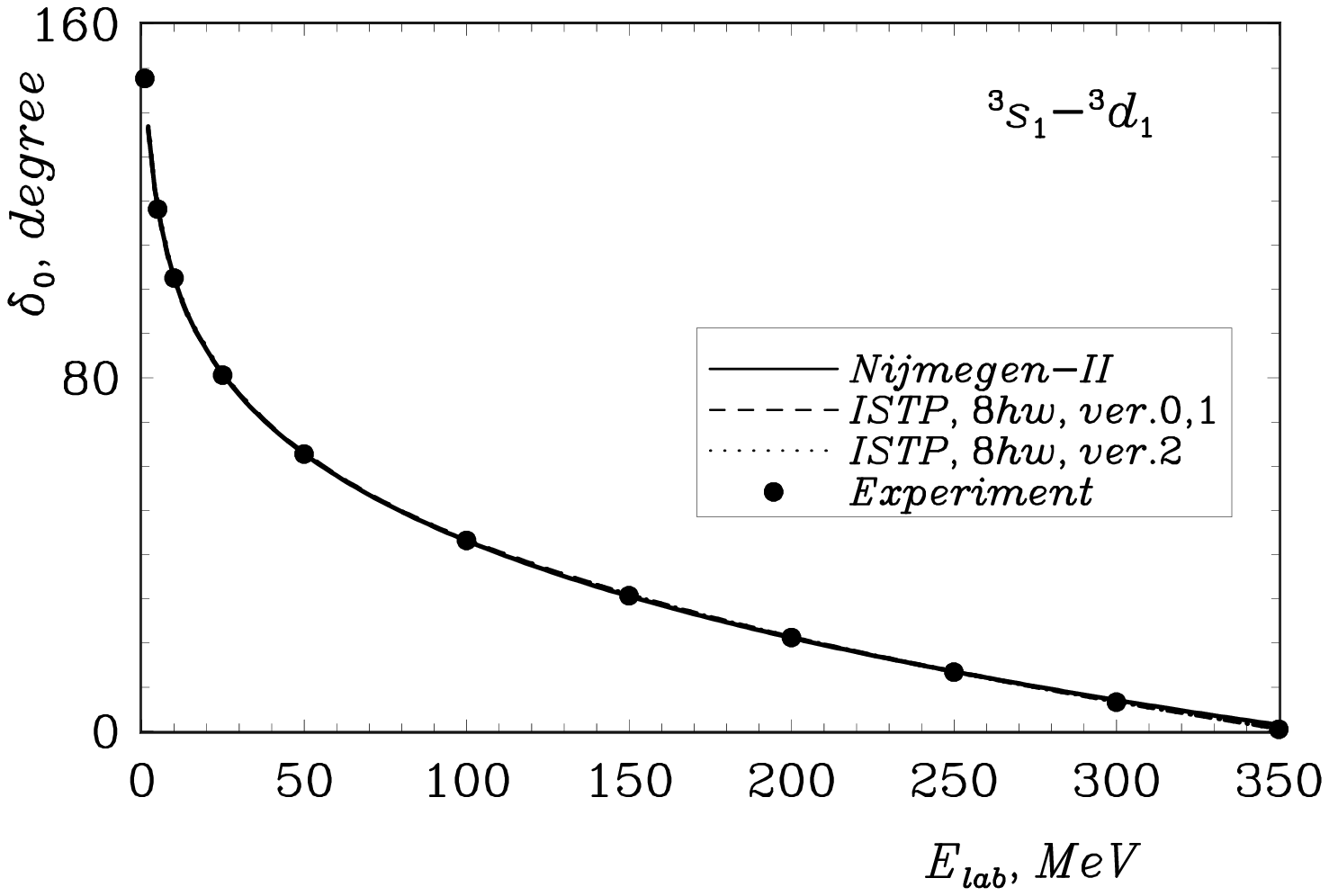,width=0.5\textwidth}}
\caption{${^3s_1}$  $np$ scattering phase shifts $\delta_s$ (coupled
$sd$ waves).
Filled circles --- experimental data of Ref.~\cite{Stocks}; solid
line~--- realistic meson exchange Nijmegen-II potential~\cite{Stocks}
phase shifts; dashed line --- Version~0 and Version~1 ISTP phase shifts; dotted
line~--- Version~2 ISTP phase shifts.} \label{phs3sd1}
\vspace{3ex}
\centerline{\psfig{figure=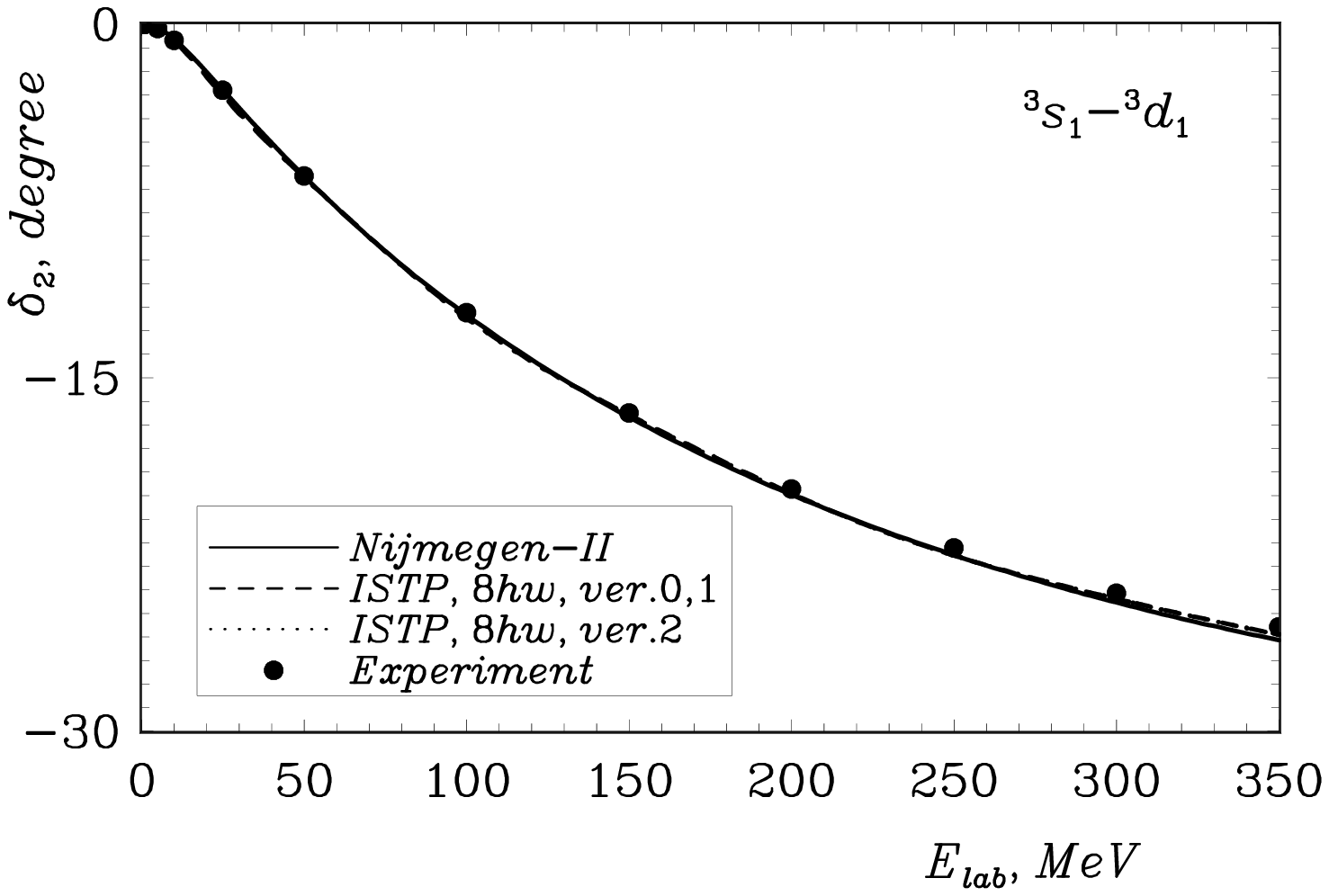,width=0.5\textwidth}}
\caption{${^3d_1}$  $np$ scattering phase shifts $\delta_d$ (coupled
$sd$ waves). See Fig.~\ref{phs3sd1} for details.} \label{phd3sd1}
\vspace{3ex}
%
\centerline{\psfig{figure=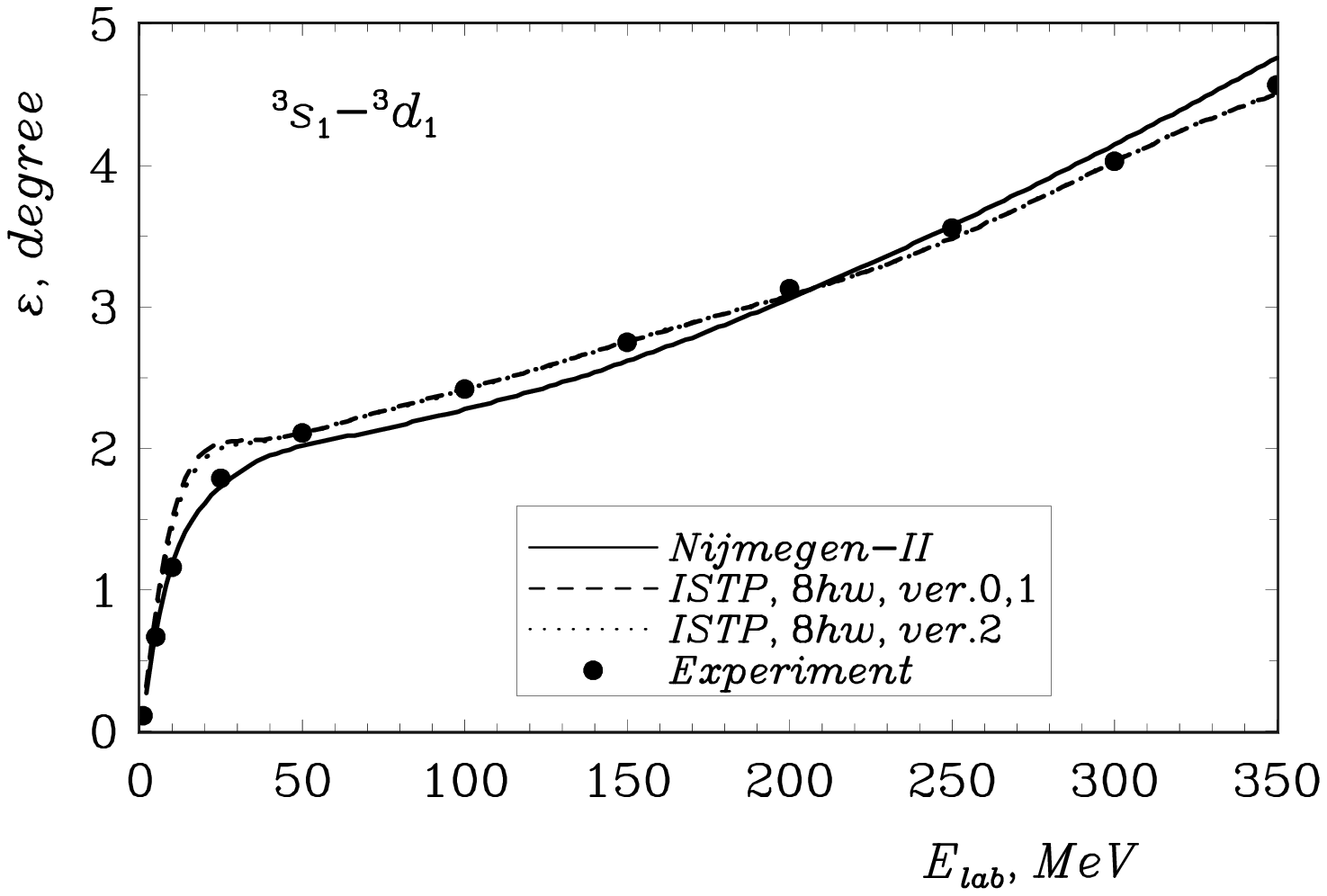,width=0.5\textwidth}}
\caption{$np$ scattering mixing parameter $\varepsilon$ in the coupled
$sd$ waves.  See Fig.~\ref{phs3sd1} for details.} \label{eps3sd1}
\end{figure}

Now we apply the inverse scattering $J$-matrix approach to the coupled
$sd$ partial waves and obtain the $8\hbar\omega$ ISTP hereafter
refered to as Version~0 ISTP. The description of the phenomenological
data by this potential
(and other ISTP versions discussed later)
is shown in
Figs.~\ref{phs3sd1}--\ref{eps3sd1}. The $np$ $s$ wave and $d$ wave phase
shifts  $\delta_s$ and $\delta_d$ are excellently reproduced up to the
laboratory energy of 350~MeV. There is a small discrepancy between the
experimental and  the Version~0 ISTP mixing parameter $\varepsilon$ at
the laboratory energy of $E_{\rm lab}\approx 25$~MeV. However, the
overall
Version~0
ISTP description of
experimental scattering data (including the
mixing parameter~$\varepsilon$)
over the full
energy interval $E_{\rm
lab}=0\div 350$~MeV is seen from
Figs.~\ref{phs3sd1}--\ref{eps3sd1} to be
competitive with the 
%
Nijmegen-II, one of the best realistic meson exchange potentials.

The Version 0 ISTP is constructed by fitting the experimental
scattering data, the deuteron ground state energy $E_d$, the
$s$ wave asymptotic normalization constant ${\mrs A}_s$ and
$\displaystyle\eta=\frac{{\mrs A}_d}{{\mrs A}_s}$. However, there are
other important deuteron observables known experimentally
such as
the deuteron rms radius ${\langle r^2\rangle}^{-1/2}$ and the
probability of the $d$ state.
Various deuteron properties obtained
with the Version~0 ISTP (and other ISTP versions discussed later) are
compared in Table~\ref{tabl}
with the predictions obtained with Nijmegen-II potential and with
recent compilations of the experimental data~\cite{deSwart,Petrov}. It
is seen from the table that the Version~0 ISTP overestimates the
deuteron rms radius and
underestimates the $d$ state probability.
%

\begin{table*}
\caption{Deuteron property predictions obtained with various
$8\hbar\omega$ ISTP versions  and with
Nijmegen-II potential in comparison with recent compilations
\cite{deSwart,Petrov}. }\label{tabl}
\begin{ruledtabular}
\begin{tabular}{cccccc} 
Potential & $E_{d}$, MeV &\parbox{2.1cm}{$d$ state probability, \%}
& \parbox{1.6cm}{rms radius, fm}
                   & 
                   ${\mrs A}_s$, fm$^{-1/2}$
&$\displaystyle\eta=\frac{{\mrs A}\vphantom{'}_d}
          {{\mrs A}_{s\vphantom{_a}}}$\\ \hline
Version 0 & $-2.224575$ & 0.4271 & 1.9877& 0.8845 &  0.0252\\
Version 1 & $-2.224575$ & 5.620 & 1.9997 &0.8845 & 0.0252\\
Version 2 & $-2.224575$ & 5.696 & 1.968  &0.8629 &0.0252\\
 Nijmegen-II & $-2.224575$ & 5.635 & 1.968 &0.8845 & 0.0252\\
Compilation \cite{deSwart} & $-2.224575(9)$ & 5.67(11) &1.9676(10)
&0.8845(8) &0.0253(2)          \\[.7ex]
Compilation \cite{Petrov} &$-2.224589$
&--- &$\left\{\hspace{-7pt}\parbox{1.6cm}{1.9635$\vphantom{'}$
                              1.9560
                              1.950$\vphantom{_q}$}\right.$
& 0.8781  &0.0272
\end{tabular}
\end{ruledtabular}
\end{table*}

\begin{figure}[b]
\centerline{\psfig{figure=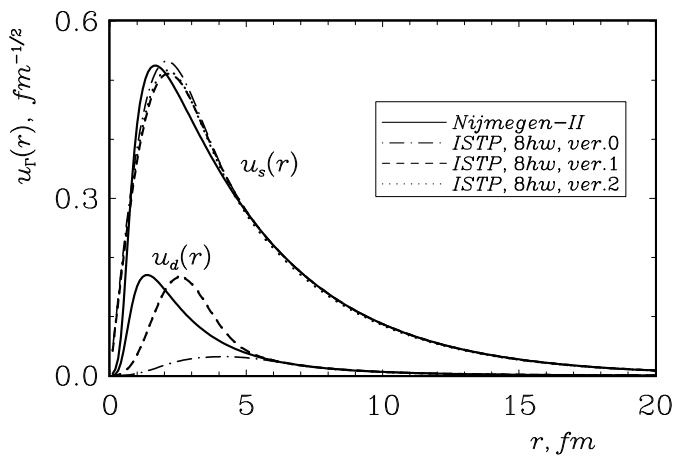,width=0.5\textwidth}}
\caption{Radial deuteron wave functions. Solid
line~--- realistic meson exchange Nijmegen\mbox{-}II potential~\cite{Stocks}
wave functions; dot-dash line~---  Version~0  ISTP wave functions;
dashed line --- Version~1 ISTP wave functions; dotted
line~--- Version~2 ISTP wave functions.} \label{wfdeut}
\end{figure}

The deuteron wave functions can be calculated by utilizing the
$J$-matrix formalism at the negative energy $E_d$ as is discussed in
Ref.~\cite{Halo-b,Halo-c}. The plots of the deuteron wave functions are
presented in Fig.~\ref{wfdeut}. It is seen that the Version~0 ISTP $s$
wave component is very close to
that of Nijmegen-II.
The  Version~0
ISTP $d$ wave component coincides with
that of Nijmegen-II
at large
distances since both potentials provide the same ${\mrs A}_d$ value;
however at the distances less than 5~fm the  Version~0 ISTP  $d$ wave
component is
suppressed. We note also that the Version~0
ISTP scattering wave functions (not shown in the figures below)
are significantly different from those of Nijmegen-II at short
distances.
%

Our conclusion is that the  Version~0 ISTP does not seem to be a
realistic $NN$ potential.

To improve the description of the deuteron properties,  it
appears
natural to apply to our Version~0 ISTP a phase equivalent
transformation that leaves
unchanged the scattering observables $\delta_s$, $\delta_d$,
$\varepsilon$, the deuteron ground state energy $E_d$ and the deuteron
asymptotic normalization constants ${\mrs A}_s$ and ${\mrs A}_d$. The
phase equivalent transformation discussed in Refs.~\cite{PHT,Halo-b,Halo-c}
is very convenient for our purposes since it is defined in the
oscillator basis. This transformation
gives rise to an
ambiguity of
the potential fit within the inverse scattering $J$-matrix approach,
which have been mentioned several times already. We now need
to discuss this
in more detail.

This phase equivalent transformation is based on the unitary
transformation
\begin{subequations}
\label{Unitary}
\begin{equation}
U= \sum_{\Gamma=s,d} \;\sum_{\Gamma'=s,d}\;
\sum_{n=0}^{\infty}\;\sum_{n'=0}^{\infty} \,
| n \Gamma \rangle\: U_{n n'}^{\Gamma \Gamma '}\: \langle n'\Gamma ' | ,
\label{Unit}
\end{equation}
where the unitary matrix $\left[U\right]$
with matrix elements~$U_{n n'}^{\Gamma \Gamma '}$ should be of the
form \cite{PHT,Halo-b,Halo-c}
\begin{equation}
\left[U\right] = \left[U_{0}\right]  \oplus  \left[I\right]
            = \left[\begin{array}{c|c}
                          \left[U_{0}\right] &  0  \\ \hline
                               0             & \left[I\right]
                       \end{array}\right]
\label{PhEq2}
\end{equation}
\end{subequations}
and $\left[I\right]$ is the infinite 
unit matrix. The
unitary transformation~(\ref{Unitary}) is applied to the
infinite 
Hamiltonian matrix $\left[H\right]$ in the
oscillator basis~$\left\{|n\Gamma\rangle\right\}$:
\begin{equation}
\lbrack \widetilde{H}]=\left[ U\right] \left[ H\right] \left[ U^{+}\right] .
\label{PhEq1}
\end{equation}
The transformed Hamiltonian $\widetilde{H}$ is defined  through its
(infinite) matrix $[\widetilde{H}]$ with matrix elements
$\widetilde{H}^{\Gamma\Gamma'}_{nn'}\equiv\langle n\Gamma|
\widetilde{H}|n'\Gamma'\rangle$.
That is,
the matrix $[\widetilde{H}]$
is obtained by means of the
unitary transformation~(\ref{PhEq1})
%
{\em
in the original
basis~$\left\{|n\Gamma\rangle\right\}$ and not in the transformed basis
$\{|\widetilde{n\Gamma}\rangle\}\equiv U\{|n\Gamma\rangle\}$.}
Clearly the spectra of the Hamiltonians $H$ and $\widetilde{H}$ are
identical. If the submatrix $\left[U_{0}\right]$ is small enough, the
unitary transformation~(\ref{PhEq1}) leaves unchanged the
last components $\langle N_\Gamma\Gamma|\lambda\rangle$ of the
eigenvectors $\langle n\Gamma|\lambda\rangle$ obtained by solving the
algebraic problem (\ref{Alge-m}) and hence it leaves unchanged  the functions
${\cal G}_{\Gamma\Gamma'}$ that completely determine the $K$-matrix,
the $S$-matrix, the phase shifts $\delta_s$ and $\delta_d$, the mixing
parameter $\varepsilon$, the asymptotic normalization constants
${\mrs A}_s$ and ${\mrs A}_d$, etc.

The potential $\widetilde{V}$ entering the Hamiltonian $\widetilde{H}$,
phase equivalent to the initial potential $V$ entering  the
Hamiltonian $H$, can be expressed as
\begin{subequations}
\label{PHTV}
\begin{gather}
\widetilde{V}=V+\Delta V,
\label{DV-non-loc}
\intertext{where}
\Delta {V}= 
\widetilde{H}-{H}.
\label{PhEq3}
\end{gather}
\end{subequations}

We should improve the tensor component of the $NN$ interaction to
increase the $d$ state probability in the deuteron
and reduce the rms radius.
%
%
Therefore the
only non-trivial submatrix $\left[U_{0}\right]$  of the
matrix~(\ref{PhEq2}) should couple the oscillator components
$|ns\rangle$ and $|n'd\rangle$ of different partial waves. We
take
the simplest form of  the  submatrix $\left[U_{0}\right]$:
a $2\times 2$ matrix coupling the $|0s\rangle$ and $|0d\rangle$
basis functions. In other words, the non-trivial matrix elements
$U^{\Gamma\Gamma'}_{nn'}$ constitute a $2\times 2$ rotation matrix with a
single continuous parameter~$\vartheta$:
\begin{subequations}
\label{UPh}
\begin{gather}
\left[U_{0}\right]=\begin{bmatrix}
   U^{ss}_{00}&  U^{sd}_{00} \\
    U^{ds}_{00}     &    U^{dd}_{00}
\end{bmatrix}
=\begin{bmatrix}
  \cos\vartheta & +\sin\vartheta \\
             -\sin\vartheta & \cos\vartheta
\end{bmatrix} ,
\label{PhEqTr2}
\intertext{while all the
remaining
matrix elements}
\label{Ul}
U_{n n'}^{\Gamma \Gamma '}=\delta_{nn'}\,\delta_{\Gamma\Gamma'}
\qquad \text{for}\;\; n>0\;\; \text{or}\;\; n'>0.
\end{gather}
\end{subequations}

Varying the parameter $\vartheta$ of the transformation
(\ref{PhEq1})--(\ref{UPh}), we
obtain a family of phase equivalent potentials and examine which of
them provides the better description of the deuteron properties and $np$
scattering wave functions. The best
result
seems to be the potential obtained
with $\vartheta=-14^\circ$. This potential is hereafter referred to as
Version~1 ISTP.

\begin{figure}
\setlength{\unitlength}{747sp}
\begin{picture}(10841,10841)(-7,-9994)
\put( 76,-5161){\framebox(5925,5925){}}
\put(6001,-9961){\framebox(4800,4800){}} \thinlines
\put(6001,-5161){\framebox(4800,5925){}} \put(
76,-9961){\framebox(5925,4800){}} \thicklines \put( 76,764){\line(
1,-1){5925}} \put(6001,-5161){\line( 1,-1){4800}}
\put(6001,764){\line( 1,-1){4762.500}} \put(10801,-5161){\line(-1,
1){4800}} \put(1201,764){\line( 1,-1){4762.500}} \put(
76,-361){\line( 1,-1){4762.500}} \put(6001,-6361){\line(
1,-1){3600}} \put(7201,-5161){\line( 1,-1){3600}} \put(
76,-5161){\line( 1,-1){4762.500}} \put(1201,-5161){\line(
1,-1){4800}} \put(3826,-436){\makebox(0,0)[lb]{\smash{$S$}}}
\put(9601,-6136){\makebox(0,0)[lb]{\smash{$D$}}}
\put(8551,-436){\makebox(0,0)[lb]{\smash{$SD$}}}
\put(3601,-6511){\makebox(0,0)[lb]{\smash{$DS$}}} \thinlines \put(
76,-6436){\circle*{250}} \put(7201,764){\circle*{250}}
\end{picture}
\caption{Structure of the Version~1 and Version~2 ISTP matrix.
The location of non-zero matrix is schematically illustrated by
solid lines and  filled circles. }\label{sd54pot} 
\end{figure}
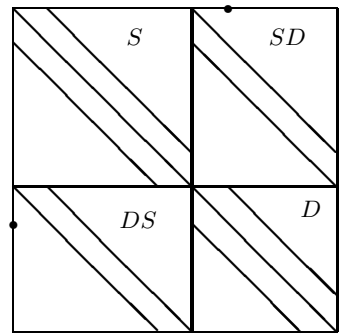

As a result of the transformation~(\ref{PhEq1})--(\ref{UPh}), the
potential energy matrix
acquires
two additional non-zero matrix elements
$V^{sd}_{01}=V^{ds}_{10}$. These additional matrix elements are
schematically illustrated by filled circles in Fig.~\ref{sd54pot}. The
non-zero matrix elements of the Version~1 ISTP are given in
Table~\ref{potsd1} (in $\hbar\omega=40$~MeV units).

\begin{table*}
\caption{Non-zero matrix elements elements in $\hbar\omega$ units of
the Version~1 ISTP matrix in the $sd$ coupled waves.} \label{potsd1}
\begin{ruledtabular}
\begin{tabular}{>{$}c<{$}>{$}c<{$}>{$}c<{$}>{$}c<{$}} 
 \multicolumn{3}{c}{$V^{ss}_{nn'}$ matrix elements} \\
  n & V_{nn}^{ss}  & V_{n,n+1}^{ss}=V^{ss}_{n+1,n}  \\ \cline{1-3} 
 0&   -0.4576704509056663 &  0.2111262515300726  \\
 1&    -0.2783240605925930 & 0.07816883400308394 \\
 2&     -0.01153153008556052 & -0.05346707187864697\\
 3&     0.1514476294157645 & -0.05592826862748490  \\
 4&    0.03632278173824943 &  \\[1.5ex] 
 \multicolumn{3}{c}{$V^{dd}_{nn'}$ matrix elements} \\  
 n & V_{nn}^{dd}  & V_{n,n+1}^{dd}=V^{dd}_{n+1,n} \\ \cline{1-3} 
0&    0.008456639591855719 &  -0.08337354364629734\\
1 &   0.3220439073711110 & -0.1788388098603870\\
2&     0.3084931588662858 & -0.09304409937329216\\
3&  0.06118166034551464 &\\[1.5ex] 
  \multicolumn{4}{c}{$V^{sd}_{nn'}=V^{ds}_{n'n}$ matrix elements} \\  
  n & V_{n,n-1}^{sd}=V^{ds}_{n-1,n}  & V_{nn}^{sd}=V_{nn}^{ds}
              &  V_{n,n+1}^{sd}=V^{ds}_{n+1,n} \\ \hline
 0&     &  -0.4824076895869836 & 0.2540123500192352  \\
 1&     -0.06899752955786595 & -0.06136692873982898 &  \\
 2&    0.06774418012432615 & -0.08068524598671112 &  \\
 3&     0.04913873244946831 &  -0.02041291263896068 & \\
 4&   -0.001715094993409672 &  &   
\end{tabular}
\end{ruledtabular}
\end{table*}

The deuteron properties obtained with the  Version~1 ISTP are
presented in Table~\ref{tabl}.  The $d$ state probability is
improved by the phase equivalent
transformation. However  the phase equivalent transformation
produces an
increase of the deuteron rms radius;
so
this
observable becomes even worse than
that given
by the Version~0 ISTP.
We found
it impossible to
obtain an
exact
description of all deuteron properties by means of the phase
equivalent transformation (\ref{PhEq1}) with the simplest
matrix~(\ref{UPh}).

The deuteron wave functions provided by the Version~1 ISTP are shown
in Fig.~\ref{wfdeut}. The Version~1 ISTP $s$ wave component is seen to
be very close to
that of the Nijmegen-II.
The maximum of the  Version~1
ISTP $d$ wave component is seen to be shifted to larger
distances as compared with
that of the Nijmegen-II. Of course, the
shape of the $d$ wave component of the
wave function cannot be determined experimentally.
%
Hence the shape of
the Version~1 ISTP  deuteron wave functions look realistic though
these wave functions result in the slightly  overestimated
deuteron rms radius.
%

\begin{figure}
\centerline{\psfig{figure=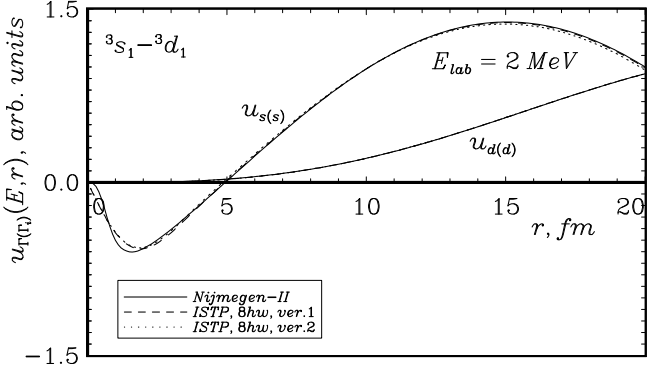,width=0.5\textwidth}}
\caption{Large components $u_{s(s)\protect\vphantom{'}}(E,r)$ and
$u_{d(d)\protect\vphantom{'}}(E,r)$ of the coupled $sd$ waves
$np$ scattering wave function at the laboratory energy
$E_{\rm lab}=2$~MeV.
See Fig.~\ref{wfdeut} for details.} \label{wf02sd}
\end{figure}

\begin{figure}
\centerline{\psfig{figure=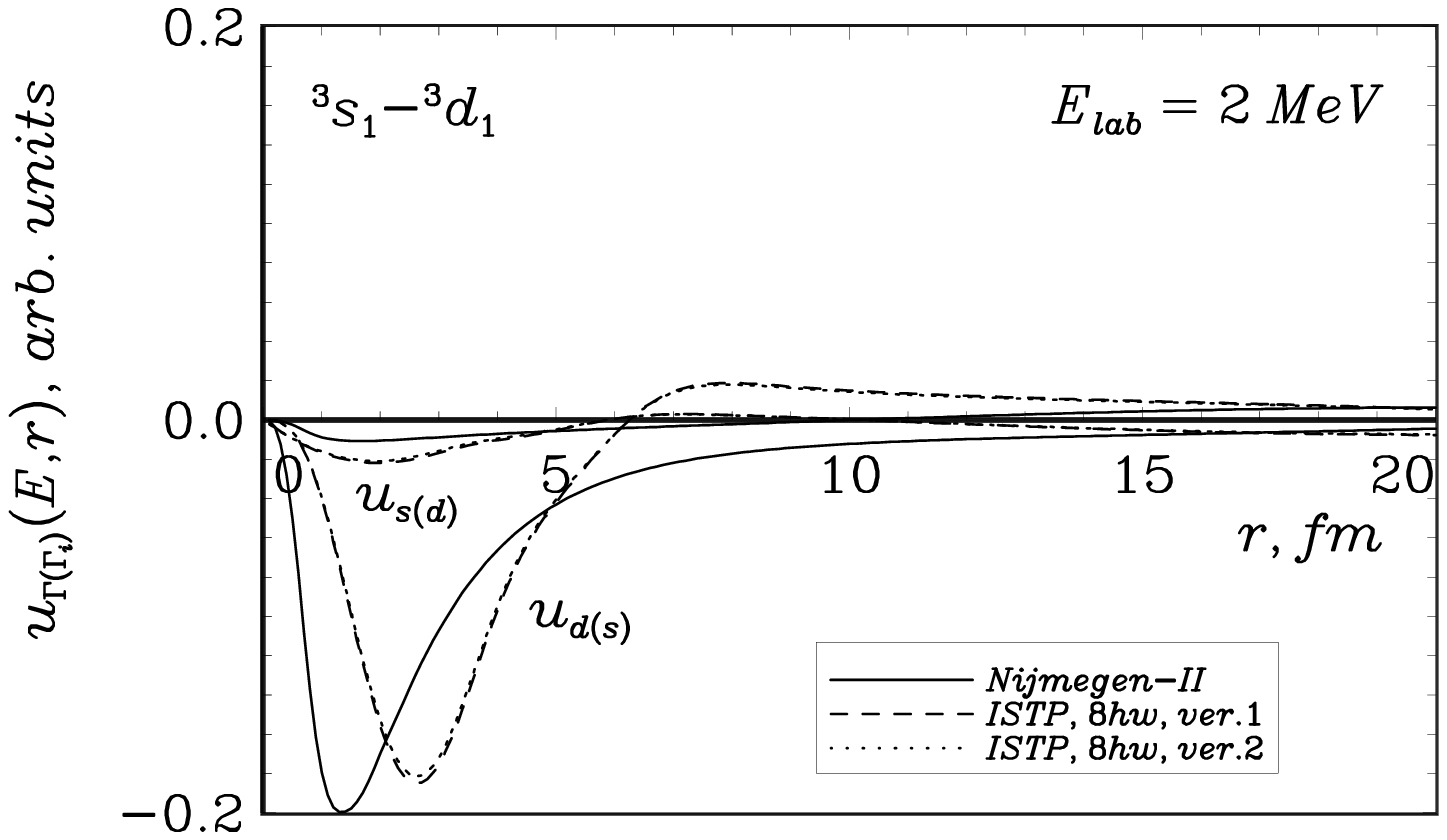,width=0.5\textwidth}}
\caption{Small components $u_{s(d)\protect\vphantom{'}}(E,r)$ and
$u_{d(s)\protect\vphantom{'}}(E,r)$ of the coupled $sd$ waves
$np$ scattering wave function at the laboratory energy
$E_{\rm lab}=2$~MeV.
See Fig.~\ref{wfdeut} for details.} \label{wf02sds}
\end{figure}

\begin{figure}
\centerline{\psfig{figure=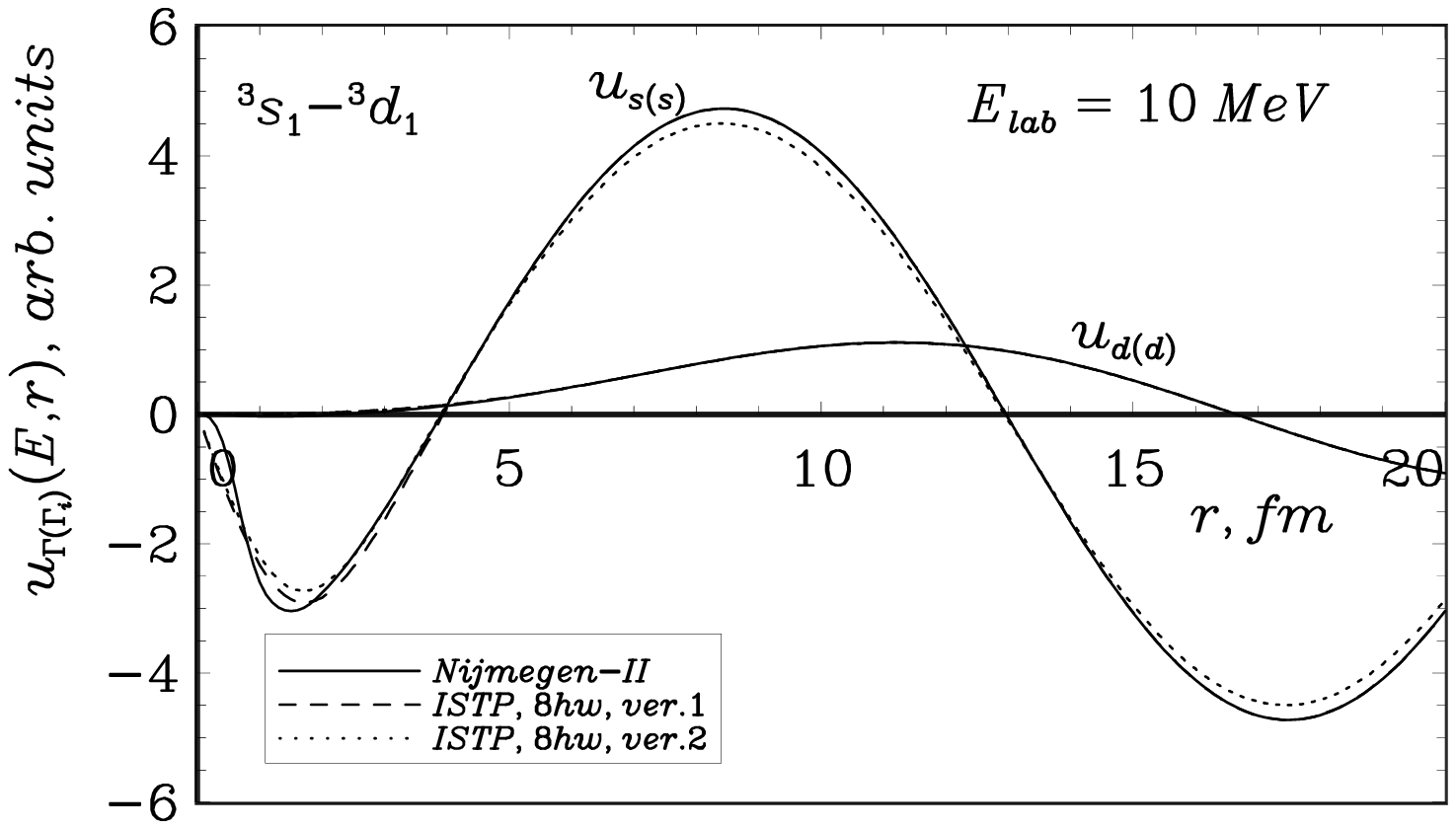,width=0.5\textwidth}}
\caption{Large components $u_{s(s)\protect\vphantom{'}}(E,r)$ and
$u_{d(d)\protect\vphantom{'}}(E,r)$ of the coupled $sd$ waves
$np$ scattering wave function at the laboratory energy
$E_{\rm lab}=10$~MeV.
See Fig.~\ref{wfdeut} for details.}
\label{wf10sd}
\end{figure}

\begin{figure}
\centerline{\psfig{figure=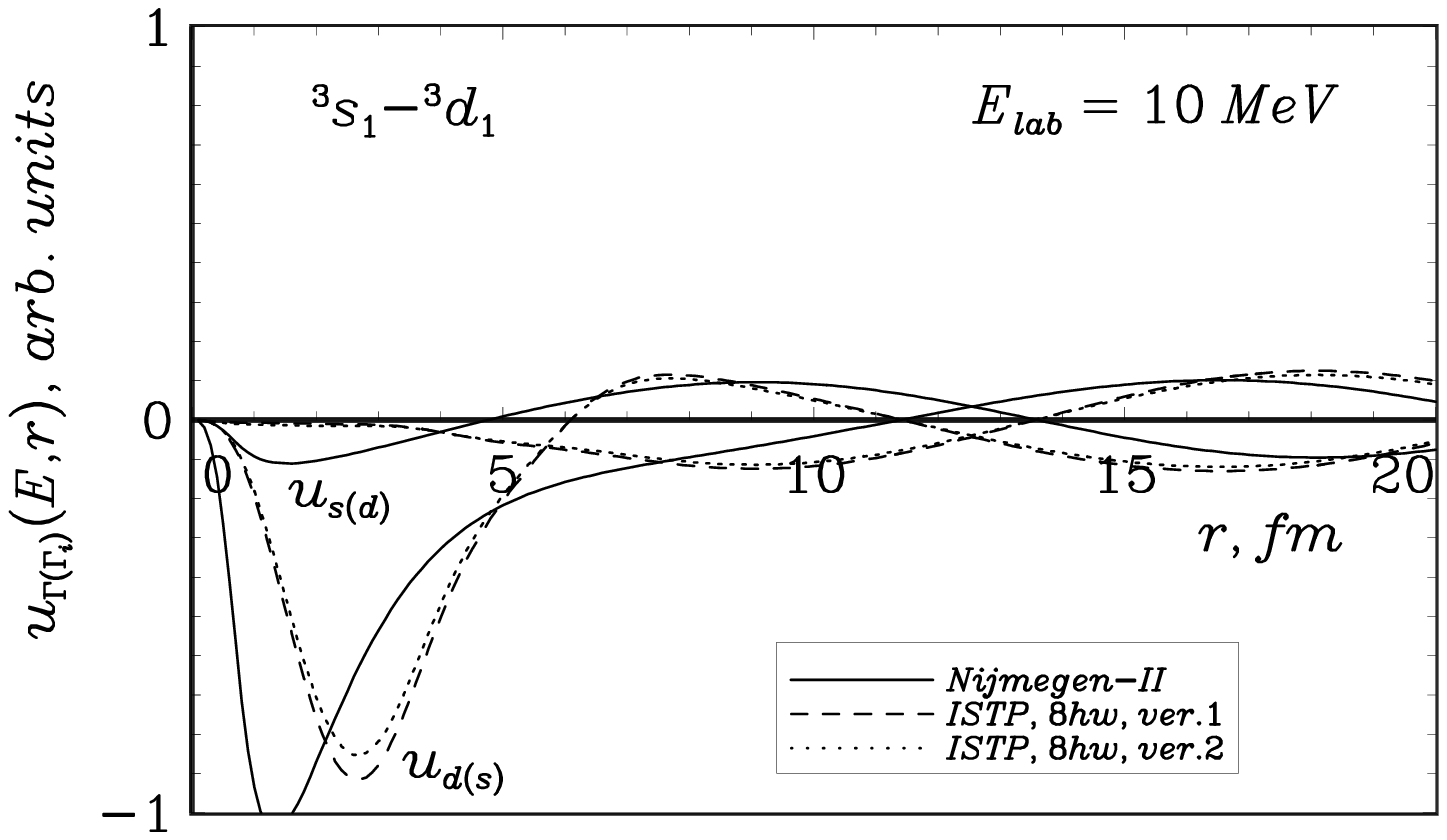,width=0.5\textwidth}}
\caption{Large components $u_{s(d)\protect\vphantom{'}}(E,r)$ and
$u_{d(s)\protect\vphantom{'}}(E,r)$ of the coupled $sd$ waves
$np$ scattering wave function at the laboratory energy
$E_{\rm lab}=10$~MeV.
See Fig.~\ref{wfdeut} for details.}
\label{wf10sds}
\end{figure}

\begin{figure}
\centerline{\psfig{figure=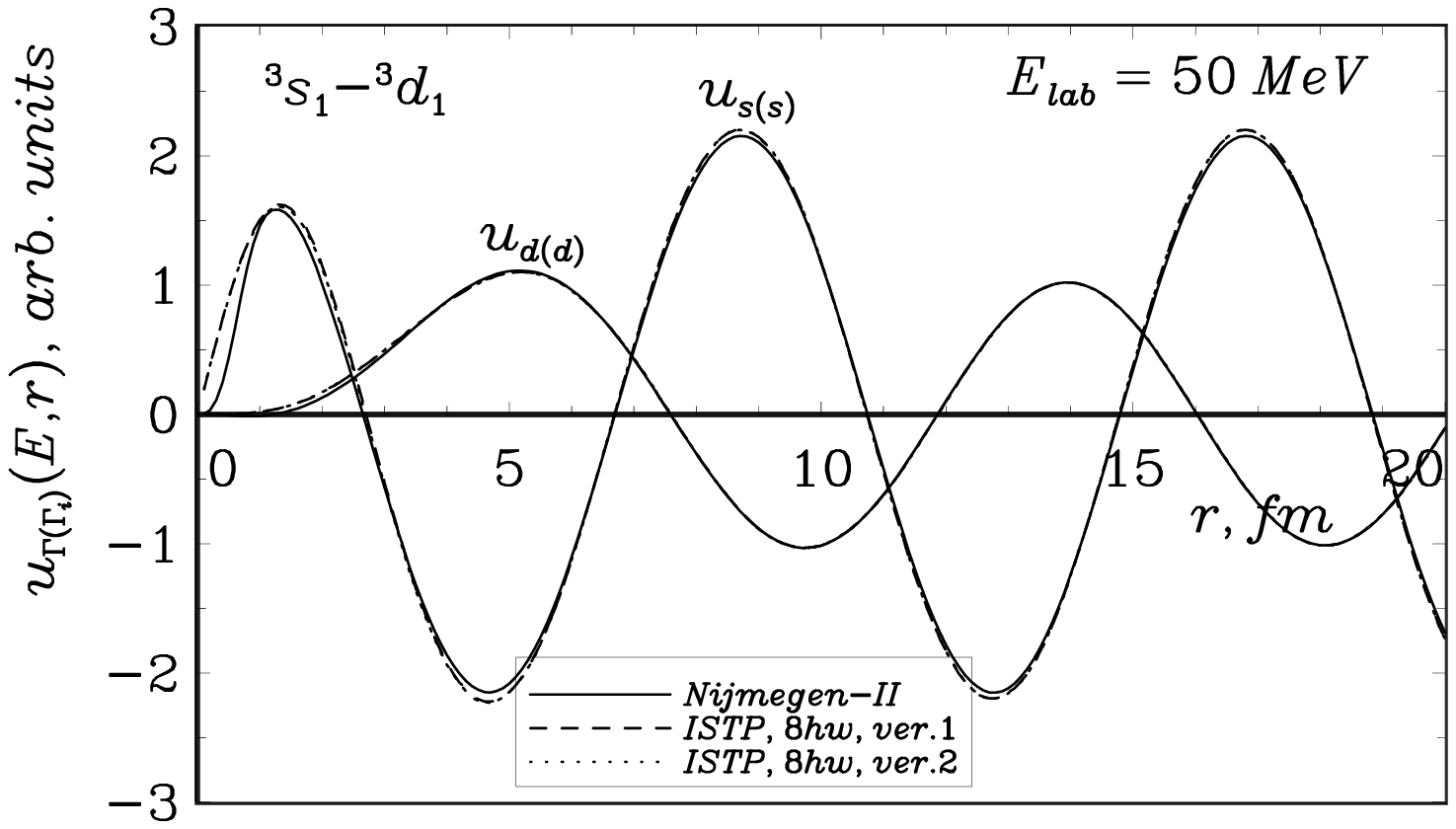,width=0.5\textwidth}}
\caption{Large components $u_{s(s)\protect\vphantom{'}}(E,r)$ and
$u_{d(d)\protect\vphantom{'}}(E,r)$ of the coupled $sd$ waves
$np$ scattering wave function at the laboratory energy
$E_{\rm lab}=50$~MeV.
See Fig.~\ref{wfdeut} for details.}
\label{wf50sd}
\vspace{4ex}
\centerline{\psfig{figure=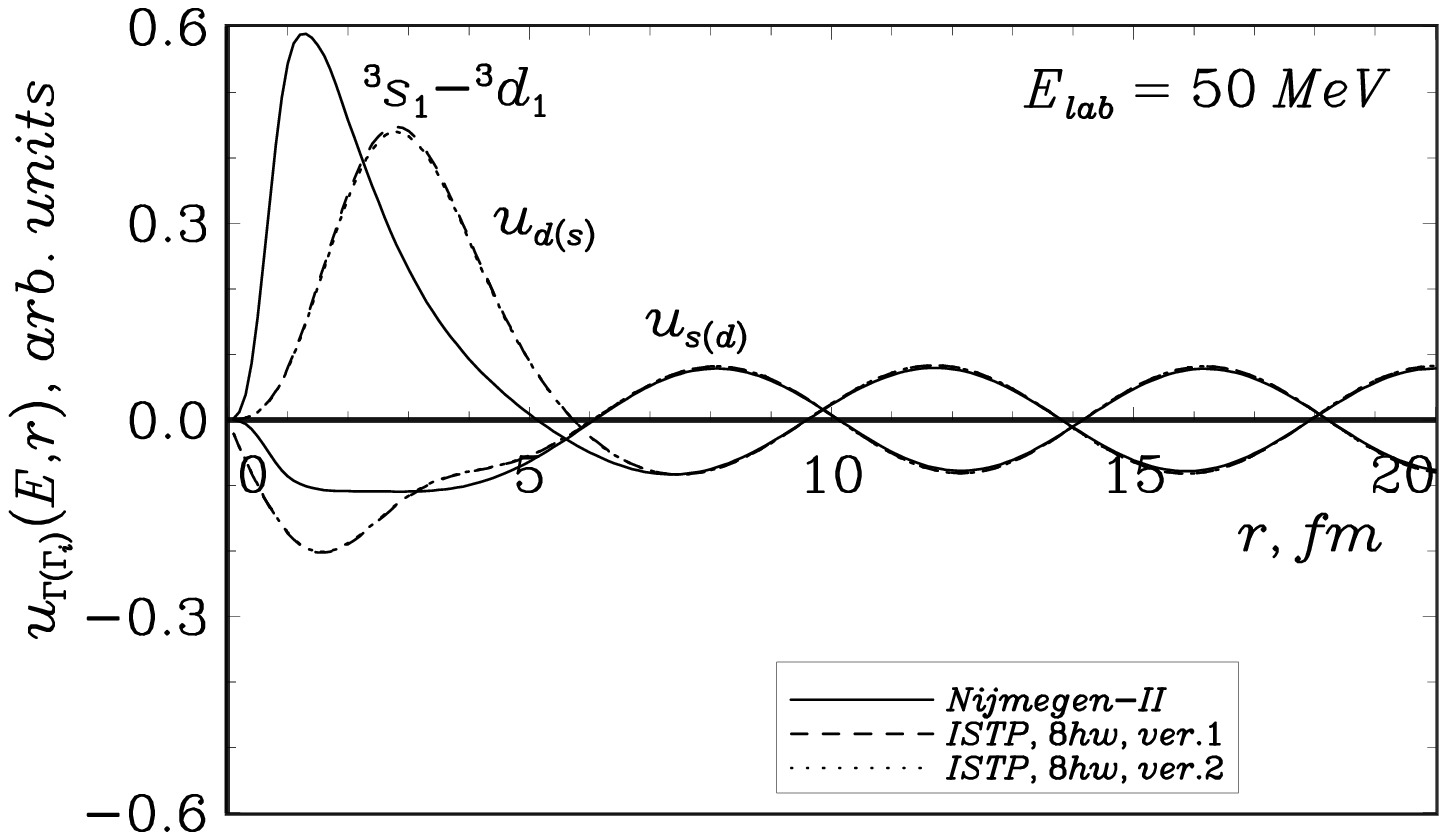,width=0.5\textwidth}}
\caption{Small components $u_{s(d)\protect\vphantom{'}}(E,r)$ and
$u_{d(s)\protect\vphantom{'}}(E,r)$ of the coupled $sd$ waves
$np$ scattering wave function at the laboratory energy
$E_{\rm lab}=50$~MeV.
See Fig.~\ref{wfdeut} for details.}
\label{wf50sds}
%
\vspace{4ex}
\centerline{\psfig{figure=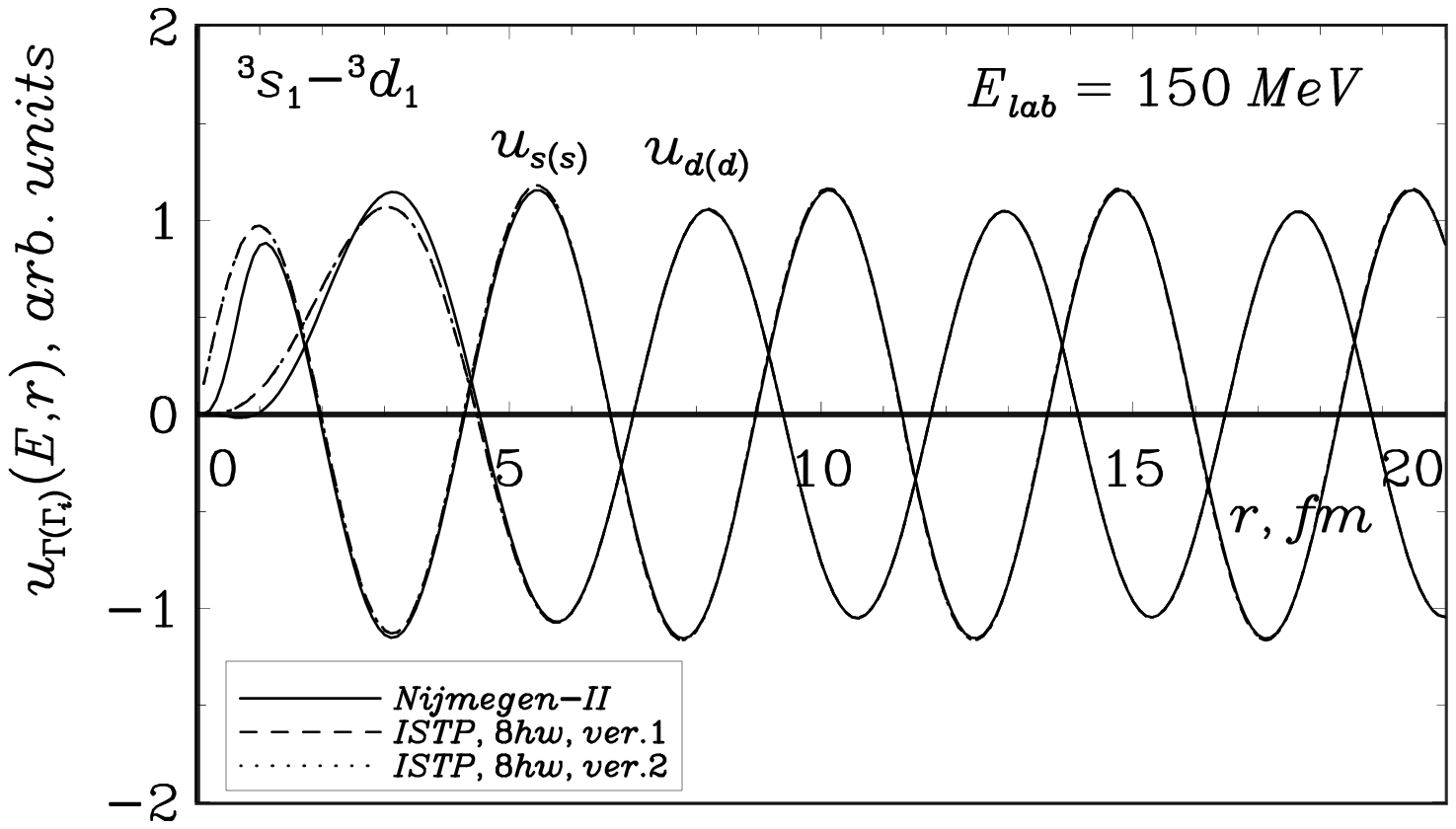,width=0.5\textwidth}}
\caption{Large components $u_{s(s)\protect\vphantom{'}}(E,r)$ and
$u_{d(d)\protect\vphantom{'}}(E,r)$ of the coupled $sd$ waves
$np$ scattering wave function at the laboratory energy
$E_{\rm lab}=150$~MeV.
See Fig.~\ref{wfdeut} for details.}
\label{wf150sd}
\end{figure}

\begin{figure}
\centerline{\psfig{figure=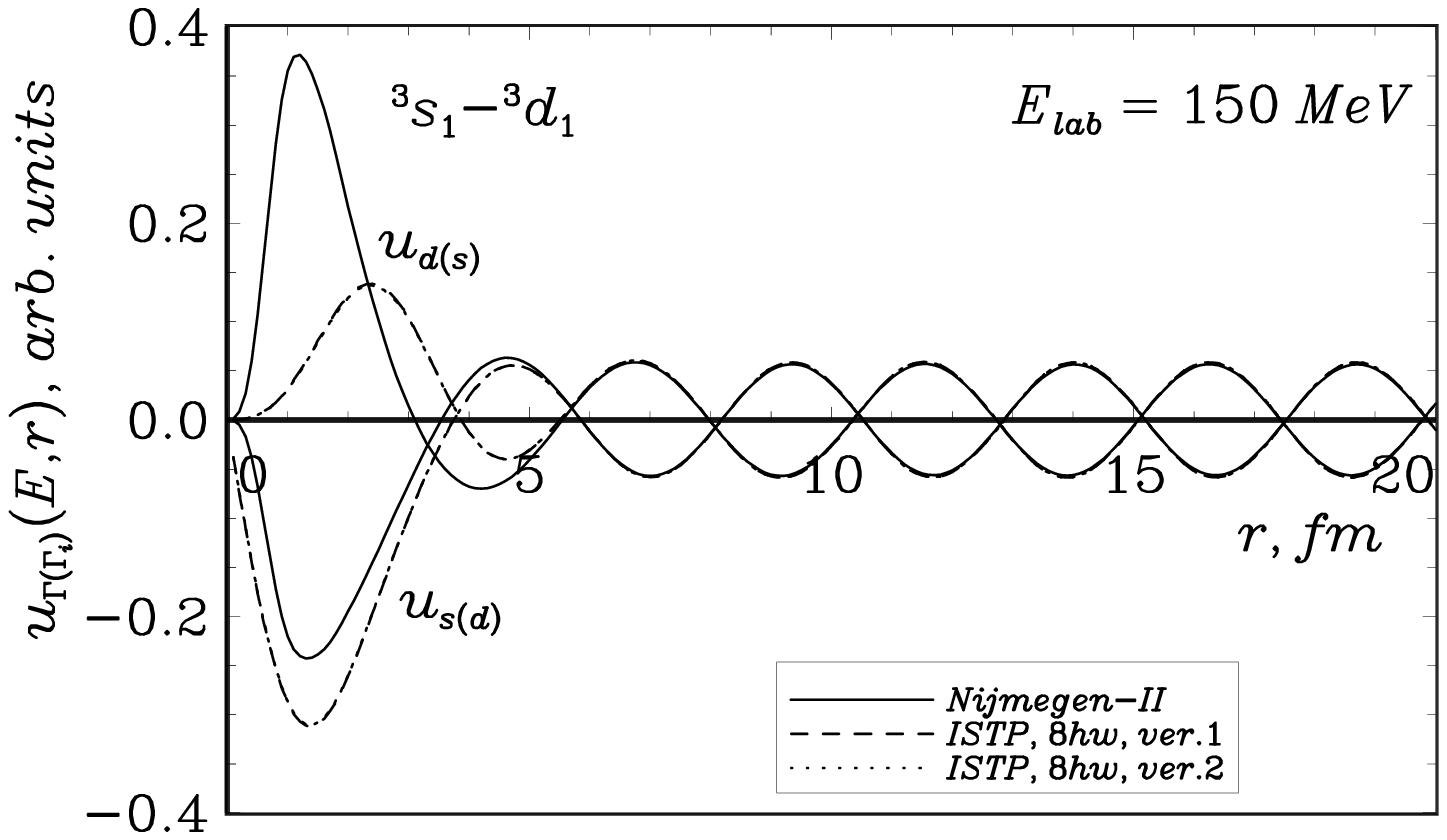,width=0.5\textwidth}}
\caption{Small components $u_{s(d)\protect\vphantom{'}}(E,r)$ and
$u_{d(s)\protect\vphantom{'}}(E,r)$ of the coupled $sd$ waves
$np$ scattering wave function at the laboratory energy
$E_{\rm lab}=150$~MeV.
See Fig.~\ref{wfdeut} for details.}
\label{wf150sds}
\vspace{4ex}
\centerline{\psfig{figure=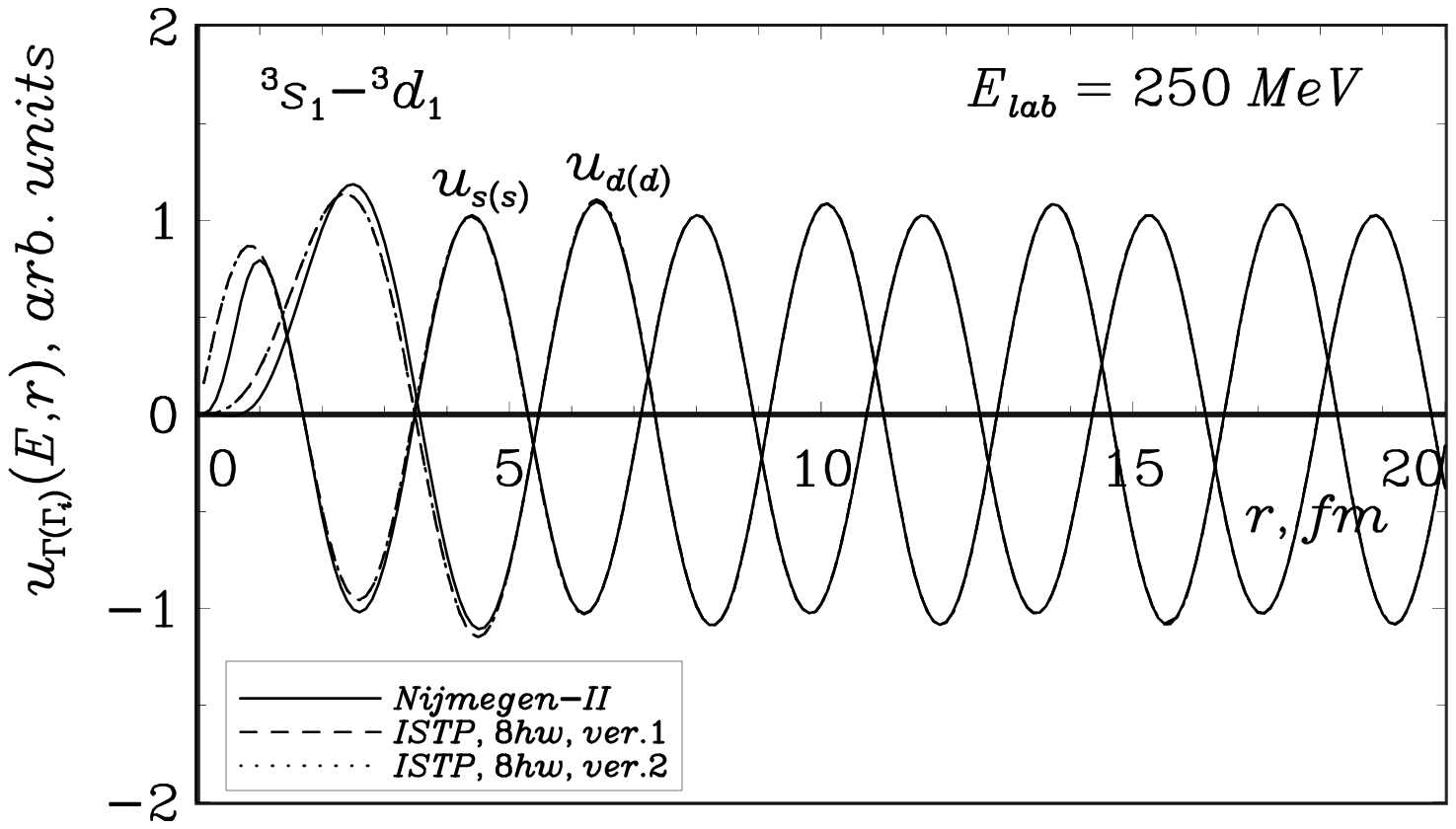,width=0.5\textwidth}}
\caption{Large components $u_{s(s)\protect\vphantom{'}}(E,r)$ and
$u_{d(d)\protect\vphantom{'}}(E,r)$  of the coupled $sd$ waves
$np$ scattering wave function at the laboratory energy
$E_{\rm lab}=250$~MeV.
See Fig.~\ref{wfdeut} for details.}
\label{wf250sd}
\vspace{4ex}
\centerline{\psfig{figure=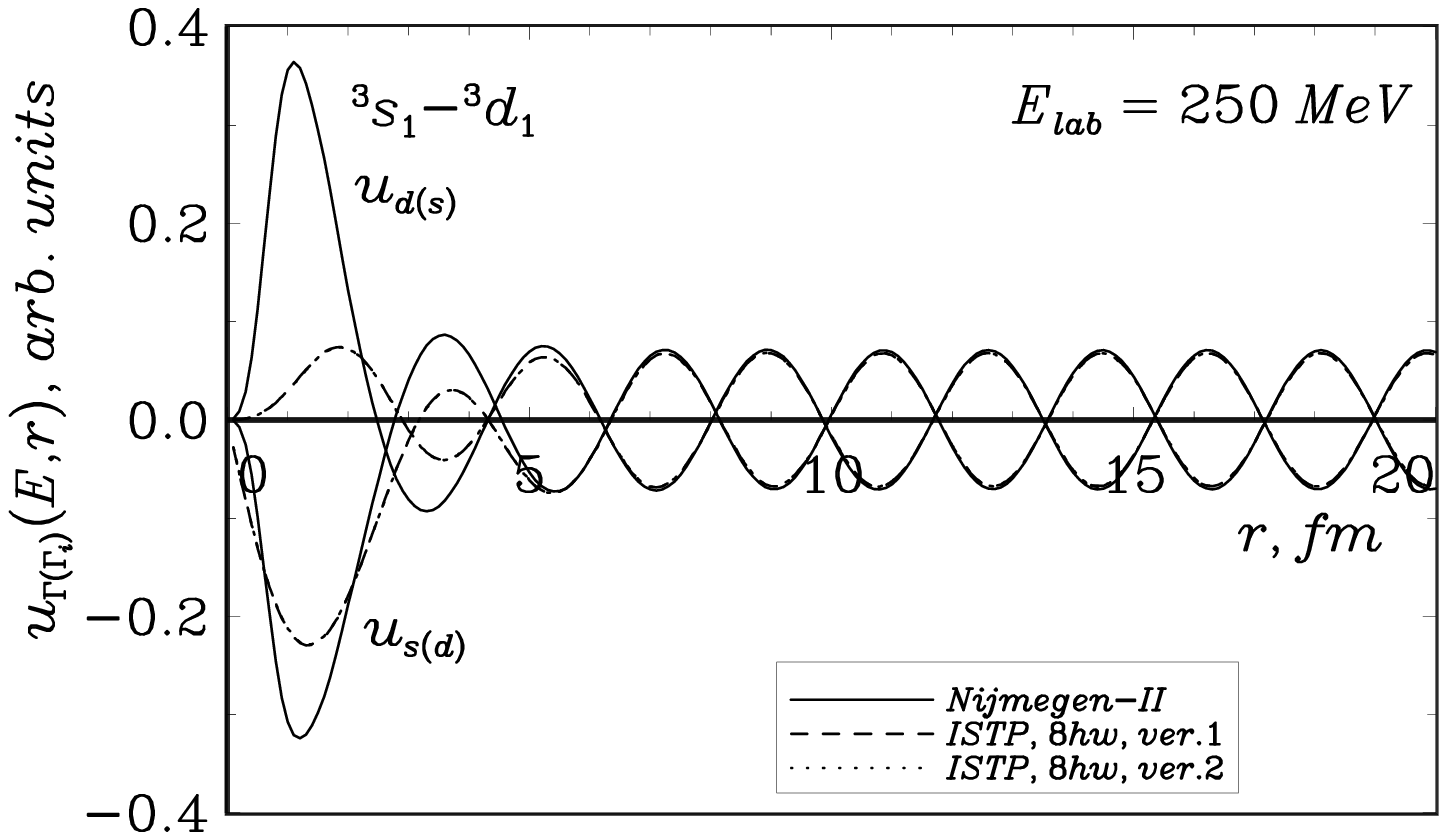,width=0.5\textwidth}}
\caption{Small components $u_{s(d)\protect\vphantom{'}}(E,r)$ and
$u_{d(s)\protect\vphantom{'}}(E,r)$  of the coupled $sd$ waves
$np$ scattering wave function at the laboratory energy
$E_{\rm lab}=250$~MeV.
See Fig.~\ref{wfdeut} for details.}
\label{wf250sds}
\end{figure}

The Version~1 ISTP $np$ scattering wave function components at the laboratory
energies $E_{\rm lab}=2$, 10, 50, 150 and 250~MeV are shown in
Figs.~\ref{wf02sd}--\ref{wf250sds} in comparison with those of  Nijmegen-II
potential.
As in the case of the coupled $pf$ partial waves, the large  components 
$u_{s(s)\protect\vphantom{'}}(E,r)$ and
$u_{d(d)\protect\vphantom{'}}(E,r)$ differ very little from the
Nijmegen-II ones but the small components are esentially different at
short distances due to the difference of the tensor interaction of
these two potential models.

%
Generally we conclude that the  Version~1 ISTP is very
close to the realistic interaction. The most important
discrepancy of this interaction is that it overestimates the
deuteron rms radius by approximately~1.5\%.

We
attempted
the phase equivalent transformation~(\ref{PhEq1})
with a more complicated matrix $\left[U\right]$ than~(\ref{UPh}).
However,
we did not manage to obtain
a
completely satisfactory interaction. It is
possible to obtain the potential providing the required values of the
deuteron rms radius and of the $d$ state probability by
increasing the dimension of the submatrix  $\left[U_0\right]$ and
introducing additional transformation parameters, but
our attempts yielded
unrealistic scattering wave functions.

To improve the $sd$-ISTP
we suggest
a slight change to
the $s$ wave asymptotic normalization constant ${\mrs A}_s$ that is
used as an input in our inverse scattering approach. The  ${\mrs A}_s$
value cannot be measured in a direct experiment. As
was
mentioned in Ref.~\cite{Petrov}, the  ${\mrs A}_s$ values discussed in
the literature vary within a broad range from 0.7592~fm$^{-1/2}$ to
0.9863~fm$^{-1/2}$. Therefore, the modified value
${\mrs A}_s=0.8629$~fm$^{-1/2}$ that we use for the construction
of the improved \mbox{$sd$-}ISTP, seems to be reasonable. We do not change
the remaining
inputs in our inverse scattering approach including
$\eta=\displaystyle\frac{{\mrs A}_d}{{\mrs A}_s}$ (and hence we modify
 ${\mrs A}_d$ together with  ${\mrs A}_s$) to obtain the ISTP of the
type shown in Fig.~\ref{pf43pot} and apply to it the phase equivalent
transformation~(\ref{PhEq1}) with the parameter~$\vartheta=-14^\circ$
of the matrix~(\ref{UPh}). This potential is referred to as Version~2
ISTP. This potential has the structure schematically depicted in
Fig.~\ref{sd54pot} and its matrix elements are listed in Table~\ref{potsd2}.

\begin{table*}
\caption{Non-zero matrix elements elements in $\hbar\omega$ units of
the Version~2 ISTP matrix in the $sd$ coupled waves.}
\label{potsd2}
\begin{ruledtabular}
\begin{tabular}{>{$}c<{$}>{$}c<{$}>{$}c<{$}>{$}c<{$}} 
 \multicolumn{3}{c}{$V^{ss}_{nn'}$ matrix elements} \\
  n & V_{nn}^{ss}  & V_{n,n+1}^{ss}=V^{ss}_{n+1,n}  \\ \cline{1-3}
 0&    -0.4660631463496376 &  0.2168839488356998 \\
 1&    -0.2761680294726432 & 0.08090773569137233 \\
 2&     -0.009473803658917924 & -0.05188144310822707 \\
 3&      0.1528737342886162 & -0.05519358984226530  \\
 4&      0.03754792988022171 &     \\[1.5ex]  
 \multicolumn{3}{c}{$V^{dd}_{nn'}$ matrix elements} \\  
 n & V_{nn}^{dd}  & V_{n,n+1}^{dd}=V^{dd}_{n+1,n} \\ \cline{1-3}
0 &   0.008667454659207596 & -0.08333937455951757\\
 1&   0.3221264718049914 & -0.1788087936408669 \\
 2&  0.3085166730609980 & -0.09301260476557038\\
 3&  0.06120003719298150 & \\[1.5ex] 
  \multicolumn{4}{c}{$V^{sd}_{nn'}=V^{ds}_{n'n}$ matrix elements} \\  
  n & V_{n,n-1}^{sd}=V^{ds}_{n-1,n}  & V_{nn}^{sd}=V_{nn}^{ds}
              &  V_{n,n+1}^{sd}=V^{ds}_{n+1,n} \\ \hline
 0&     &   -0.4833085003127391 &  0.2540038307090694  \\
 1&      -0.06722102540443002 & -0.06047658569273850  & \\
 2&     0.06804449696337271 & -0.08018710645793066 &  \\
 3&       0.04940057881591606 & -0.02020564623073210  & \\
 4&    -0.001503998138989182 &  &    
\end{tabular}
\end{ruledtabular}
\end{table*}

The deuteron properties are seen from
Table~\ref{tabl} to be
well
described by the Version~2 ISTP. The
Version~2 ISTP  scattering wave functions are  very close to those of
the Version~1 ISTP 
%
 (see Figs.~\ref{wf02sd}--\ref{wf250sds}). Its
deuteron wave functions are  very close to
those of Version~1 ISTP
(see Fig.~\ref{wfdeut}) and differ from
those of Nijmegen-II
in the position of the $d$ wave component maximum.

 We 
suppose that the Version~2 ISTP can be treated as a realistic
 interaction in the coupled $sd$ partial waves.

\section{Application of $NN$ ISTP in $^3$H and $\bf ^4He$ calculations}

We employ the obtained ISTP
in the $^3$H and $^4$He calculations within the
no-core shell model~\cite{Vary,Vary3} with $\hbar\omega=40$~MeV. The same
$NN$ potentials are used to describe the neutron-neutron and
neutron-proton interactions; in the proton-proton case these potentials
are supplemented by the Coulomb interaction.

The calculations are performed in the complete $N\hbar\omega$ model
spaces with $N\leq 14$. We
use
both $7\hbar\omega$-ISTP and
$9\hbar\omega$-ISTP in odd partial waves. The $^3$H and $^4$He nuclei
are
slightly
more bound in the case when we use the  $7\hbar\omega$-ISTP in the
odd waves. However, the
differences are
very small: less than 15 keV for $^3$H
and about 40 keV for $^4$He. The sequence of levels in
the $^4$He spectrum provided by the odd wave  $7\hbar\omega$-ISTP  and
by the odd wave
$9\hbar\omega$-ISTP is the same but the energies of excited
$^4$He states are shifted down in the case of
the odd wave \mbox{$7\hbar\omega$-ISTP} by
approximately 100 keV or less. Therefore the deviations of the
\mbox{$7\hbar\omega$-ISTP} predictions from the
experimental odd wave scattering data at high enough energies seem to
produce a negligible effect in the $^3$H and $^4$He calculations. At
the same time, $7\hbar\omega$-ISTP has a smaller matrix than
$9\hbar\omega$-ISTP and hence is more convenient in
applications. Below we present only the results obtained with the
$7\hbar\omega$-ISTP in the odd partial waves.

We have
presented
various versions of ISTP in the coupled $sd$ partial
waves. The choice of ISTP in other partial waves is
fixed.
Using this
fixed
set of the non-$sd$-ISTP in combination with the
Version~M
$sd$-ISTP, we have the set of potentials that is refered to as the
Version~M
potential model in what follows.

\begin{figure}
\psfrag{1/N}{$1/N$}\psfrag{Egs,  MeV}{$E_{t}$, MeV}
\psfrag{Triton}{\Large$^3$H}
\centerline{\psfig{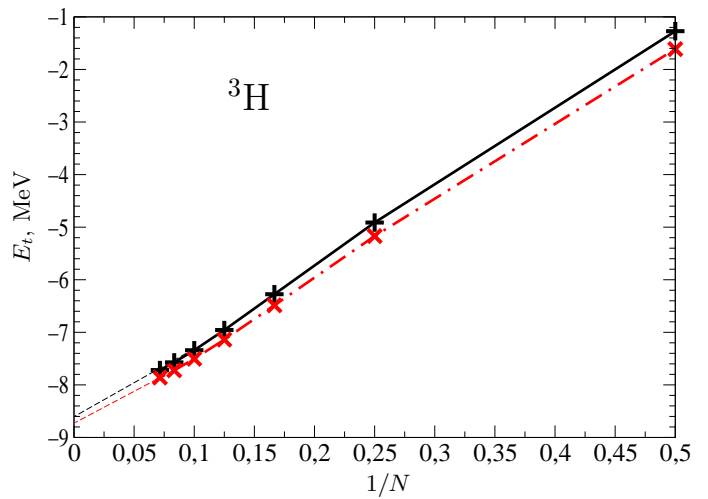}}
\caption{$^3$H ground state energy obtained in the $N\hbar\omega$ no-core
shell model calculation vs $1/N$. 
$+$~--- Version~1 potential model; 
$\times$~--- Version~2 potential model; dashed line~--- linear extrapolation to
the infinite $N\hbar\omega$ model space
based on the last two calculated points
; solid and dash-dot lines are
to guide the eye.}
\label{triton}
\end{figure}

The $^3$H ground state energies $E_t$ obtained with the Version~1 and the
Version~2 potential models in $N\hbar\omega$ model
spaces are presented in
Fig.~\ref{triton} as functions of $1/N$. It is seen that both
potential models provide very
similar
$E_t$ values.
The convergence of
the calculations with $N$
appears adequate.
The ground state energy $E_t$ is seen from the figure to be nearly
a linear function of $1/N$. Therefore it
is
natural to perform a
linear extrapolation to the infinite  $N\hbar\omega$ model space,
i.~e. to the point $1/N=0$.
The linear extrapolation using the two results at the highest N-values
yields
$E_t\approx-8.6$~MeV in the Version~1 potential model and in
$E_t\approx-8.7$~MeV in the Version~2 potential model.

\begin{figure} 
\psfrag{1/N}{$1/N$}\psfrag{Egs,  MeV}{\raisebox{2mm}{$E_{\alpha}$, MeV}}
\psfrag{Alpha}{\Large$^4$He}
\centerline{\psfig{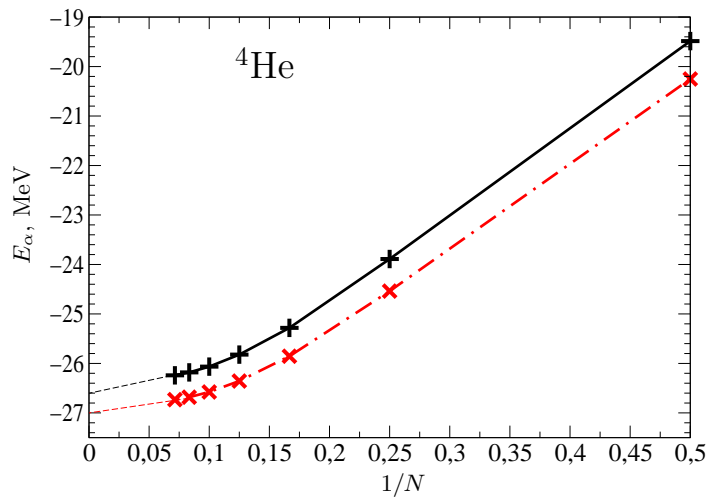}}
\caption{$^4$He ground state energy obtained in the $N\hbar\omega$ no-core
shell model calculation vs $1/N$. See Fig.~\ref{triton} for details.}
\label{helium}
\end{figure}

In Fig.~\ref{helium} we present the results of the $^4$He ground state
energy $E_\alpha$ calculations with the same potential models. In the
$^4$He case we also obtain very
similar
results with the Version~1 and the
Version~2 potential models. It is interesting that the convergence of
the $^4$He ground state energy is better than
that of $^3$H. In this case
the curves
connecting the $E_\alpha$ values deviate from the straight
lines. Nevertheless we also perform the linear extrapolations of
$E_\alpha(1/N)$ to infinite $N$ using the $E_\alpha$ values obtained in
$12\hbar\omega$ and $14\hbar\omega$ calculations and
 obtain $E_\alpha\approx-26.6$~MeV in the
Version~1 potential model and
$E_\alpha\approx-27.0$~MeV in the Version~2 potential model.

\begin{figure}
\psfrag{1/N}{$1/N$}\psfrag{Egs,  MeV}{\raisebox{2mm}{$E_{d}$, MeV}}
\psfrag{Deuteron}{\Large$^2$H}
\centerline{\psfig{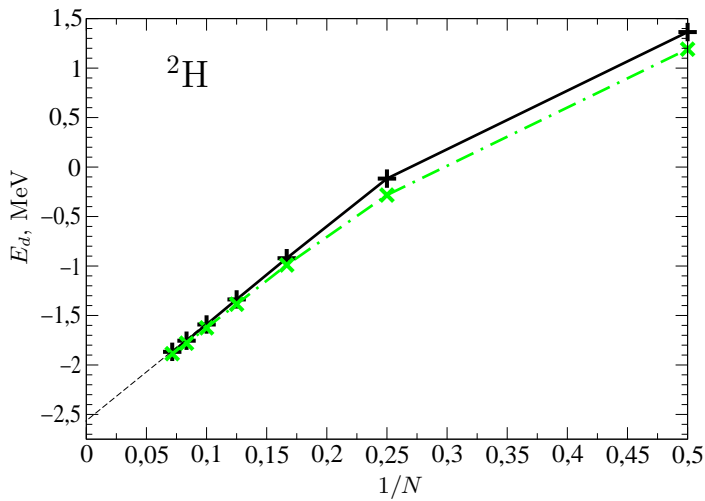}}
\caption{Deuteron ground state energy obtained in the $N\hbar\omega$ no-core
shell model calculation vs $1/N$. See Fig.~\ref{triton} for details.}
\label{deuteron}
\end{figure}

The quality of the linear extrapolation of $E_{\rm g.s.}(1/N)$ may be
tested in the deuteron calculations. In the deuteron case, we know the
exact result for the infinite $N\hbar\omega$ model space ground state
energy $E_d=-2.244575$~MeV obtained by the $S$-matrix pole calculation
with our potentials. The $E_d$ results obtained in the $N\hbar\omega$
model spaces with $N\leq 14$ with the Version~1 and Version~2 $sd$-ISTP, are
shown in Fig.~\ref{deuteron}. It is seen that $E_d(1/N)$ seems to be a
linear function in the interval $4\leq N\leq 14$. The linear
extrapolation results in $E_d\approx -2.5$~MeV that differs
from the exact energy. Therefore the linear extrapolation results can
be regarded
only as a rough estimate of the binding energy. However
in the $^4$He case we achieved a
reasonable
convergence and
by the linear extrapolation we increase the binding
energy by approximately 0.3~MeV only. Therefore our estimate of the
$^4$He binding energy seems to be accurate enough.

The differences in convergence rates for the deuteron, $^3$H and $^4$He
can be understood from the fact that $\hbar\omega=40$~MeV is more optimal
for the tighter bound $^4$He than for the lesser bound systems.

\begin{table*} 
\caption{$^3$H and $^4$He ground state energies (in MeV) obtained in
$14\hbar\omega$ no-core shell model calculations and by the linear
extrapolation to the infinite $N\hbar\omega$ model space.}\label{gstable}
\begin{ruledtabular}
\begin{tabular}{c>{$}c<{$}>{$}c<{$}>{$}c<{$}>{$}c<{$}} 
Potential
 & \multicolumn{2}{c}{$^3$H} & \multicolumn{2}{c}{$^4$He}\\ 
\raisebox{.3ex}[0pt][0pt]{model}
 &\mbox{\hspace{18pt}}14\hbar\omega\mbox{\hspace{18pt}}&\text{extrapolation}
&\mbox{\hspace{18pt}}14\hbar\omega\mbox{\hspace{18pt}}&\text{extrapolation}
\\ \hline
Version 0 & -9.091 &9.7 &-33.223 & -33.4 \\
Version 1 &  -7.718  &-8.6 & -26.241 & -26.6  \\
Version 2 &  -7.860  & -8.7 & -26.734 & -27.0 \\ 
Nature    &\multicolumn{2}{c}{$ -8.48$}  &\multicolumn{2}{c}{$-28.30$}
\end{tabular}
\end{ruledtabular}
\end{table*}

Our results of the $^3$H and $^4$He ground  state energy calculations
are summarized in Table~\ref{gstable}. We also present in the table
the results obtained with the
less realistic
Version~0 potential model.
Both $^3$H and $^4$He are essentially overbound in
this potential model. With both Version~1 and Version~2 potential
models we
obtain a reasonable
description of the  $^3$H and $^4$He
bindings. Our $^4$He results are better than the ones obtained (see
Ref.~\cite{alpha-3})  with any of the realistic
meson exchange interactions without allowing for the three-body
interactions. In the $^3$H case, we have underbinding in the
$14\hbar\omega$ model space and a small overbinding obtained by the
linear extrapolation.
Unfortunately, the difference between the $14\hbar\omega$ model space
and the linear  extrapolation results is
rather large.
Most probably
the $^3$H ground state energy curve in Fig.~\ref{triton} will
flatten out
in larger model spaces.
This will shift
the extrapolated ground state energy upwards from our
current result.
Hence the expected ground state energy in the
$N\to\infty$ limit  lies between the $14\hbar\omega$ and the present
linear extrapolation.
In other words, our linear extrapolation
and  $14\hbar\omega$ results are expected to be the lower and upper
boundaries for the exact
results, respectively.
An approximately  0.9~MeV difference
between the
$14\hbar\omega$ and the linear extrapolation ground state energies in
the $^3$H case indicates the 0.9~MeV
uncertainty
of our predictions.
The $^3$H ground state energy obtained in Faddeev calculations with
CD-Bonn $NN$ potential is $-8.012$~MeV (see~\cite{alpha-3}). All the
remaining
modern realistic meson exchange potentials predict the  $^3$H
binding energy to be less than 7.4~MeV~\cite{alpha-3}. Therefore our
$^3$H binding energy predictions are
not worse than
those
obtained with the realistic meson exchange potentials without allowing
for the three-body forces while our $^4$He  binding energy predictions
are better.

\begin{figure}
\centerline{\psfig{figure=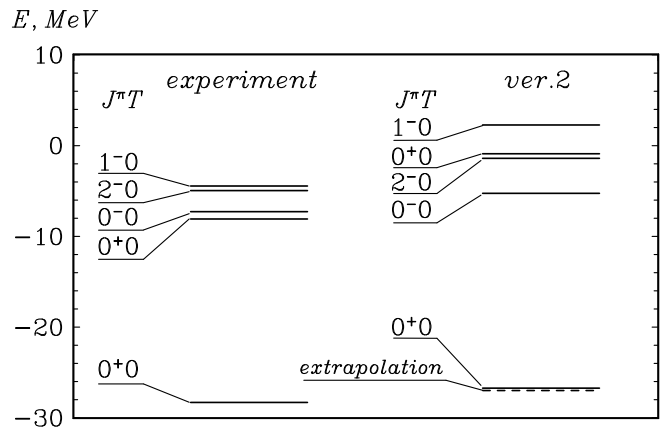,width=0.5\textwidth}}
\caption{$^4$He spectrum  obtained with Version~2 potential model
in the no-core
shell model in the $14\hbar\omega$  ($13\hbar\omega$) model space for
even (odd) parity states. Dashed line shows the result of the linear
extrapolation of the ground state energy to the infinite
$N\hbar\omega$ model space. Experimental data are taken from \cite{LANL}.}
\label{spectrum}
\end{figure}

 In Fig.~\ref{spectrum} we present the spectrum of the lowest excited
$^4$He states of each $J^\pi$. The description of the excited states
energies is reasonable
though further from experiment than
the ground state.
On the other hand, we expect the excited states to be less converged and
to drop more in larger model spaces. Of course, a full discussion of the
states above breakup must await proper extensions of the theory to the
scattering domain.

\section{Concluding remarks}

We obtained nucleon-nucleon ISTP potentials by means of the $J$-matrix
version of the inverse scattering approach. The potentials
accurately
describe the scattering data. They  are in the
form of  $8\hbar\omega$-truncated matrices in the oscillator basis with
$\hbar\omega=40$~MeV. The potential matrices are tridiagonal in the
uncoupled partial waves. In the coupled partial waves, the potential matrices
have two additional quasi-diagonals in each of the
submatrices responsible for the channel coupling.
The $sd$-ISTP of this type (Version~0)
underestimates the
deuteron $d$ state probability and overestimates the deuteron rms
radius. We designed two other $sd$-ISTP with two additional matrix
elements providing the correct description of the $d$ state
probability, one of them (Version~1) overestimates the rms radius by
approximately 1.5\% while the other one (Version~2) provides the correct
description of the deuteron rms radius. All other deuteron observables
are reproduced by all $sd$-ISTP versions.

The ISTP potentials are used in the $^3$H and $^4$He no-core shell
model calculations. Both Version~1 and Version~2 ISTP potential models
provide very good predictions for the $^3$H and $^4$He binding energies
and a reasonable $^4$He spectrum. With the less realistic Version~0 potential
model, we obtain overbound $^3$H and $^4$He nuclei. We note that there
were other attempts to design the $NN$ interaction providing the
description of the triton binding energy together with the $NN$
scattering data and the deuteron properties~\cite{Dole,Plessas}. Our
interactions are much simplier and can be directly used in the
shell model calculations of heavier nuclei.

Generally our approach is aimed at shell model
applications in heavier nuclei. However our potentials are simple
enough and can be used directly in other microscopic approaches, e.~g. in
Faddeev calculations. We hope that our interactions
minimize the need for
three-body forces. It is known~\cite{Poly} that
the three-body force effect can be reproduced in a three-body
system by the phase equivalent transformation of the two-body
interaction. This phase equivalent transformation can
also
spoil
the description of the deuteron observables, in particular, the
deuteron rms radius can be arbitrary changed by phase equivalent
transformations~\cite{RMS}. We
expect that there exist transformations minimizing the need for
three-body force effects,
that do not significantly
change the nucleon-nucleon interaction.
That is,
the deuteron properties, the
deuteron and scattering wave functions of the transformed $NN$
potential
may
remain very close to the ones
developed here while achieving improved descriptions of other nuclei.
In this context, it is worth noting that our approach does not assume
either a particular operator structure to the interaction or locality.

From this point of view, the Version~2 ISTP accurately
describing the deuteron properties and providing good predictions for
the $^3$H and $^4$He bindings, can be regarded as such
an
interaction effectively
accounting for effects that might otherwise be attributed to
three-body forces.
%
%
Clearly, additional efforts may provide superior NN interactions with less
dependence on three-body forces for precision agreement with experiment.

Finally, we suggested a new approach to the construction of the
high-quality $NN$ interaction and examined the obtained ISTP $NN$
interaction in three and four nucleon systems by means of the no-core
shell model. The $^3$H and $^4$He binding energies are surprisingly
well described. Obviousely it will be very interesting to extend these
studies on  heavier nuclei, to investigate in detail not
only their binding but the spectra of excited states as well. It is
also important to investigate more carefully the ISTP description of
the  two-nucleon system since, for example, we have deferred 
the discussion of the deuteron quadrupole moment $Q$. We just mention here
that the Version~2 ISTP prediction of $Q=0.317$~fm$^2$  is not so
far from the experimental value of $0.2875\pm 20$~fm$^2$ \cite{Qdeu}.
The phase equivalent transformations discussed above make it
possible to improve the $Q$ predictions and to  examine the effect of
such improvement  in  light 
nuclear systems. We plan to address this problem in future publications.

\medskip

This work was supported in part by the State Program ``Russian
Universities'',  by the Russian Foundation of Basic Research grant
No~02-02-17316,
by US~DOE grant No~DE-FG-02~87ER40371 and by  US~NSF grant No~PHY-007-1027.


\begin{thebibliography}{99}

\bibitem{Bonn}  R. Machleidt, F. Sammarruca, and Y. Song, Phys. Rev.
{\bf C~53}, 1483 (1996).


\bibitem{Argonne} R. B. Wiringa, V. G. J. Stoks, and R.~Schiavilla,
Phys. Rev.  {\bf C~51}, 38 (1995).

\bibitem{Stocks}  V. G. J. Stoks, R. A. M. Klomp, C. P. F. Terheggen, and
J. J. de Swart, Phys. Rev.  {\bf C~49}, 2950 (1994).

\bibitem{TM} S. A. Coon, M.~D.~Scadron, P.~C.~McNamee, B.~R.~Barrett,
D.~W.~E.~Blatt, and  B.~H.~J.~McKellar, Nucl. Phys. {\bf A~317}, 242 (1979);
J.~L.~Friar, D.~H\"uber, and  U.~van Kolck, Phys. Rev.  {\bf C~59}, 53 (1999);
 D.~H\"uber, J.~L.~Friar, A.~Nogga, H.~Witala, and  U.~van Kolck, Few
 Body Syst. {\bf 30}, 95 (2001).


\bibitem{Argonne-3}  B. S. Pudliner, V. R. Pandharipande, J. Carlson,
S.~C.~Pieper, and R.~B.~Wiringa, Phys. Rev. {\bf C~56}, 1720 (1997);
R.~B.~Wiringa, Nucl. Phys. {\bf A~631}, 70c (1998).

\bibitem{alpha-3} A. Nogga, H. Kamada, and W.~Gl\"ockle,
Phys. Rev. Lett. {\bf 85}, 944 (2000).


\bibitem{trinucl}  A. Picklesimer, R. A. Rice, and R. Brandenburg, Phys.
Rev. Lett. {\bf 68}, 1484 (1992); Phys. Rev. {\bf C~45}, 547 (1992);
Phys. Rev. {\bf C~45}, 2045 (1992); Phys. Rev. {\bf C~45}, 2624
(1992); Phys. Rev.   {\bf C~46}, 1178 (1992).

\bibitem{vanKolck} S.~R.~Beane, P.~F.~Bedaque, W.~C.~Haxton,
D.~R.~Phillips, M.~J.~Savage, 
in: M.~Shifman (Ed.), {\em At the Frontier of Particle Physics},
Vol.~1, 133 (World Scientific); nucl-th/0008064.

\bibitem{Refs:Machleidt} D.~R.~Entem, R.~Machleidt, Phys. Lett. 
{\bf B~524}, 93 (2002).


\bibitem{Bench} H. Kamada, A. Nogga, W.~Gl\"ockle, E.~Hiyama, M.~Kamimura,
K.~Varga, Y.~Suzuki, M.~Viviani, A.~Kievsky, S.~Rosati, J.~Carlson,
S.~C.~Pieper, R.~B.~Wiringa, P.~Navr\'atil, B.~R.~Barrett, N.~Barnea,
W.~Leidemann, and G.~Orlandini,  Phys. Rev. {\bf C~64}, 044001 (2001).


\bibitem{Vary} D.~C. Zheng, J. P. Vary, and B.~R.~Barrett, Phys. Rev.
{\bf C~50}, 2841 (1994); D.~C.~Zheng, J.~P.~Vary,  B.~R.~Barrett,
W.~C.~Haxton, and C.~L.~Song, Phys. Rev. {\bf C~52}, 2488 (1995).



\bibitem{Vary3}  P.~Navr\'atil, J.~P.~Vary, and B.~R.~Barrett,
Phys. Rev. Lett. {\bf 84}, 5728 (2000);
Phys. Rev. {\bf C~62}, 054311 (2000).


\bibitem{Ztmf1}
S.~A.~Zaytsev, Teoret. Mat. Fiz. {\bf 115}, 263 (1998) [Theor.
Math. Phys. {\bf 115}, 575 (1998)].

 \bibitem{Zret} S.~A.~Zaytsev, in
{\em Proc. XIV Int. Workshop on High Energy Physics and Quantum
Field Theory, Moscow 1999)} (Ed. B.B.Levchenko and
V.I.Savrin), 666 (Moscow, MSU-Press, 2000); 
Teoret. Mat. Fiz. {\bf 121}, 424 (1999) [Theor. Math. Phys. {\bf
121}, 1617 (1999)].


\bibitem{Zret2} S.~A.~Zaitsev and  E.~I.~Kramar J. Phys. {\bf G~27},
2037 (2001).


\bibitem{PHT} Yu.~A.~Lurie and A.~M.~Shirokov,  Izv. Ros. Akad. Nauk,
Ser. Fiz. {\bf 61}, 2121 (1997) [Bull.
Rus. Acad. Sci., Phys. Ser. {\bf 61}, 1665 (1997)].

\bibitem{Halo-b}  Yu.~A.~Lurie and A.~M.~Shirokov, 
to be published in A.~D.~Alhaidari, E.~J.~Heller, H.~A.~Yamani, and
M.~S.~Abdelmonem (eds.), 
{\em $J$-matrix method and its applications}
(Nova Science Publishers, Inc.).

\bibitem{Halo-c}  Yu.~A.~Lurie and A.~M.~Shirokov, nucl-th/0312028; to
be published in Ann. Phys.

\bibitem{Dole} P. Doleschall and I. Berb\'ely, Phys. Rev. {\bf C~62},
054004 (2000).
\bibitem{Plessas} P. Doleschall, I. Berb\'ely,  Z. Papp, and
W.~Plessas,   Phys. Rev. {\bf C~67},
064005 (2003).

\bibitem{Kuo-rev}  S. K. Bogner, T. T. S. Kuo, and A.~Schwenk,
Phys. Rep. {\bf 386}, 1 (2003).


\bibitem{Kuo2} S.~Bogner, T. T. S. Kuo, L.~Coraggio, A.~Covello, and
N.~Itaco, Phys. Rev. {\bf C~65}, 051301 (2002). 




\bibitem{Yamani} H.~A.~Yamani, L.~ Fishman,  J. Math. Phys.,
{\bf 16}, 410 (1975).


\bibitem{Broad} J. T. Broad and W. P. Reinhardt, Phys. Rev. {\bf A~14}, 2159
(1976); J. Phys. {\bf B~9}, 1491 (1976).

\bibitem{St-Uzh} A.~M.~Shirokov, Yu. F. Smirnov, and L.~Ya.~Stotland,
in {\em Proc. XIIth Europ. Conf. on Few-Body Phys.,  Uzhgorod, USSR,
1990} (Ed. V.~I.~Lengyel and M.~I.~Haysak), 173 (Uzhgorod, 1990).
\bibitem{St-izv} Yu. F. Smirnov, L. Ya. Stotland, and  A.~M.~Shirokov,
Izv. Akad. Nauk SSSR, Ser. Fiz.  {\bf  54}, No~5, 897 (1990)
[Bull. Acad. Sci. USSR, Phys. Ser. {\bf  54}, No~5, 81 (1990)].

\bibitem{Konov} D. A. Konovalov and I. A. McCarthy, J. Phys. {\bf B27}, L407
(1994); J. Phys. {\bf B27}, L741 (1994).

\bibitem{Fil} G. F. Filippov and I. P. Okhrimenko, Yad. Fiz. {\bf 32}, 932
(1980) [Sov. J. Nucl. Phys. {\bf 32}, 480 (1980)]; G.~F.~Filippov, Yad.
Fiz. {\bf 33}, 928 (1981) [Sov. J. Nucl. Phys. {\bf 33}, 488 (1981)].

\bibitem{NeSm} Yu. F. Smirnov and Yu. I. Nechaev, Kinam {\bf 4}, 445 (1982);
Yu. I. Nechaev and Yu.~F.~Smirnov, Yad. Fiz. {\bf 35}, 1385 (1982) [Sov. J.
Nucl. Phys. {\bf 35}, 808 (1982)].

\bibitem{FVCh} G. F. Filippov, V. S. Vasilevski, and L.~L.~Chopovski, Fiz.
Elem. Chastits At. Yadra {\bf 15}, 1338 (1984); {\bf 16}, 349 (1985)
[Sov. J. Part. Nucl. {\bf 15}, 600 (1984); {\bf 16}, 153 (1985)].
\bibitem{Rev} G.~F.~Filippov, Rivista Nuovo Cim. {\bf 12}, 1 (1989).
\bibitem{M1} V.~A.~Knyr, A.~I.~Mazur, and Yu.~F.~Smirnov, Yad. Fiz. {\bf 52},
754 (1990) [Sov. J. Nucl. Phys. {\bf 52}, 483 (1990)].
\bibitem{M2} V.~A.~Knyr, A.~I.~Mazur, and Yu.~F.~Smirnov, Yad. Fiz.
{\bf 54}, 1518 (1991) [Sov. J. Nucl. Phys. {\bf 54}, 927 (1991)].


\bibitem{SmSh} Yu. F. Smirnov and A. M. Shirokov, Preprint ITF-88-47R
(Kiev, 1988); A. M. Shirokov, Yu. F. Smirnov, and S. A. Zaytsev,
in {\em Modern Problems in Quantum Theory} (Ed. V.~I.~Savrin and
O.~A.~Khrustalev),  
(Moscow, 1998), 184;  Teoret. Mat. Fiz. {\bf 117}, 227 (1998)
[Theor. Math. Phys. {\bf 117}, 1291 (1998)].

\bibitem{Mikh} T. Ya. Mikhelashvili, Yu.~F.~Smirnov, and A.~M.~Shirokov,
Yad. Fiz. {\bf 48}, 969 (1988) [Sov. J. Nucl. Phys. {\bf 48}, 617
(1988)]; J. Phys. {\bf G~16}, 1241 (1990).
 
\bibitem{Zeit} D. E. Lanskoy, Yu. A. Lurie, and A. M. Shirokov,
Z. Phys. {\bf A~357}, 95 (1997).

\bibitem{Bang}
J.~M.~Bang, A.~I.~Mazur, A.~M.~Shirokov, Yu.~F.~Smirnov, and
S.~A.~Zaytsev, Ann. Phys. (NY), {\bf 280}, 299 (2000).


\bibitem{Abram} M. Abramowitz and I. A. Stegun (eds.), {\em Handbook on
Mathematical Functions} (Dover, New York, 1972).


\bibitem{Babikov} V. V. Babikov, {\em Method of Phase Functions in
Quantum Mechanics} (Nauka Publishers, Moscow,  1976).

\bibitem{Stapp} H.~P.~Stapp, T.~I.~Ypsilantis, and N.~Metropolis,
Phys. Rev. {\bf 105}, 302 (1957).


 \bibitem{BZP} A. I. Baz, Ya. B. Zeldovitch, and A.~M.~Perelomov,
{\em Scattering,
Reactions and Decays in Non-relativistic Quantum Mechanics} (Nauka
Publishers, Moscow, 1971).

\bibitem{BBD} L. D. Blokhintsev, I.~Borbely, and \'E.~I.~Dolinskii,
Sov. J. Part. Nucl. 8, 485 (1977). 


\bibitem{deSwart} J. J. de Swart, C. P. F. Terheggen, and
V.~G.~J.~Stoks, {\em Invited talk by J.~J.~de~Swart at the
3$^{\mbox{rd}}$ Int. Symp. ``Dubna Deuteron~95'', Dubna,
Russia, July 4--7, 1995};
nucl-th/9509032.

\bibitem{Petrov} V. A. Babenko and N. M. Petrov, 
Yad. Fiz. {\bf 66}, 1359 (2003) [Phys. At. Nucl. {\bf 66}, 1319 (2003)]. 

\bibitem{LANL} LANL T-2 Nucl. Information Service,
http://t2.lanl.gov/data/map.html.


\bibitem{Poly} W. N. Polyzou and W. Gl\"ockle, Few-Body Syst. {\bf 9},
97 (1990).

\bibitem{RMS}  W. N. Polyzou, Phys.Rev. {\bf C~58}, 91 (1998).

\bibitem{Qdeu} R. V. Reid, Jr. and M.~L.~Vaida, Phys. Rev. Lett. 
{\bf 29},  494 (1972).




\end{thebibliography}
\end{document}